\documentclass[useAMS,usenatbib]{mn2e}
\usepackage{graphicx}
\usepackage{rotating}
\usepackage{lscape}
\usepackage{makeidx}
\usepackage{listings}
\usepackage{longtable}
\usepackage{array}
\usepackage{aas_macros}
\usepackage{subfigure}
\usepackage{times}
\usepackage{hyperref}
\usepackage{tikz}
\title[The Largest X-ray Selected Sample of $z>3$ AGNs]{The Largest X-ray Selected Sample of $z>3$ AGNs: C-COSMOS \& ChaMP}

\author[E. Kalfountzou et al.]
         {E. Kalfountzou$^{1,2}$\thanks{Email: ekalfountzou@cfa.harvard.edu}, F. Civano$^{3,1}$, M. Elvis$^{1}$, M. Trichas$^{4}$, P. Green$^{1}$
\\ 
\footnotesize
	$^1$ Harvard Smithsonian Center for Astrophysics, 60 Garden St., Cambridge, MA 02138, USA\\
	$^2$ Centre for Astrophysics, Science $\&$ Technology Research Institute, University of Hertfordshire, Hatfield, Herts, AL10 9AB, UK\\
	$^3$ Yale Center for Astronomy and Astrophysics, 260 Whitney ave, New Haven, CT \\
	$^4$Airbus Defence $\&$ Space, Gunnels Wood Road, Stevenage, SG1 2AS, UK}

\begin{document}
\date{Received May 29, 2014; accepted August 25, 2014}

\pagerange{\pageref{firstpage}--\pageref{lastpage}} \pubyear{2014}
\maketitle
\label{firstpage}

\begin{abstract}
\\
We present results from an analysis of the largest high-redshift ($z>3$) X-ray-selected active galactic nucleus (AGN) sample to date, combining the {\it Chandra} C-COSMOS and ChaMP surveys and doubling the previous samples. The sample comprises 209 X-ray-detected AGN, over a wide range of rest frame 2-10 keV luminosities log\,$L_{X}=43.3 - 46.0~ {\rm erg~s^{-1}}$. X-ray hardness rates show that $\sim$39\% of the sources are highly obscured, $N_{H}>10^{22}{\rm cm^{-2}}$, in agreement with the $\sim$37\% of type-2 AGN found in our sample based on their optical classification. For $\sim$26\% of objects have mismatched optical and X-ray classifications. Using the $1/V_{max}$ method, we confirm that the comoving space density of all luminosity ranges of AGNs decreases with redshift above $z > 3$ and up to $z \sim 7$.  With a significant sample of AGN ($N=27$) at $z>4$, it is found that both source number counts in the 0.5 -2 keV band  and comoving space density are consistent with the expectation of a luminosity dependent density evolution (LDDE) model at all redshifts, while they exclude the luminosity and density evolution (LADE) model. The measured comoving space density of type-1 and type-2 AGN shows a constant ratio between the two types at $z>3$. Our results for both AGN types at these redshifts are consistent with the expectations of LDDE model.
\end{abstract}


\section{INTRODUCTION}

Active galactic nuclei (AGN) evolution at high redshifts, before their density peak, illuminates the role of AGN in the formation and co-evolution of galaxies and their central supermassive black holes (SMBHs) during the time of rapid SMBH growth.  The so-called `downsizing' evolution has been revealed for both AGN \citep[e.g.][]{Ueda2003,Hasinger2005,Aird2010} and galaxies \citep[e.g.][]{Cowie1996,Kodama2004,Damen2009}. Supporting this idea, X-ray surveys have shown that the number density of luminous AGN peaks at higher redshifts than less luminous ones \citep[e.g.][]{Ueda2003,Aird2010}. This sort of cosmological `co-evolution' scenario is inferred from the tight correlation exists locally between SMBH mass and galactic bulge properties \citep[e.g.][]{Magorrian1998, Ferrarese2000, Gebhardt2000,McConnell2013}.

To elucidate the co-evolution of SMBH and galaxies \citep[e.g.][]{Granato2001,Granato2004,Croton2006,Hopkins2006,Menci2008,Trichas2009,Trichas2010,Kalfountzou2011,Kalfountzou2012,Kalfountzou2014}, the accretion activity in the Universe has to be studied both at high redshifts and for low luminosities. This requires large samples of AGN spanning wide ranges of properties.  While many optical surveys have investigated the space density of high-redshift AGN \citep[e.g.][]{Richards2006, Jiang2009, Willott2010,Glikman2011,Ikeda2011,Ross2013}, the results are still controversial due to their inevitable incompleteness, especially at the faint luminosity end due to the host contamination, and the bias against obscured sources. Contrary to the optical surveys, X-ray observations are less contaminated by the host galaxy emission  and include AGN populations with a wide range of neutral hydrogen column density.

For the investigation of the absorption evolution \citep[e.g.][]{Ueda2003,Hasinger2008,Draper2010}, X-ray selected samples include all types of AGN (e.g. type-1/unobscured and type-2/obscured) and provide reduced obscuration bias in comparison with optically selected AGN. Although X-ray surveys have inferred the existence of an anti-correlation between the obscured AGN fraction and the luminosity, several of these studies have suggested an increase of the fraction toward higher redshift from $z = 0$ to $z \sim 2$ with limited samples at $z>3$ \citep[e.g.][]{LaFranca2005, Ballantyne2006,Treister2006,Ballantyne2008,Hiroi2012}. 

However, the evolution of AGN is still rife with uncertainty. On the basis of hard X-ray surveys many studies agreed that the XLF of AGN is best described by a luminosity dependent density evolution (LDDE) model \citep[e.g.][]{Ueda2003,Gilli2007, Silverman2008,Ueda2014}. \cite{Aird2010} preferred instead a luminosity and density evolution model (LADE). In LADE, the shift in the redshift peak of the AGN space density versus X-ray luminosity is much weaker than in LDDE models, yet gives a similarly good fit to their data.  While the $z<2$ downsizing behavior is common to both models, quite different numbers of AGN are predicted at higher redshifts ($z \geq 3$). 

X-ray surveys (2-10 keV) are now sensitive enough to sample the bulk of the $z>3$ AGN population. Two studies have been performed on high-redshift AGN exploiting the deep X-ray surveys in the Cosmological Evolution Survey (COSMOS) field carried out with XMM-Newton ($N_{\rm AGN}=40$; \citealp{Brusa2009}) and {\it Chandra} ($N_{\rm AGN}=81$; \citealp{Civano2011}), limited to $2 - 10$ keV luminosities $L_{\rm 2-10 keV} > 10^{44.2}~{\rm erg~ s^{-1}}$ and $10^{43.5}~{\rm erg~ s^{-1}}$, respectively. A more recent study based on the 4 Ms {\it Chandra} Deep Field South (CDF-S, \citealp{Xue2011}) was able to investigate the evolution of $z>3$ AGN down to $L_{X} \sim 10^{43}~{\rm erg~ s^{-1}}$ ($N_{\rm AGN}=34$; \citealp{Vito2013}). These results are consistent with a decline of the AGN space density at $z > 3$, but the shape of this decline remains highly uncertain at $z > 4$. 
To overcome these limitations, in this work we combined the two largest samples of $z > 3$ X-ray selected AGN with spectroscopic redshifts, both derived from Chandra X-ray Observatory (Weisskopf et al. 2002) surveys:  the wide but shallow ChaMP survey \citep{Kim2007a,Green2009}, and the deeper but narrower C-COSMOS survey \citep{Elvis2009}. This combination results in the largest X-ray AGN sample with $N_{\rm AGN}=211$ at $z>3$ and $N_{\rm AGN}=27$ at $z>4$. At the same time, by combining two surveys with different flux limits, we are able to determine the density evolution of both low luminosity ($L_{X} < 10^{44}~{\rm erg~ s^{-1}}$) and high luminosity AGN. Our sample includes both obscured and unobscured AGN, and their separate evolution has been determined.

The paper is structured as follows. In Section 2, we discuss the the data sets used in this work and the selection of the high-$z$ sample. In Section 3, we present the optical and X-ray properties of the selected high-$z$ AGN sample and we explain the AGN type classification using X-ray or optical data. In Sections 4 and 5, the number counts and space density of the sample are compared with model predictions. Section 6 summarizes the conclusions. A cosmological model with $\Omega_{o} = 0.3$, $\lambda_{o} = 0.7$, and a Hubble constant of $70~{\rm  km~s^{-1}~ Mpc^{-1}}$ is used throughout \citep{Spergel2003}. Errors are quoted at the $1\sigma$ level.

\section{Sample selection}

\begin{figure}
\hspace{-0.5cm}
\includegraphics[scale=0.26]{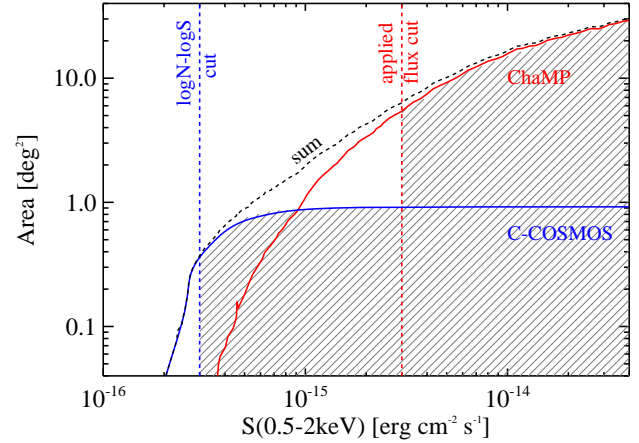}
\caption{Sky area vs. X-ray flux sensitivity curves for the C-COSMOS (blue solid line) and ChaMP/SDSS (red solid line) samples and the total area (black dashed line). The vertical blue dashed line indicates the flux corresponding to 10\% of the total C-COSMOS area (see section 4). The vertical red dashed line indicates the ChaMP X-ray flux limit with $>75$\% completeness from SDSS/UKIDSS/WISE (see section 2.2). The total area, after the applied cuts, used for this work is represented by the shadowed grey area.}
\label{fig:Area_flux}
\end{figure}

The high redshift AGN sample used in this work has been selected from the C-COSMOS X-ray catalog, combining the spectroscopic and photometric information available from the identification catalog of X-ray C-COSMOS sources \citep{Civano2011,Civano2012} and the ChaMP X-ray catalog using only the 323 ChaMP obsids overlapping with SDSS DR5 imaging. In Figure~\ref{fig:Area_flux}, we show the sky coverage (the area of a survey that is sensitive to sources above a given X-ray flux) using the observed soft band (0.5-2 keV) source detections for the two surveys, and their sum. This corresponds to 2-8 keV rest frame for $z>3$. 

A schematic diagram of the sample selection with the detailed number of sources for each step is presented in Figure~\ref{fig:sample_diagram}.

\begin{figure*}
\begin{picture}(500,540)(0,20)

\setlength{\fboxsep}{10pt}  
\put(160,560){\makebox(0,0)[t]{\fcolorbox{black}{blue!20}{\shortstack{\large  C-COSMOS\\ \large 1,761\\X-ray detections}}}}
\setlength{\fboxsep}{5pt}   
\put(349,560){\makebox(0,0)[t]{\fcolorbox{black}{red!20}{\shortstack{\large ChaMP\\ \large 15,202\\X-ray detections\\(exclude target sources)}}}}
\put(197,535){\line(1,0){108}}
\put(250,535){\line(0,-1){40}}
\put(22,500){\line(1,0){440}}

\put(22,500){\line(0,-1){20}}
\setlength{\fboxsep}{10pt}   
\put(22,480){\makebox(0,0)[t]{\fcolorbox{black}{blue!20}{\large{\shortstack{870\\$z_{spec}$}}}}}
\put(100,500){\line(0,-1){20}}
\put(100,480){\makebox(0,0)[t]{\fcolorbox{black}{blue!20}{\large{\shortstack{891\\$z_{phot}$}}}}}
\put(175,500){\line(0,-1){20}}
\setlength{\fboxsep}{9pt}   
\put(175,480){\makebox(0,0)[t]{\fcolorbox{black}{red!20}{\large{\shortstack{1,547\\$z_{spec}$}}}}}
\put(250,500){\line(0,-1){20}}
\put(250,480){\makebox(0,0)[t]{\fcolorbox{black}{red!20}{\large{\shortstack{1,654\\$z_{phot}$}}}}}
\put(350,500){\line(0,-1){20}}
\setlength{\fboxsep}{4pt}   
\put(350,480){\makebox(0,0)[t]{\fcolorbox{black}{red!20}{\shortstack{\large 7,759\\SDSS (S); UKIDSS (U);\\WISE (W) counterparts}}}}

\put(462,500){\line(0,-1){20}}
\put(462,480){\makebox(0,0)[t]{\fcolorbox{black}{red!20}{\shortstack{\large 4,242\\without (S,U,W)\\counterparts}}}}
\put(462,444){\line(0,-1){23}}
\setlength{\fboxrule}{1pt}
\put(462,420){\makebox(0,0)[t]{\fcolorbox{black}{red!20}{\shortstack{1,164\\$>$ flux cut}}}}
\setlength{\fboxrule}{0.2pt}

\put(22,442){\line(0,-1){22}}
\put(100,442){\line(0,-1){22}}
\put(22,420){\line(1,0){78}}
\put(61,420){\line(0,-1){20}}
\setlength{\fboxsep}{8pt}   
\put(61,400){\makebox(0,0)[t]{\fcolorbox{black}{blue!20}{\shortstack{122 high-$z$\\AGN}}}}
\put(61,367){\line(0,-1){217}}
\setlength{\fboxsep}{4pt}   
\put(61,345){\line(1,0){20}}
\put(100,355){\makebox(0,0)[t]{\fcolorbox{black}{blue!20}{\shortstack{32\\$z_{spec}>3$}}}}
\put(61,303){\line(-1,0){20}}
\put(19,313){\makebox(0,0)[t]{\fcolorbox{black}{blue!20}{\shortstack{30\\$z_{phot}$+$1\sigma>$3}}}}
\put(61,261){\line(1,0){20}}
\put(100,271){\makebox(0,0)[t]{\fcolorbox{black}{blue!20}{\shortstack{45\\$z_{phot}>3$}}}}
\put(61,214){\line(-1,0){24}}
\put(19,229){\makebox(0,0)[t]{\fcolorbox{black}{blue!20}{\shortstack{15\\optical\\dropouts}}}}

\put(175,442){\line(0,-1){22}}
\setlength{\fboxsep}{6pt}   
\setlength{\fboxrule}{1pt}
\put(175,420){\makebox(0,0)[t]{\fcolorbox{black}{red!20}{\shortstack{1,123\\$>$ flux\\cut}}}}
\setlength{\fboxrule}{0.2pt}
\put(175,383){\line(0,-1){20}}
\setlength{\fboxsep}{4pt}   
\put(175,363){\makebox(0,0)[t]{\fcolorbox{black}{red!20}{\shortstack{44\\$z_{spec}>3$}}}}
\put(175,339){\line(0,-1){189}}

\put(250,442){\line(0,-1){22}}
\setlength{\fboxsep}{6pt}   
\put(250,420){\makebox(0,0)[t]{\fcolorbox{black}{red!20}{\shortstack{1,611\\$good\ge 0.0$}}}}
\put(250,390){\line(0,-1){20}}
\setlength{\fboxrule}{1pt}
\put(250,370){\makebox(0,0)[t]{\fcolorbox{black}{red!20}{\shortstack{1,495\\$>$ flux cut}}}}
\setlength{\fboxrule}{0.2pt}
\put(250,341){\line(0,-1){20}}
\put(250,321){\makebox(0,0)[t]{\fcolorbox{black}{red!20}{\shortstack{28\\high-$z$}}}}
\put(250,293){\line(0,-1){143}}
\setlength{\fboxsep}{4pt}   
\put(250,275){\line(1,0){20}}
\put(285,285){\makebox(0,0)[t]{\fcolorbox{black}{red!20}{\shortstack{15\\$z_{phot}>3$}}}}
\put(250,250){\line(-1,0){20}}
\put(210,260){\makebox(0,0)[t]{\fcolorbox{black}{red!20}{\shortstack{13\\$z_{phot}$+$1\sigma>$3}}}}

\put(350,443){\line(0,-1){22}}
\setlength{\fboxsep}{6pt}   
\setlength{\fboxrule}{1pt}
\put(350,420){\makebox(0,0)[t]{\fcolorbox{black}{red!20}{\shortstack{5,683\\$>$ flux cut}}}}
\setlength{\fboxrule}{0.2pt}
\put(350,391.5){\line(0,-1){12}}
\put(350,380){\makebox(0,0)[t]{\fcolorbox{black}{red!20}{\shortstack{1,508\\point-like\\good photometry}}}}
\put(350,340){\line(0,-1){105}}
\setlength{\fboxsep}{3pt}   
\put(350,325){\line(1,0){13}}
\put(380,335){\makebox(0,0)[t]{\fcolorbox{black}{red!20}{\shortstack{177\\S+U+W}}}}
\put(350,313){\line(-1,0){12}}
\put(327,323){\makebox(0,0)[t]{\fcolorbox{black}{red!20}{\shortstack{526\\S+W}}}}
\put(350,301){\line(1,0){13}}
\put(373,311){\makebox(0,0)[t]{\fcolorbox{black}{red!20}{\shortstack{44\\S+U}}}}
\put(350,289){\line(-1,0){12}}
\put(327,299){\makebox(0,0)[t]{\fcolorbox{black}{red!20}{\shortstack{16\\U+W}}}}
\put(350,277){\line(1,0){11}}
\put(370,287){\makebox(0,0)[t]{\fcolorbox{black}{red!20}{\shortstack{367\\S}}}}
\put(350,265){\line(-1,0){11}}
\put(330,275){\makebox(0,0)[t]{\fcolorbox{black}{red!20}{\shortstack{359\\W}}}}
\put(350,253){\line(1,0){11}}
\put(368,263){\makebox(0,0)[t]{\fcolorbox{black}{red!20}{\shortstack{19\\U}}}}

\setlength{\fboxsep}{6pt}   
\put(350,238){\makebox(0,0)[t]{\fcolorbox{black}{red!20}{\shortstack{15\\high-$z$}}}}
\put(350,210){\line(0,-1){60}}
\setlength{\fboxsep}{4pt}   
\put(350,192){\line(1,0){20}}
\put(385,203){\makebox(0,0)[t]{\fcolorbox{black}{red!20}{\shortstack{13\\$z_{phot}>3$}}}}
\put(350,178){\line(-1,0){20}}
\put(310,188){\makebox(0,0)[t]{\fcolorbox{black}{red!20}{\shortstack{2\\$z_{phot}$+$1\sigma>$3}}}}

\put(61,150){\line(1,0){289}}

\put(210,170){\makebox(0,0)[t]{\fcolorbox{black}{white!20}{\shortstack{\large 209\\high-$z$\\X-ray selected\\AGN}}}}
\put(210,125){\line(0,-1){69}}
\setlength{\fboxsep}{5pt}   
\put(210,107){\line(1,0){20}}
\put(250,120){\makebox(0,0)[t]{\fcolorbox{black}{white!20}{\shortstack{\large 76\\$z_{spec}>3$}}}}
\put(210,90){\line(-1,0){20}}
\put(170,103){\makebox(0,0)[t]{\fcolorbox{black}{white!20}{\shortstack{\large 73\\$z_{phot}>3$}}}}
\put(210,72){\line(1,0){20}}
\put(262,85){\makebox(0,0)[t]{\fcolorbox{black}{white!20}{\shortstack{\large 45\\$z_{phot}+1\sigma>3$}}}}
\put(210,56){\line(-1,0){20}}
\put(170,69){\makebox(0,0)[t]{\fcolorbox{black}{white!20}{\shortstack{\large 15\\optical\\dropouts}}}}

\bigskip
\bigskip
\smallskip

\end{picture}
\label{fig:sample_diagram}
\smallskip
\bigskip
\smallskip
\bigskip
\caption{Schematic flow diagram of the high-$z$ sample selection.}
\end{figure*}

\subsection{The C-COSMOS sample} \label{section:Cosmos_sample}

The \textit{Chandra}-COSMOS survey (C-COSMOS; \citealp{Elvis2009}; \citealp{Civano2012}) covers the central 0.9 ${\rm deg^{2}}$ of the COSMOS field up to a depth of 200 ksec in the inner 0.5 ${\rm deg^{2}}$, with the ACIS-I CCD imager \citep{Garmire2003} on board {\it Chandra}. The C-COSMOS X-ray source catalog comprises 1,761 point-like X-ray sources detected down to a maximum likelihood threshold detml=10.8 in at least one band. This likelihood threshold corresponds to a probability of $\sim 5\times 10^{-5}$ that a catalog source is instead a background fluctuation \citep{Puccetti2009}. Given this likelihood threshold, the flux limit reached in the survey is $5.7\times10^{-16}~{\rm erg~ cm^{-2}~s^{-1}}$ in the Full band (0.5-10 keV), $1.9\times10^{-16}~{\rm erg~ cm^{-2}~s^{-1}}$ in the Soft band (0.5-2 keV) and $7.3\times10^{-16}~{\rm erg~ cm^{-2}~s^{-1}}$ in the Hard band (2-10 keV). 

The $z>3$ C-COSMOS sample, as presented by \citep{Civano2011}, comprises 107 X-ray detected sources with available spectroscopic (32) and photometric (45) redshifts plus 30 sources with a formal $z_{phot} < 3$ but with a broad photometric redshift probability distribution, such that $z_{phot}  + 1\sigma_{phot} > 3$. All of the spectroscopic C-COSMOS sources have a quality flag 3 (2 sources) or 4 corresponding respectively to a secure redshift with two or more emission or absorption lines and a secure redshift with two or more emission or absorption lines with a good-quality, high S/N spectrum (see \citealp{Lilly2007,Lilly2009} for thorough explanation of quality flags). Tuned photometric redshifts for the C-COSMOS sources have been computed and presented in \cite{Salvato2011}. Due to the large number of photometric bands and the sizable spectroscopic training sample spanning a large range in redshift and luminosity the estimated photometric redshifts are expected to be quite robust at $z>2.5$ even at the fainter magnitudes ($i_{AB} > 22.5$). The COSMOS photometric redshifts for X-ray selected sources have an accuracy of $\sigma_{\Delta z / (1+z_{spec})}=0.015$ with a small fraction of outliers ($<6$\%), considering the sample as a whole at $i<22.5$. At fainter magnitudes, the dispersion increases to $\sigma_{\Delta z / (1+z_{spec})}=0.035$ with $\sim 15$\% outliers, still remarkably good for an AGN sample. For the $z>3$ C-COSMOS sample, an accuracy of $\sigma_{\Delta z / (1+z_{spec})}=0.014$ is achieved with only 3 catastrophic outliers ($<9$\%). The spectral energy distributions (SEDs) of the sources with photometric redshift larger than 3 have been visually inspected together with the photometric fitting and the probability distribution of all the possible solutions.

There are 91 sources selected in the 0.5-2 keV band, 14 in the 2-10 keV, and 4 in the 0.5-10 keV bands. There are 15 C-COSMOS sources without a counterpart in the optical bands, but with a K-band and IRAC (7), only IRAC (6) or no infrared detection (2). Given the small number of bands in which these objects are detected, no photometric redshift is available for them. In X-ray selected samples, non-detection in the optical band has been often assumed to be a proxy for high redshift \citep[e.g.][]{Koekemoer2004}, or for high obscuration, or a combination of both. Four of the 15 sources have no detection in the soft band suggesting high obscuration, possibly combined with high redshift. More details about the sample selection can be found in \cite{Civano2011} and are also presented in Figure~\ref{fig:sample_diagram}.

\subsection{The ChaMP sample}

The \textit{Chandra} Multi-wavelength Project (ChaMP) is a wide-area non-continuous X-ray survey based on archival X-ray images of the high Galactic latitude ($|b| > 20$ deg) sky observed with ACIS on \textit{Chandra}. The flux levels (in ${\rm erg~ cm^{-2}~s^{-1}}$) reached in the survey are $9.4\times10^{-16} - 5.9\times10^{-11}$ in the Full ($0.5 - 8$ keV), $3.7\times10^{-16} - 2.5\times10^{-11}$ ($0.5 - 2$ keV) in the Soft and $1.7\times10^{-15} - 6.7\times10^{-11}$ ($2 - 8$ keV) in the Hard band, respectively.  The ChaMP survey includes a total of 392 fields, omitting pointings from dedicated serendipitous surveys like C-COSMOS, the {\it Chandra} Deep Fields, as well as fields with extended ($>3'$) bright optical or X-ray sources. The list of {\it Chandra} pointings avoids any overlapping observations by eliminating the observation with the shorter exposure time. The survey has detected a total of $>$19,000 X-ray sources \citep{Kim2007a, Green2009} over $33~{\rm deg^{2}}$ with $\sim$15,350 X-ray sources positionally matched to SDSS optical counterparts \citep{Green2009}. 

The study of the X-ray detected AGN properties requires accurate
estimation of redshifts, luminosities and source classification thus,
good quality spectra or, when not available, multi-band
photometry. Hence for our X-ray analysis we chose only the 323 fields
overlapping with SDSS DR5 imaging for which the sensitivity curve is
given by \cite{Green2009}, to determine accurate number
counts. Optical spectroscopy of ChaMP X-ray sources was described by
\cite{Trichas2012}, where redshifts and classifications for a total of
1,569 {\it Chandra} sources are presented. Since the ChaMP is a {\it
  Chandra} archival survey, most ChaMP fields contain targeted sources
selected by the target's PI, and those targets are likely to be biased
toward special X-ray populations such as bright AGN. Of the targeted
sources $\sim90$\% have a secure spectroscopic redshift with 33 of
them having at $z>3$ and 29 at $z>4$ (see \citealp{Trichas2012}). The
high rate of high-redshift detected sources clearly shows the strong
selection biases that could affect our analysis if we included the
targeted sources. Therefore, we exclude all targeted sources (153) to
reduce bias in sample properties and source number counts.

For SDSS point sources with $i < 21$ and without available spectroscopy, efficient photometric selection of quasars is possible using a nonparametric Bayesian classification based on kernel density estimation as described in \cite{Richards2009}. To select high-$z$ candidates without available spectroscopic or photometric redshift, SDSS detection is required in at least the $i-$ and $z-$ bands, to detect Lyman dropouts \citep[e.g.][]{Steidel1996}.

Searching the ChaMP catalog for X-ray sources within $4''$ of the optical SDSS quasar coordinate (95\% of the matched sample has an X-ray/optical position difference of less than $3''$; see \citealp{Green2009}), yields $9,727$ unique matches ($\sim$63\% of the total ChaMP X-ray selected sample). We additionally searched for cross-matches in the Wide-field Infrared Survey Explorer (WISE; \citealp{Wright2010}) and UKIRT (UK Infrared Telescope) Infrared Deep Sky Survey (UKIDSS; \citealp{Warren2000, Hewett2006, Maddox2008}).\footnote{The UKIDSS project is defined in \cite{Lawrence2007}. UKIDSS uses the UKIRT Wide Field Camera (WFCAM; \citealp{Casali2007}) and a photometric system described in \cite{Hewett2006}. The pipeline processing and science archive are described in \cite{Hambly2008}.}

For a source to be included in the {\it WISE} All Sky Source Catalog
\citep{Wright2010}, a SNR$ >$5 detection was required for one of the
four photometric bands, $W1$, $W2$, $W3$ or $W4$, with central
wavelengths of roughly 3.4, 4.6, 12, and 22 $\mu$m, and angular
resolutions of 6.1, 6.4, 6.5 and 12.0 arcsec. Because of the different
spatial resolutions, 6.0 arcsec (WISE; $W1$) and 1-2 arcsec (SDSS), we
use $6''$ as the matching radius for WISE counterparts \citep{Wu2012}.  

Similarly, we searched the UKIDSS Large Area Survey (LAS;
\citealp{Lawrence2007}) Data Release 10 for NIR counterparts to
ChaMP X-ray sources. The photometric system is described in
\cite{Hewett2006}, and the calibration is described in
\cite{Hodgkin2009}. We used the LAS $Y J H K$ Source table, which
contains only fields with coverage in every filter and merges the
data from multiple detections of the same object. The X-ray source
catalogs were then matched within $3''$ of the X-ray position
separately to each UKIDSS band:  $Y$ (0.97-1.07 $\mu$m), $J$
(1.17-1.33 $\mu$m), $H$ (1.49-1.78 $\mu$m) and $K$ (2.03-2.37 $\mu$m)
recovering also the areas with coverage in a single UKIDSS band. The
individual band lists were then combined. For objects not
detected in a UKIDSS band we use the $5\sigma$ detection 
limits provided in \cite{Dye2006} of $Y = 20.23$, $ J = 19.52$, $H = 18.73$,
and $K = 18.06$. Matching the ChaMP catalog to WISE and UKIDSS, we
find 1,103 additional WISE and/or UKIDSS counterparts which do not
have a SDSS counterpart (the detailed numbers are reported in
Figure~\ref{fig:sample_diagram}). 

In summary, $\sim$70\% of the total ChaMP X-ray sample have SDSS and
UKIDSS/WISE photometry ($9,727$ SDSS and/or WISE and/or UKIDSS and
$1,103$ WISE and/or UKIDSS). The limited fraction of optical matches
shows how optical counterparts of faint X-ray sources are fainter than
the SDSS magnitude limit ($i=21.0$). SDSS quasars were identified to
$i< 19.1$ for spectroscopy by their UV-excess colors, with an
extension for $z > 3$ quasars to $i = 20.2$ using $ugri$ color
criteria \citep{Richards2002}.  

\begin{figure}
\hspace{-0.9cm}
\includegraphics[scale=0.62]{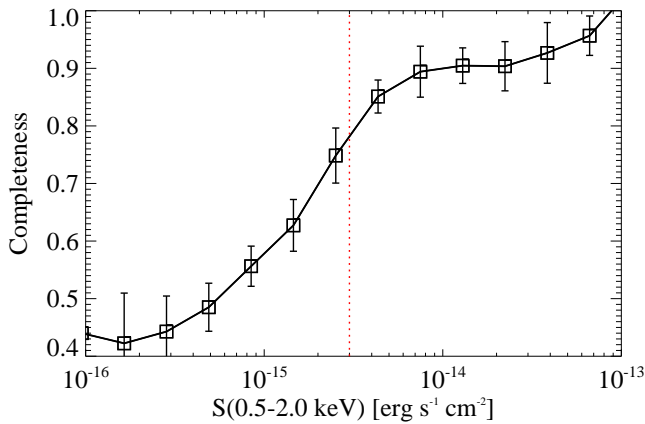}
\caption{Optical and/or infrared ChaMP survey completeness as a function of 0.5-2 keV X-ray flux. The red dotted line indicates the ChaMP X-ray flux cut with $>75$\% completeness from SDSS/UKIDSS/WISE.}
\label{fig:completeness}
\end{figure}

Based on the X-ray limits, the identification completeness of ChaMP X-ray sources falls rapidly for objects with fainter optical counterparts. Figure~\ref{fig:completeness} shows the optical (SDSS $i-$band counterparts) and infrared (WISE and UKIDSS counterparts) completeness of the X-ray selected sample as a function of the soft (0.5-2 keV) X-ray flux. This incompleteness can severely bias determination of the number counts and space density, particularly at high redshifts \citep[e.g.][]{Barger2005}. 

To address this issue, we set a relatively high X-ray flux limit in
ChaMP, where spectroscopic completeness is higher, and photometric
coverage allows good photometric redshifts. We use a soft flux limit
for ChaMP at $S_{0.5-2~{\rm keV}}>3\times 10^{-15}~{\rm
  erg~s^{-1}~cm^{-2}}$ as at these brighter fluxes the completeness is
higher than $\ge 75$\% (see Figure~\ref{fig:completeness}). The completeness fraction as a function of flux has been taken into account for the estimation of the number counts and comoving space density (see Sections~\ref{section:logNlogS} and \ref{section:space_density}). For sources not detected in the soft band, the 0.5-2 keV flux has been computed by converting the 2-10 keV flux using $\Gamma=1.8$ (see Section~\ref{section:xray_properties}). One of the main advantages of our compilation is that we do not miss the faint high redshift population, since this is recovered by the C-COSMOS survey. In this way, the ChaMP sample is used for the determination of the bright
end of the luminosity function at high redshifts.  

\subsubsection{Spectroscopic redshifts}	
We compiled secure spectroscopic redshifts for a total of 1,547 sources. We have used 1,056 sources (excluding target sources) from existing ChaMP spectroscopy \citep{Trichas2012} for the selected ChaMP fields. Additional spectroscopic redshifts are given in the SDSS-III ($N=91$; \citealp{Noterdaeme2012}) and SDSS-DR10 quasar catalogs ($N=145$; \citealp{Paris2014}). We also searched the literature by cross-correlating optical positions with the NASA Extragalactic Database (NED), using a $2''$ match radius where we found 255 more sources with spectroscopic redshift. 

The high redshift spectroscopic sample consists of 44 sources with
$z>3$. All of these sources have a soft band X-ray detection, and only
3 sources lack a hard band detection. Among them, there are 7 sources
with $z>4$ and 1 source with $z=6.016\pm0.005$ \citep{Jiang2007}. All but 6 of them have SDSS optical spectra with mean S/N $>4.5$
(none of them has S/N $<2.0$) with at least 2 broad emission lines
(Ly$\alpha$ and CIV) significantly detected. For 5 of the remaining
sources, redshifts have been obtained by \cite{Trichas2012} while
for the source with the highest spectroscopic redshift ($z=6.016$)
we have used the estimate from \cite{Jiang2007}. For 30 sources of
the ChaMP spectroscopic sample there are available photometric
redshifts derived by \cite{Richards2009} (see
 Section~\ref{section:SDSSphoto}) with an accuracy of $\sigma_{\Delta z / (1+z_{spec})}=0.013$ and only one catastrophic outlier.

\subsubsection{SDSS Photometric redshifts}	\label{section:SDSSphoto}

For the sources without spectroscopic redshifts, we derived
photometric redshifts. The criteria used in SDSS DR6 have now been
refined to include objects redder than ($u - g$)$ = 1.0$ which may
well be high-$z$ quasars. The resulting catalog of $\sim1$ million
photometrically identified quasars and their photometric redshifts
from  SDSS Data Release 6 (DR6)  is described in
\cite{Richards2009}. Only point sources (type=6) with $i-$band
magnitudes between 14.5 and (de-reddened)  21.3 (${\rm 
  psfmag}\_{i}>14.5$ \& ${\rm psfmag}\_{i}-{\rm
  extinction}\_{i}<21.3$; where ${\rm psfmag}$ are the point-spread-function magnitudes).  They estimate the overall efficiency of the catalog to be better than 72\%, with subsamples (e.g., X-ray detected objects) being as efficient as 97\%. At the faint limit of the catalog some additional galaxy contamination is expected. 

There are 1,611 sources with SDSS high quality photometric redshifts and no spectroscopic redshifts in ChaMP (i.e. those with $good\geq0.0$, where $good$ is the quality flag; 6=most robust; -6=least robust; \citealp{Richards2009}). Among them there are 14 sources with $z_{phot}>3$  and one with $z_{phot}>4$, above the adopted ChaMP flux limit. All of these sources are detected in both soft and hard band. The SDSS photo-$z$ code also gives a probability of an object being in a given redshift range. In this way, we have not only the most likely redshift but also the probability that the redshift is between some minimum and maximum value, which is crucial for dealing with catastrophic failures. The redshift probability distribution for each source is taken into account for the estimation of the number counts and comoving space density (see Sections~\ref{section:logNlogS} and \ref{section:space_density}). As for C-COSMOS selection of high-z sources, we also included 13 sources having $z_{photo}+1\sigma_{zphoto}>3$ and $z_{photo}<3$. This adds another 10 objects to the main sample, all of them detected in both soft and hard bands.

\subsubsection{High-$z$ candidate selection and photometric redshift estimation}

For the remaining 7,759 without a spectroscopic or photometric SDSS redshift, we selected the high-redshift AGN candidates using their optical and/or their infrared colors. Most of these sources ($\sim70$\%), despite being included in SDSS DR6 catalog, were rejected from \cite{Richards2009} selection criteria. The remaining sources come from later SDSS data releases. 

Following the same morphological criteria as \cite{Richards2009}, a candidate is required to be unresolved in images taken through the two redder filters (e.g. $g$ and $r$ for $z\sim3$ selection). This minimizes contamination from low-$z$ galaxies since even type-2 AGN at $z>3$ appear point like.  However, we avoid using any faint flux cut in order to insure that we do not miss faint high-$z$ candidates since non-detection can imply high-$z$ dropouts. We reject sources with flags indicating that their photometry may be problematic (e.g., blending of close pairs of objects, objects too close to the edge of the frame, objects affected by a cosmic-ray hit). Overall, we reject 5,079  non-point like sources or with problematic photometry. This number ($\sim65$\%) is in good agreement with the rejected number of sources by \cite{Richards2009} using the same criteria which explain the lack of a photometric redshift for these sources. 

Photometric redshift criteria must strike a quantifiable balance between completeness and efficiency, i.e., a probability can be
assigned both to the classification and the redshift. Using the SDSS, UKIDSS and WISE\footnote{We use the colors related to WISE W3 and W4  magnitudes only for sources lacking SDSS and/or UKIDSS detections
  because WISE uncertainties are substantially larger \citep{Wu2004}.}
photometric data can help us to select quasar candidates more
efficiently than using each survey individually (see
Table~\ref{Table:photo_reliab}). The photometric redshift reliability,
defined by \citep{Wu2012} as the fraction of the sources with the
difference between the photometric and spectroscopic redshifts smaller
than 0.2 is given in Table~\ref{Table:photo_reliab}. The highest
reliability can be reached only in the UKIDSS surveyed area, which is
much smaller (4,000 sq. deg) than the sky coverage of both SDSS and
WISE surveys.  

\begin{table}
\caption{Photometric redshift reliability defined by \citep{Wu2012} and number of sources for ChaMP sources without spectroscopic or SDSS photometric redshifts.}
\centering
\begin{minipage}{7.3cm}

\begin{tabular} {c c c c }

\hline

Surveys\footnote{S=SDSS; W=WISE; U=UKIDSS} & Reliability (\%)  & $N_{\rm obj}$\footnote{Number of point-like objects in each combination of surveys.} & $N_{\rm obj-lim}$\footnote{Number of point-like objects in each combination of surveys with $S_{0.5-2~{\rm keV}}>3\times 10^{-15}~{\rm erg~s^{-1}~cm^{-2}}$.}  \\
\hline

W		&		&	955	& 359	\\
U		&		&	27	& 19	\\
U+W	&	67.4		&	19 &  16	\\	
S		&	70.4	& 637	& 367		\\	
S+W	 & 77.2	&  733	&  526	\\	
S+U	&	84.8		&  71	&  44	\\	
S+U+W & 87.0 	& 238	&  177	\\	
\hline

Total	&		&	2,680	&	1,508  	\\	
\hline

\end{tabular}
\label{Table:photo_reliab}
\end{minipage}
\end{table}

\cite{Richards2002} used a 3D multi-color space to select high
redshift QSO candidates in SDSS:  $griz$ ($g$-$r$, $r$-$i$, $i$-$z$)
for candidates with $z > 3.0$. Following the SDSS group, we search for
high-$z$ candidates in three redshift intervals ($z\simeq3.0-3.5$,
$z\simeq3.5-4.5$, $z\simeq4.5$). The details of the selection criteria
are given in the Appendix. Our selection criteria require that our
sources lie outside of a $2\sigma$ region surrounding the stellar
locus. We still expect the sample to be contaminated by stars and
low-$z$ galaxies. For this reason, we use some additional criteria
described by \cite{Richards2002} to exclude objects in color regions
containing predominantly white dwarfs, A stars and unresolved red-blue
star pairs.  During the color selection process, no specific line is
drawn between optically selected quasars (type-1 AGN) and type-2
AGN. Taking into account that both type-1 and type-2 AGN are
unresolved in optical images at $z>3$ and type-2 AGN should lie
outside the stellar locus due to their red optical colors, we expect that the above criteria efficiently select both high-redshift AGN populations. We found 53 SDSS-detected high-$z$ candidates. 

To increase the reliability of the photometric estimation, we also
combine the SDSS selection with the redder baselines from UKIDSS and
WISE, where the contamination of the stellar locus and low-redshift
galaxies is lower. We used the combination of UKIDSS and SDSS colors
in the $Y-K$ versus $g-z$ color-color diagram suggested by
\citep{Wu2010} to efficiently separate quasars with redshift $z<4$
from stars. Similarly, \cite{Wu2012} suggested that $z-W1$ and $g-z$
colors could be used to separate stars
from quasars. Based on these criteria, we have rejected 10 sources
associated with stars based on both SDSS-UKIDSS and SDSS-WISE
color-color diagrams.  For sources detected only by UKIDSS we used the
$i=21.3$ upper limit and a $Y-K$ versus $i-Y$ color-color diagram to
separate stars and low-$z$ galaxies from high-$z$ candidates. We found
4 high-$z$ candidates. In the case of sources detected only by WISE
there is no efficient way detailed in the literature to separate
high-$z$ quasars from stars. 

Photometric redshifts have been estimated for the high-$z$ candidates
by comparing the observed colors with theoretical color-redshift
relations derived from samples with known redshifts
\citep{Richards2002,Wu2010,Wu2012}.  A standard $\chi^{2}$
minimization method is used to estimate the most probable photometric
redshifts. Here the $\chi^{2}$ is defined as (see \citealp{Wu2004}):

\begin{equation}
\chi^2=\sum_{ij}{\frac{[(m_{i,cz}-m_{j,cz})-(m_{i,ob}-m_{j,ob})]^2}
{\sigma_{m_{i, ob}}^2+\sigma_{m_{j,ob}}^2}},
\end{equation}
where the sum is obtained for all four SDSS colors and/or WISE and/or
UKIDSS colors, $m_{i,cz} - m_{j,cz}$ is the color in the
color-redshift relations, $m_{i,ob}-m_{j,ob}$ is the observed color of
a quasar, and $\sigma_{m_{i,ob}}$ and $\sigma_{m_{j,ob}}$ are the
uncertainties of observed magnitudes in two bands. The uncertainty in
the measurement was obtained by mapping the $\Delta \chi ^{2}$
error. Since the above studies are dominated by optically selected
quasars, we would expect that the photometric redshifts uncertainties
in type-1 AGN are smaller. However, since the Ly$\alpha$ break enters
the $g$ band at $z\sim 3.5$, the $g-r$ colors quickly redden with
redshift for both populations. \cite{Alexandroff2013} found that $g-r$
colors are indistinguishable at a 84\% confidence level between type-1
and type-2 quasars at $z>2$ suggesting that even in the case of type-2
AGN the photometric redshifts are reliably estimated. Overall, we
found 8 sources with $z>3$ at greater than $1\sigma$ significance, 4 sources with $z>3$ but lower than $1\sigma$ significance, and 2 sources with $z_{phot}+1\sigma_{phot}>3$.

\subsubsection{The ChaMP high-$z$ sample}

The total $z >3$ ChaMP sample includes 87 sources with $z > 3$. Among
them there are 44 sources with secure spectroscopic redshift, 15
sources with SDSS $z_{phot}>3$ and 13 sources with SDSS
$z_{phot}+1\sigma_{phot}>3$ available from \cite{Richards2009}, and 15
sources with estimated photometric redshifts based on optical/infrared
color - redshift relations (13 with $z_{phot}>3$ and 2 with
$z_{phot}+1\sigma_{phot}>3$).

\section{The C-COSMOS \& ChaMP $z > 3$ AGN Sample}

\begin{figure}
\includegraphics[trim=10 5 1 1, clip, scale=0.38]{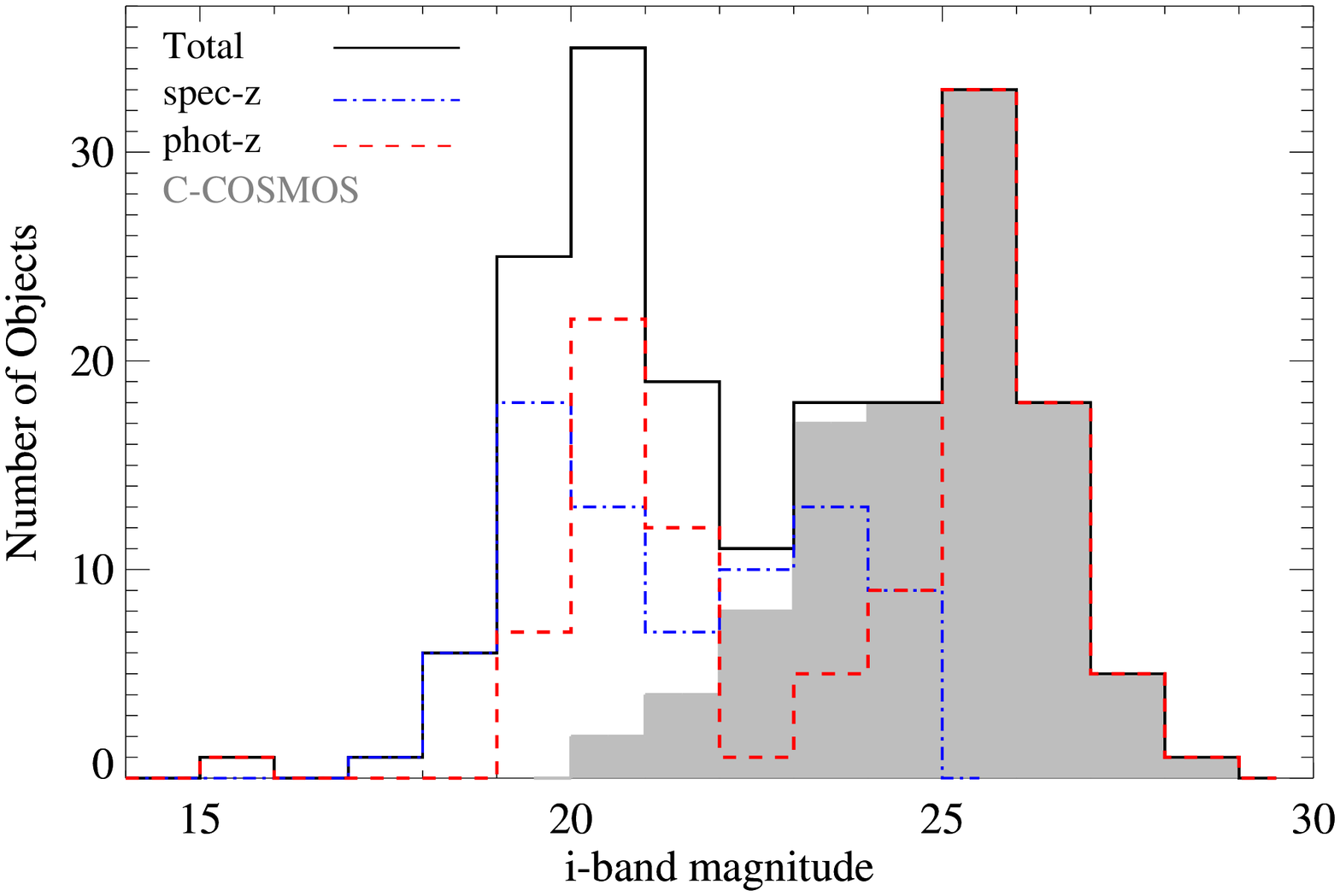}
\includegraphics[trim=10 5 1 20, clip, scale=0.38]{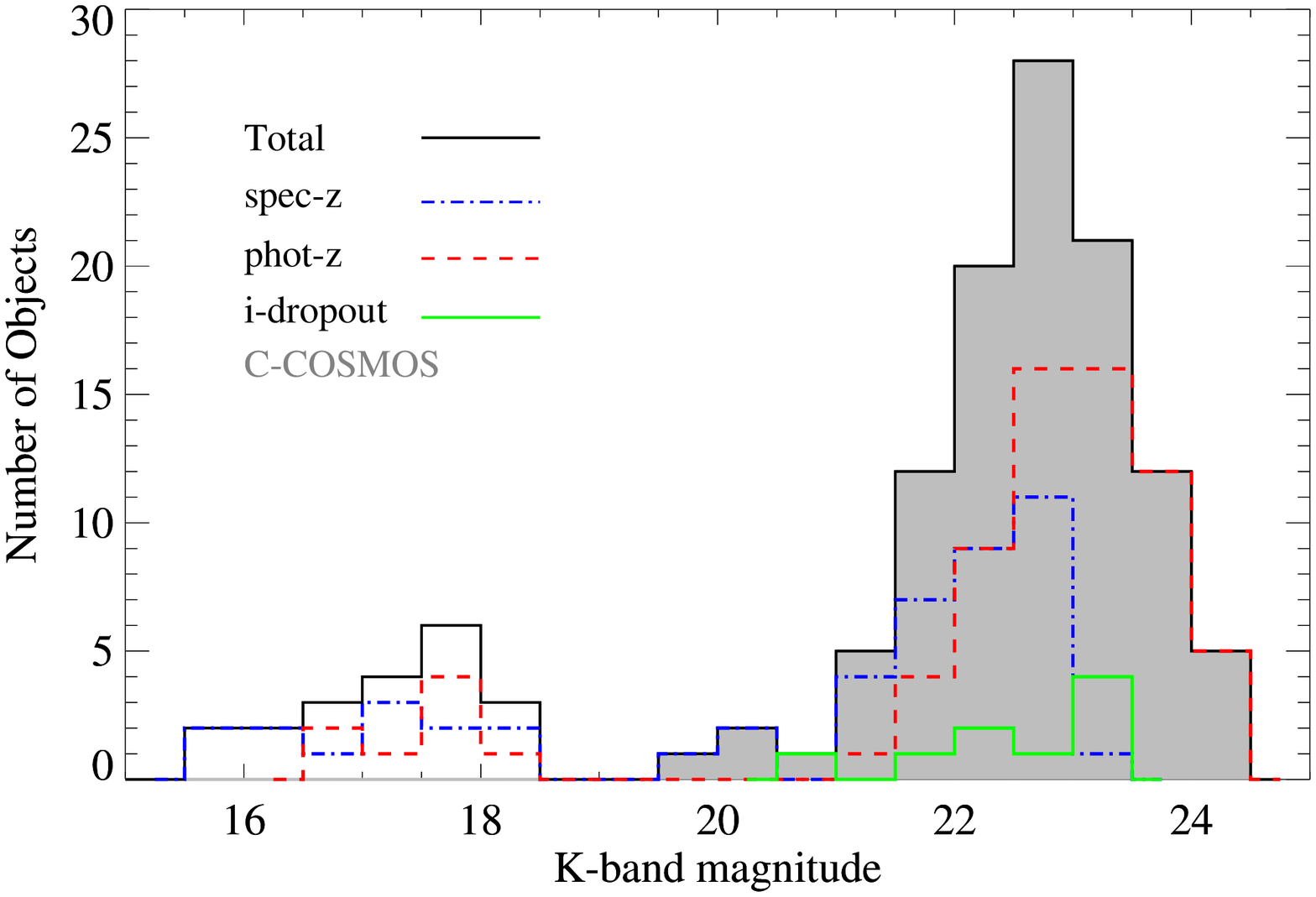}
\includegraphics[trim=10 3 1 20, clip, scale=0.38]{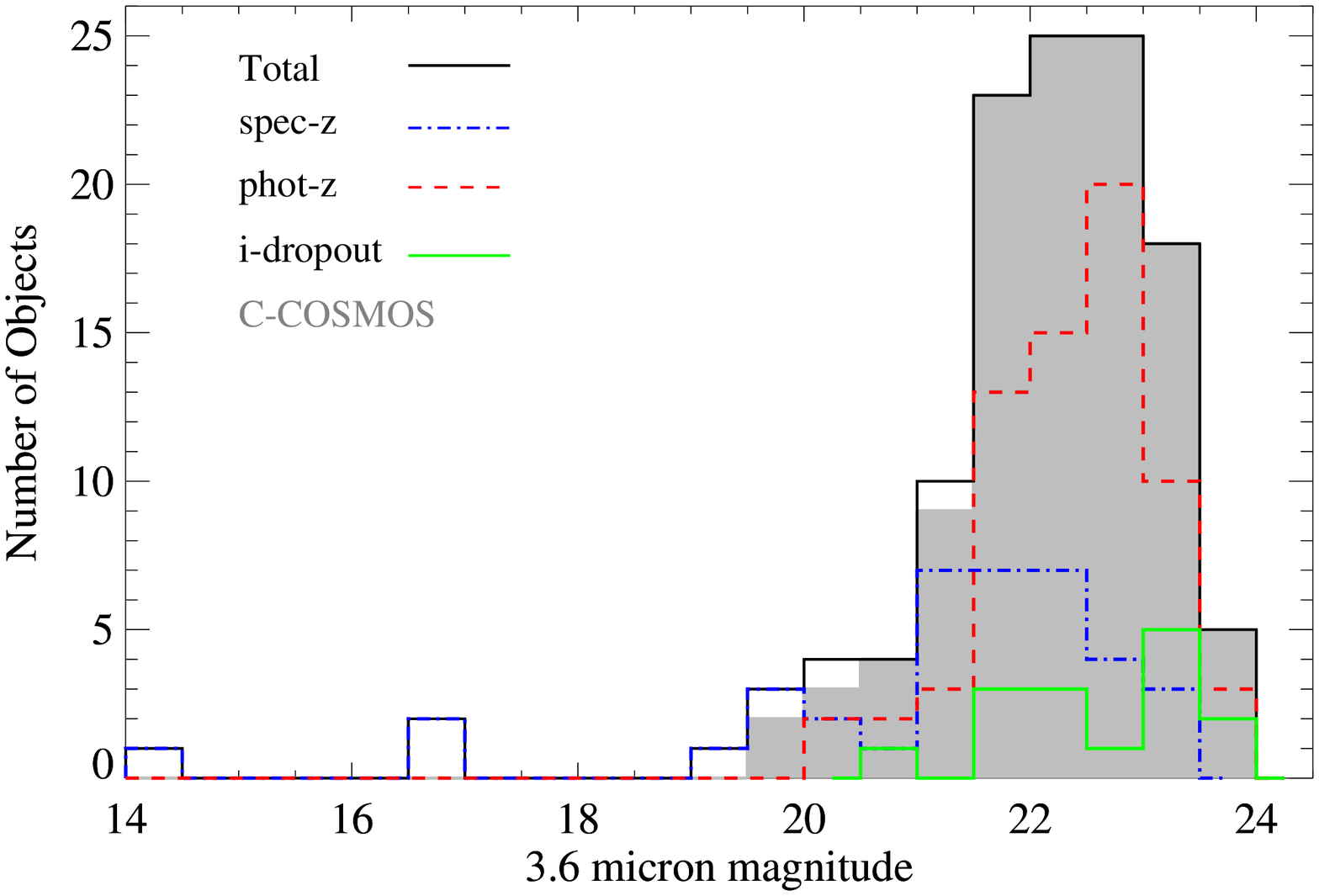}
\caption{Observed AB magnitude distribution of all the $i$-band,
  $K$-band, and 3.6 $\mu$m band (from top to bottom) high-$z$
  objects. Black solid, blue dot-dashed, red dashed and green solid
  lines represent the total, spectroscopic, photometric redshift and
  $i$-dropout samples, respectively. The $i$-band dropouts are not included in the $i$-band histogram.} 
\label{fig:mag_distri}
\end{figure}

In summary, we have assembled a sample of X-ray selected AGN at $z >
3$ in the C-COSMOS and ChaMP on the basis of both spectroscopic and
photometric redshifts. The total sample includes 209 sources with $z
>3$. Of these, 45 are selected to be at $z >3$ from their broad
$P(z)$. There are also 15 C-COSMOS sources considered to be at $z >3$
on the basis of their optical non-detection these are included only in
the derivation of the upper boundary of the $\log N - \log S$
curve. The properties of the sample members are given in
Table~\ref{Table:Total_properties} (Appendix) and the detailed numbers
are given in Figure~\ref{fig:sample_diagram}. Figure~\ref{fig:mag_distri} shows the optical and near-infrared
  ($i$, $K$, and 3.6 $\mu$m) observed magnitude distributions for the
  total high-$z$ population and for sources with spectroscopic and
  photometric redshifts, separately. Sources selected as
  $i$-dropouts are also presented.

\begin{figure}
\hspace{-0.6cm}
\includegraphics[scale=0.47]{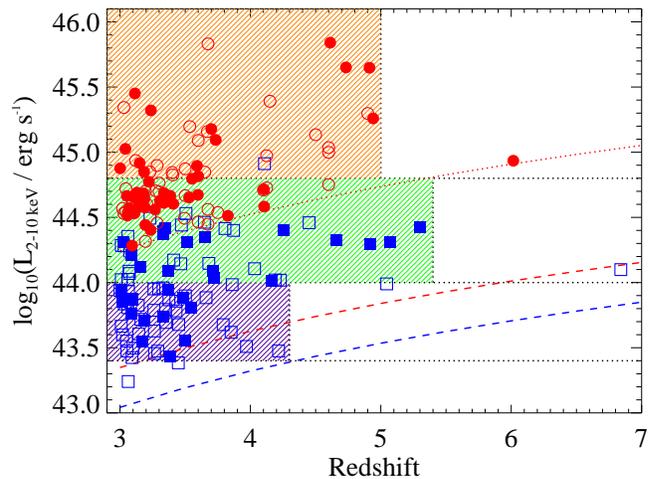}
\caption{The hard X-ray luminosity (computed with $\Gamma=1.8$)
  redshift plane for the objects in our sample. Blue squares =
  C-COSMOS sample. Red circles = ChaMP sample. Filled = spectroscopic
  redshift. Open = photometric redshift. The dashed lines represent
  the 2-10 keV luminosity limit of the surveys computed from the 0.5-2
  keV limiting flux. The dotted red line represents the completeness
  flux cut we have adopted at $3\times10^{-15}{\rm
    erg~cm^{-2}~s^{-1}}$.  The dotted black lines correspond to the
  flux limits we imposed for the computation of the space density and
  their associated areas, purple ($43.4 < \log L_{X}< 44.0$), green
  ($44.0 < \log L_{X}< 44.7$), orange ($\log L_{X}> 44.7$). } 
\label{fig:Lx_z}
\end{figure}

The hard (2-10 keV rest frame) X-ray luminosity versus 
plane is shown in Figure~\ref{fig:Lx_z} together with the flux
limit of the C-COSMOS and ChaMP surveys (dashed line) and the applied
flux cut for ChaMP (dotted line). Luminosities were computed from
in every case assuming an intrinsic $\Gamma=1.8$.

The C-COSMOS \& ChaMP high-$z$ sample is a factor of 4-5 larger than all the previous individual X-ray selected samples at $z>3$ \citep[e.g.][]{Brusa2009, Hiroi2012, Vito2013}. Most importantly, this is the first time that a significant sample of 29 X-ray selected AGNs at $z>4$ is assembled. At these redshifts previous studies had a maximum of 9 sources. The $z > 3$ X-ray selected AGNs sample also covers more than a factor of 2 of soft (2-10 keV rest frame) X-ray luminosity, and includes a significant number of both broad-line and non-broad line AGN. 

To discuss the obscured AGN fraction requires each object in our sample to be classified as obscured or unobscured. There are two commonly-adopted methods for classification; one is based on the optical emission line widths (`optical type') or, if a spectrum is not available, by the type of template that best fits the optical-infrared spectral energy distribution (SEDs) of the sources. The other is based on the column densities, $N_{H}$, in the X-ray spectra (`X-ray type') or, if an X-ray spectrum is unavailable, by the hardness ratio (HR) \citep[e.g.][]{Hasinger2001}. X-ray absorption should typically correlate with optical AGN type. In the unified scheme \citep[e.g.][]{Lawrence1982,Antonucci1993,Urry1995} as the narrow emission line AGNs are viewed through the dusty torus, and hence have higher absorption column densities than broad emission line AGNs. In fact, evidence has been mounting over the years that  the optical- and X-ray-based classifications often give contrasting results \citep{Lawrence2010,Lanzuisi2013, Merloni2014}. 

\subsection{Optical Types} \label{sec:optical_properties}

The optical type of the sources has been determined of the measured  full width at half maximum (FWHM) of the permitted emission lines. Those objects with emission lines having FWHM $>1,000~{\rm km~ s^{-1}}$ \citep[e.g.][]{Stern2012} are classified as `optical broad-line' (BLAGN), and all others as `optical nonbroad-line' (non-BLAGN) (i.e., they show narrow emission lines or absorption lines only; as done by \citealp{Civano2011,Civano2012}). 

In the C-COSMOS spectroscopic $z>3$ sample, 21 of 32 sources are classified as BLAGN. These are mainly associated with the brighter optical sources ($i_{AB} \sim 22 - 23$) of the spectroscopic sample (see Figure~\ref{fig:imag_xflux}). At fainter optical magnitudes ($i_{AB} >23$), equal numbers of broad-line and non-broad line AGNs are found. The classification for the 75 AGN in C-COSMOS with photometric redshifts is obtained by the \cite{Salvato2011}  photometric fitting method fitting the SED via $\chi^{2}$  minimization with code  LePhare\footnote{http://www.cfht.hawaii.edu/~arnouts/LEPHARE/lephare.html}. More details on the fitting can be found in \cite{Salvato2011}. Briefly, two libraries of   templates were used, depending on morphology, optical variability, and X-ray flux of the source. The first library (defined in \citealp{Salvato2009}, Table 2) consists of AGN templates, hybrid (host + AGN) templates, and a few normal galaxies  and was used for all the point-like optical sources and for the extended sources with an X-ray flux brighter than $8\times  10^{-15}~{\rm erg~ cm^{-2}~ s^{-1}}$. The second library (as defined  in \citealp{Ilbert2009}) includes only normal galaxy templates and it was used for the remaining sources (i.e., extended and with X-ray flux $<8\times 10^{-15}~{\rm erg~ cm^{-2}~ s^{-1}}$). The flowchart  in Figure 6 of \cite{Salvato2011} summarizes the procedure. \cite{Civano2012}, according to this fitting, divide the sources into obscured AGN, galaxies and unobscured AGN. About 40 per cent (28 sources) of the photometric sample is best fitted with an unobscured quasar template, and 47 sources with an obscured quasar template. For 29 AGN with spectroscopic identification, the photometric and spectroscopic types match. Given the mismatch rate of $\sim 9$\%, we estimate that $\sim 7$ out of the 75 AGNs could have been assigned the wrong SED classification. 
 
In the ChaMP $z>3$ spectroscopic sample, as expected at these fluxes \citep[e.g.][]{Brusa2009}, and due to the predominantly SDSS spectroscopic target selection, only 2/44 sources are non-BLAGN. The characterization of these sources based on their SED fittings  has been obtained by \cite{Trichas2012}. In order to be in agreement  with the spectroscopic ChaMP sample, we followed the same SED fitting method for the characterization of the 43 sources without a spectroscopic classification. According to this fitting, 11 of 43 sources are best fitted with an obscured quasar template (non-BLAGN).  More details on the fitting can be found in \cite{Trichas2012} and \cite{Ruiz2010}. Briefly, a total of 16 templates has been used including QSO, Seyfert-2 galaxies, starburst galaxies, absorption line galaxies and composite templates that are known to harbor both an AGN and a starburst. The \cite{Ruiz2010} model has been adopted, which fits all SEDs using a $\chi^{2}$ minimization technique within the fitting tool Sherpa \citep{Freeman2001}. The fitting allows for two additive components, one associated with the AGN emission and the other associated with the starburst emission. The fit with the lowest reduced $\chi^{2}$ has been chosen as the best-fit model. 


\subsection{X-ray Types}  \label{section:xray_properties}

Most sources in our sample have a low number of detected counts (median $\sim 25$ in the 0.5-8.0 keV full band). In this count regime, spectral fit results are not reliable, especially if more
than one free parameter is fit; even if the fit converges the uncertainties on the parameters are large. For these
reasons, we use the Bayesian Estimation of Hardness Ratios (BEHR)
method \citep{Park2006} to derive X-ray spectral type. Hardness count
ratios (HR), defined as ${\rm HR} = (C_{\rm HB}-C_{\rm SB})/(C_{\rm
  HB}+C_{\rm SB})$, where $C_{\rm SB}$ and $C_{\rm HB}$ are the counts
in the soft band and hard band, respectively.  

BEHR is particularly powerful in the low-count Poisson regime, because
it computes a realistic uncertainty for the HR, regardless of whether the
X-ray source is detected in both energy bands. Sources with
unconstrained upper or lower limits due to non-detections (14 
hard-only and 49 soft-only detections) have been computed by
converting the $3\sigma$ flux upper limit in the undetected band into
counts.  

\begin{figure}
\hspace{-0.6cm}
\includegraphics[scale=0.460]{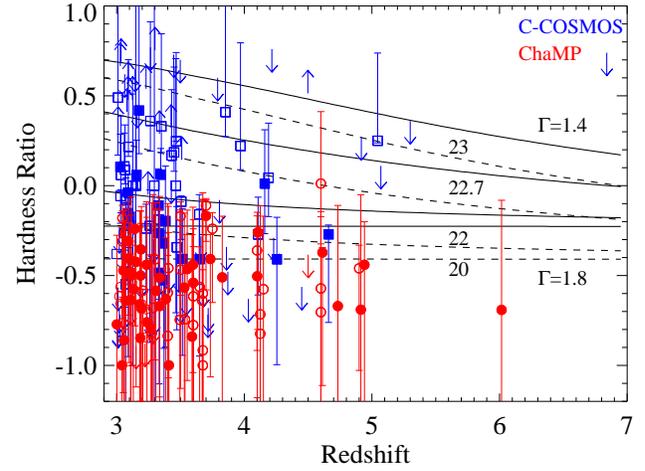}
\caption{Hardness ratio versus redshift. Blue squares = C-COSMOS
  sample. Red circles = ChaMP sample. Filled = spectroscopic
  redshift. Open = photometric redshift. Sources with no hard band or
  soft band detection are shown with arrows. Four curves of constant $N_{H}$ ($10^{20}$, $10^{22}$, $5\times10^{22}$ and $10^{23}~{\rm cm^{-2}}$) are reported for $\Gamma=1.8$ (dashed lines) and $\Gamma=1.4$ (solid lines). }
\label{fig:HR_z}
\end{figure}

To estimate the column density, curves of constant $N_{H}$ as a function of redshift have been derived for two spectral slope values, $\Gamma=1.4$ and $\Gamma=1.8$. The flatter spectral slope has been chosen to be consistent with the assumptions adopted in producing the original X-ray catalogs \citep{Kim2007a,Puccetti2009}. The steeper value is more representative of the intrinsic value if the spectrum is not affected by obscuration \citep{Nandra1994}. The relationship
between HR and redshift of our C-COSMOS and ChaMP AGN samples is shown
in Figure~\ref{fig:HR_z}. Curves of $N_{H}=10^{20}$, $10^{22}$,
$5\times10^{22}$ and $10^{23}~{\rm cm^{-2}}$ are reported for
$\Gamma=1.8$ (dashed lines) and $\Gamma=1.4$ (solid lines). We observe
that C-COSMOS sample tend to be more obscured as expected due to the
fainter X-ray sensitivity limit, than the ChaMP sample
\citep{Lawrence1982, Ueda2003,Hasinger2008, Brusa2010,Burlon2011}.  

Though the two samples (C-COSMOS and ChaMP) of $z>3$ AGN show
different trends regarding their obscuration, the large HR errors and
the similarity in this redshift range of the curves with widely
different $N_H$ values for the same spectral slope, do not allow an
accurate estimate of the column density for each source to be
made. Using the CIAO\footnote{http://cxc.harvard.edu/ciao/} spectral
analysis package, {\it
  Sherpa}\footnote{http://cxc.harvard.edu/sherpa/}, we have simulated
X-ray spectra for AGN populations at $3 < z < 7$ in order to quantify
the evolution of X-ray spectral slopes due to the k-correction of the
observed AGN spectra toward high-$z$. Based on these simulations, we
find that the HR distribution for the ChaMP sample peaks at $\Gamma
\sim1.8 - 2.0$ while the HR distribution for C-COSMOS sample peaks at
$\Gamma \sim1.4 - 1.9$. Hereafter, to better constrain the column
density and for the purpose of comparison with previous studies, we
fixed the photon index to $\Gamma = 1.8$ and converted all source
fluxes. 

In the present work, we adopt the $N_{H}=10^{22}~{\rm cm^{-2}}$ limit for the X-ray unobscured ($N_{H}<10^{22}~{\rm cm^{-2}}$) and obscured ($N_{H}>10^{22}~{\rm cm^{-2}}$) classification of the sources. Based on previous studies (e.g. \citealp{Ueda2003, Hiroi2012, Lanzuisi2013}), the adopted criterion provides a good agreement with the optical classification of the AGN.

\subsection{X-ray/Optical flux ratio}  \label{section:XO_ratio}

\begin{figure*}
\includegraphics[width=8.7 cm]{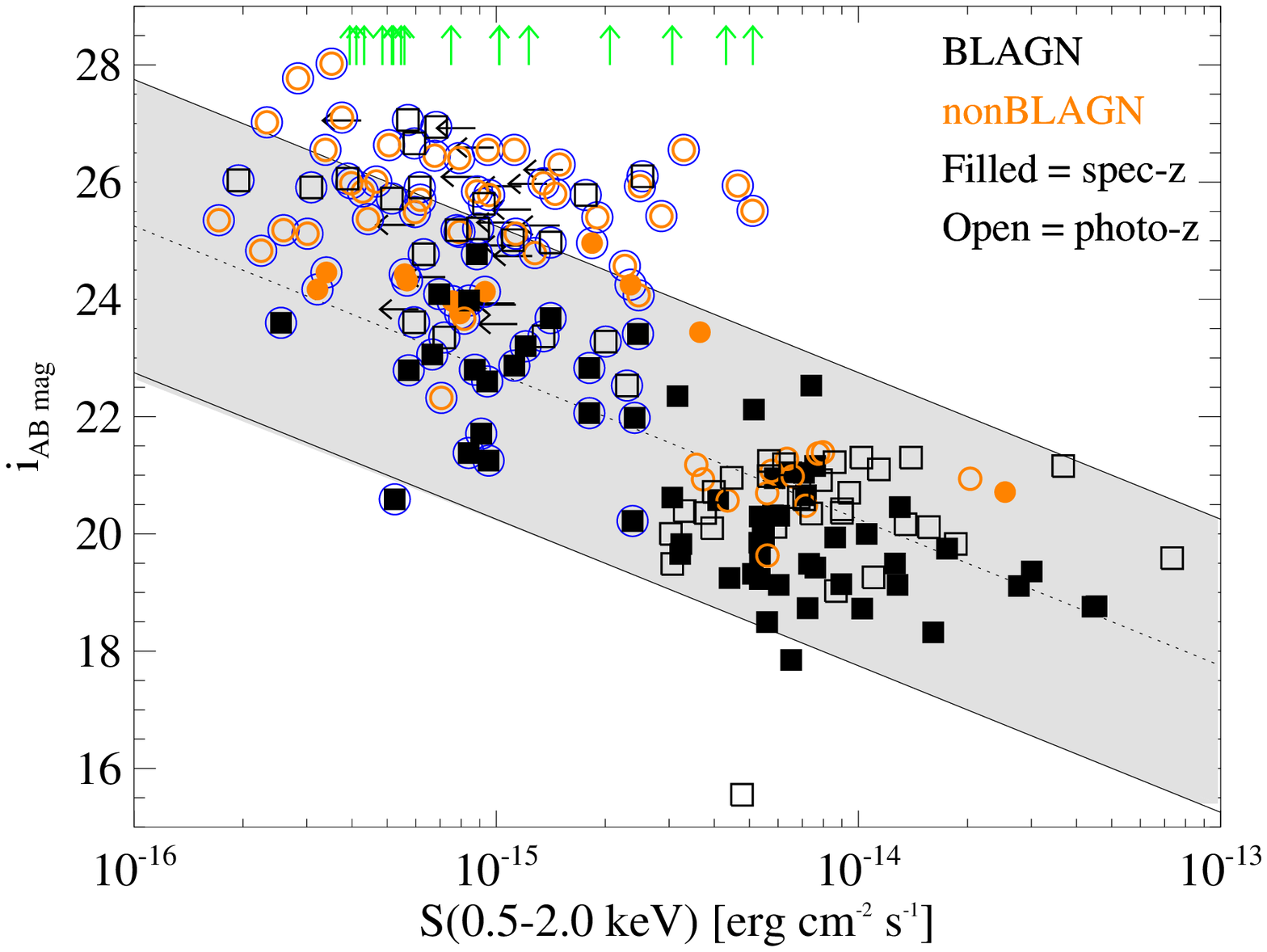}
\includegraphics[width=8.7 cm]{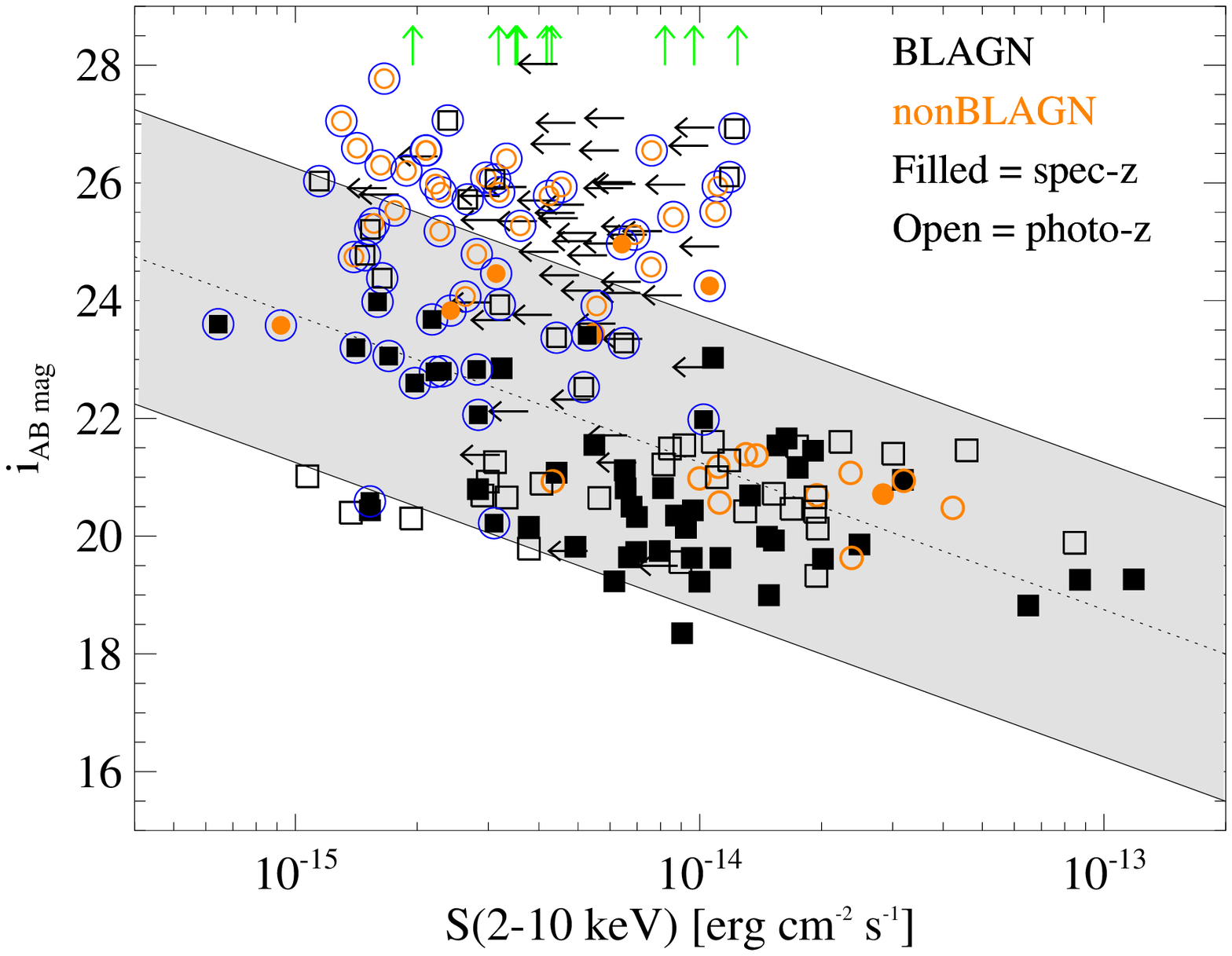}
\caption{X-ray flux (soft-left, hard-right) vs. the $i$-band magnitude for all the X-ray sources with an $i$-band counterpart. The grey shaded region represents the locus of AGNs along the correlation ${\rm X/O} = 0 \pm 1$. Sources with secure spectroscopic redshifts are represented by filled symbols and sources with a photometric redshift by open symbols. Orange circles and black squares represent non-BLAGN and BLAGN, respectively. Green upper limits represent $i$-band dropouts and black left pointing arrows represent soft and hard X-ray flux upper limits for undetected sources in each band. The C-COSMOS sample is represented by the open big blue circles.}
\label{fig:imag_xflux}
\end{figure*}

The X-ray/Optical (X/O) flux ratio is a redshift dependent quantity for obscured AGN, given that the k-correction is negative in the optical band and positive for the X-rays \citep{Comastri2003,Fiore2003,Brusa2010}. As a result, obscured sources have higher X/O  at high redshift. On the other hand, unobscured sources have similar k-corrections in the two bands, and the distribution in X/O is not correlated with the redshift \citep{Civano2012}. Usually, the $r$- or $i$-band flux is used \citep[e.g.][]{Brandt2005}  while a soft X-ray flux was used originally used for this relation with the majority of luminous spectroscopically identified AGNs in the Einstein and ASCA surveys characterized by ${\rm X/O }= 0\pm 1 $ \citep[e.g.][]{Schmidt1998,Stocke1991,Lehmann2001}. The same relation has been used also in the hard band, without really accounting for the X-ray band used or the change in spectral slope \citep[e.g.][]{Alexander2001,Fiore2003,Brusa2003,Civano2005,Laird2009,Xue2011}.

Figure~\ref{fig:imag_xflux} shows the distribution of X-ray soft (left) and hard (right) flux versus optical magnitude to illustrate the parameter space spanned by the broad-line and nonbroad-line populations. The X/O ratio \citep{Maccacaro1988} is defined as: 
\begin{equation}
{\rm X/O}=\log_{10}(f_{X}/f_{\rm opt})=\log_{10}(f_{X})+C+m_{\rm opt}/2.5
\end{equation}
where $f_{X}$ is the X-ray flux in a given energy range, $m_{\rm opt}$, is the magnitude at the chosen optical wavelength, and $C$ is
a constant which depends on the specific filter used in the optical observations. For both X-ray bands, the ${\rm X/O}=\pm1$ locus (grey area) has been defined using as $C(i) = 5.91$ \citep{Civano2012}, which was computed taking into account the width of the $i$-band filters in Subaru, CFHT (Canada-France-Hawaii Telescope), or for bright sources SDSS. In the hard band, the locus is plotted taking into account the band width and the spectral slope used to compute the X-ray fluxes ($\Gamma = 1.8$). The majority of BLAGNs with a secure spectroscopic redshift, follow the trend of $-1<\log_{10}(f_{X}/f_{i})<1$.  However, given the variation in $\alpha_{\rm OX}$ with luminosity \citep[e.g.][]{Vignali2003,Young2010,Trichas2013}, there can be some shift in the locations of QSOs with luminosity within the so-called BLAGN region. This shift is consistent with the X/O relation being originally calibrated on soft-X-ray-selected sources, bright in the optical and also in the X-rays. This might explain the mild shift between the ChaMP and C-COSMOS BLAGN.

Apart from the AGN population found in the BLAGN region , there is also a significant population that lie at $\log_{10}(f_{X}/f_{i})>1$ suggesting obscured nuclei. The main characteristics of this sample are: 1) lack of spectroscopic redshifts (open symbols), 2) non-BLAGN optical classification (green symbols) and 3) low X-ray luminosities ($10^{43}~{\rm erg~ s^{-1}} < L_{2-10{\rm
    keV}} < 10^{44}~{\rm erg~s^{-1}}$) with $N_{H}>10^{22}~{\rm  cm^{-2}}$ for $\sim 65\%$ of them, which is consistent with previous studies finding that mild obscuration is common at these luminosities \citep[e.g.][]{Silverman2010}. Furthermore, nearly 75\% of all the  
sources with X/O$ > 1$ are obscured, thus confirming that selections
based on high X/O ratio are efficient in finding samples of obscured
AGN. 

\subsection{Comparison of optical and X-ray types}  \label{section:xray_vs_opt_class}

In order to compare our optical classification to the expected obscuration of BLAGNs and nonbroad-line AGNs based on the unified scheme we have separated our total sample into broad-line and nonbroad-line AGNs (as described in Section~\ref{sec:optical_properties}). 

In the case of the BLAGN, the X-ray classification criterion ($N_{H}=10^{22}~{\rm cm^{-2}}$) gives 28/124 X-ray obscured sources for $\Gamma=1.8$. Half of these sources have a spectroscopic redshift and all but 3 come from C-COSMOS sample. If we also take into account the HR errors, then for the lower HR limits, 11 BLAGN are classified as X-ray obscured sources and 41 are classified X-ray obscured sources for the upper HR limits. In the non-BLAGN subset, the above criterion gives 49/71 X-ray obscured sources (detected in both soft and hard bands) for $\Gamma=1.8$. The 27 soft band sources in non-BLAGN sample with no detection in the hard band (reported as downward arrows in Figure~\ref{fig:HR_z}) have very high upper limits on the HR, due to the conservative flux upper limit computed by \cite{Puccetti2009}, but most of them do not thereby satisfy the $N_{H}>10^{22}~{\rm cm^{-2}}$ criterion.  

For the total sample, we find agreement between the optical and X-ray classification for $\sim74$\%:  $\sim77$\%for the BLAGN and $\sim69$\% for the non-BLAGN. These rates are consistent with recent studies \citep[e.g.][]{Lanzuisi2013,Merloni2014}. Possible explanation can be a misclassification of faint type-1 with strong optical/IR contamination from host galaxy light.

\begin{table}
\caption{Comparison of optical and X-ray types. The upper and lower limits have been estimated taking into account only the error ranges in HR.}
\centering
\begin{minipage}{7.3cm}

\begin{tabular} {| c | c c  |}

\hline

Number 	&		Unobscured&		Obscured \\
of sources	&		$N_{H}<10^{22}~{\rm cm^{-2}}$	&		$N_{H}\geq10^{22}~{\rm cm^{-2}}$	\\
\hline
	&	&	\\
BLAGN		&	$96^{+16}_{-14}$	&		$28^{+14}_{-16}$	\\
	&	&	\\
\hline
	&	&	\\
non-BLAGN		&	$22^{+13}_{-9}$	&	$49^{+9}_{-13}$		\\
	&	&	\\
\hline

\end{tabular}
\label{Table:XO_class}
\end{minipage}
\end{table}

\begin{figure}
\hspace{-0.7cm}
\includegraphics[width=9.5 cm]{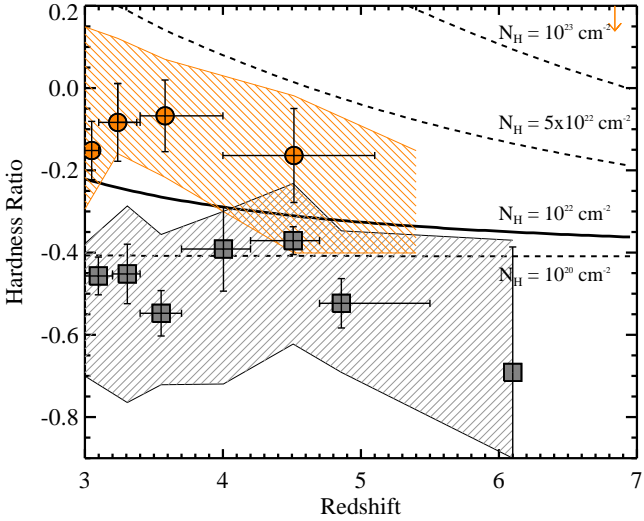}
\caption{The mean hardness ratio as a function of redshift for BLAGN  (black squares) and non-BLAGN (orange circles) $z>3$ AGN  subsamples. The error bars represent the 68\% dispersion. Only
  sources with both soft and hard band detections are taken into
  account for the estimation of the mean HR in each bin. Undetected
  sources in one of these bands are included only for estimation of
  the upper and lower limits (dashed areas). The $z_{phot}=6.88$ source with an upper limt ${\rm HR_{up}}=0.74$ has been shifted down to HR=0.2 in order to be included in the figure.} 
\label{fig:HR_z_class}
\end{figure}

To improve the statistics and gain information on the average properties of the two subclasses, we compared their mean HR values as a function of redshift (Figure~\ref{fig:HR_z_class}). Despite the $\sim30$\% misclassification for the individual sources, the mean properties of the BLAGN and non-BLAGN seem to agree with the $N_{H}\sim10^{22}~{\rm cm^{-2}}$ division. These results does not change even if we use only sources with spectroscopic redshifts. The upper and lower limits detected only in the soft or the hard band were used to compute the upper and lower boundary of the shaded area. We discuss the results in Section 5. 

\section{The $\log$N-$\log$S of the $z>3$ AGN} \label{section:logNlogS}

We derived the soft band number counts of the $z > 3$ and $z > 4$ samples by folding the observed flux distribution through the sky coverage area versus flux curve of the C-COSMOS survey \citep{Puccetti2009} and the ChaMP's 323 fields \citep{Green2009}. Additionally, we have corrected the number counts for ChaMP incompleteness in the spectroscopic/photometric coverage as a function of X-ray flux (i.e. Figure~\ref{fig:completeness}). 

To minimize the error associated with the most uncertain part of the sensitivity curve, we truncate the C-COSMOS sample at the flux corresponding to 10\% of the total area (blue dashed line in Figure~\ref{fig:Area_flux}). Following \cite{Civano2011}, all the sources with a $0.5-2$ keV flux above $3\times10^{-16}~{\rm erg~ cm^{-2}~ s^{-1}}$ have been considered (73 objects out of the 81 soft band detected). The flux limit applied to the sample is consistent with the signal-to-noise ratio thresholds chosen by \cite{Puccetti2009}, on the basis of extensive simulations, to avoid the Eddington bias in the computation of the number counts of the entire C-COSMOS sample. Thus, by applying a flux limit cut, we also reduce the Eddington bias affecting our sample. For ChaMP this would be at $S_{\rm 0.5-2keV}>2\times 10^{-15}~{\rm erg~ cm^{-2}~ s^{-1}}$, below the flux limit already applied. 

\begin{figure}
\includegraphics[trim=55 135 1 1, clip, scale=0.26]{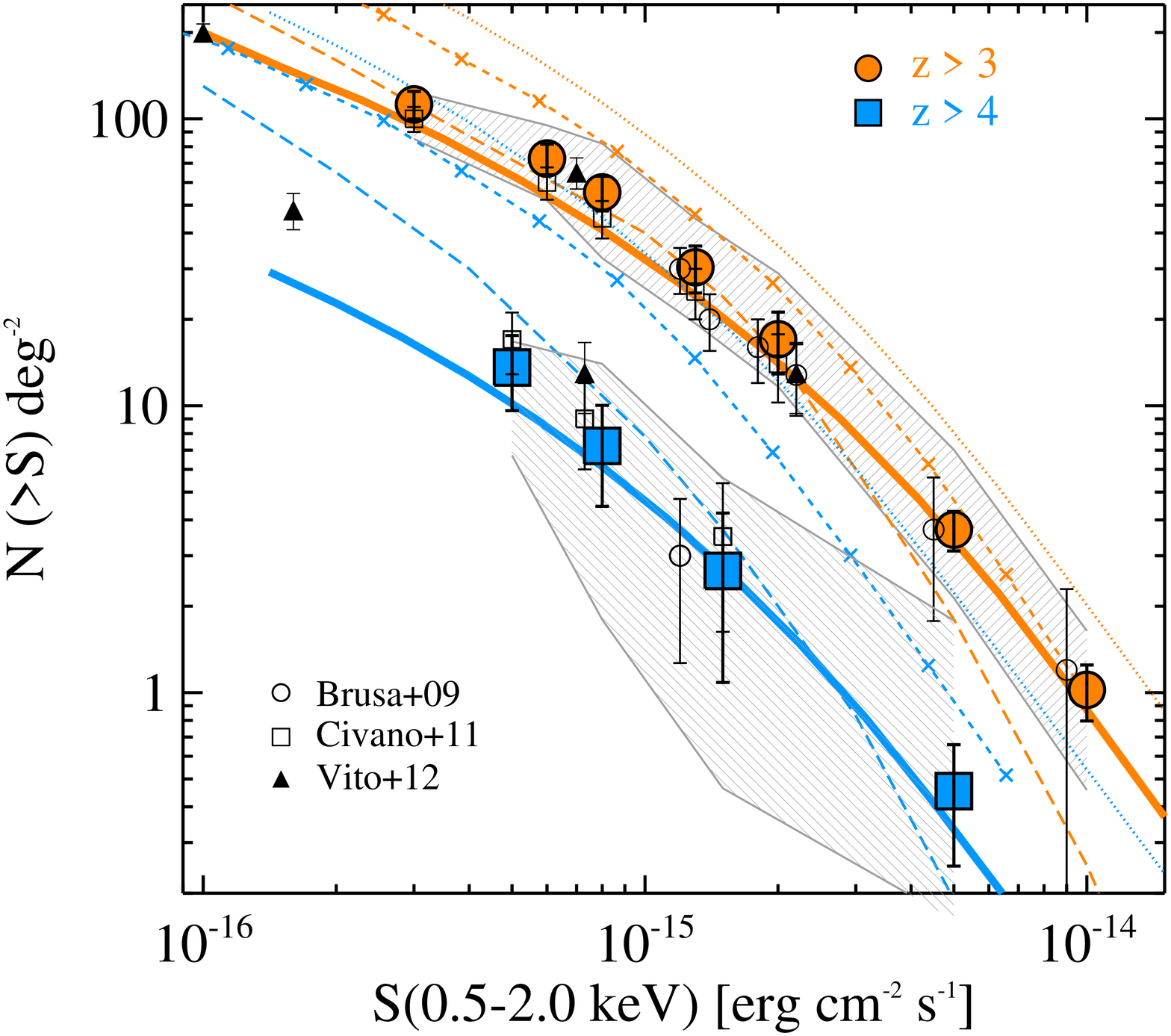}
\includegraphics[trim=55 137 1 60, clip, scale=0.26]{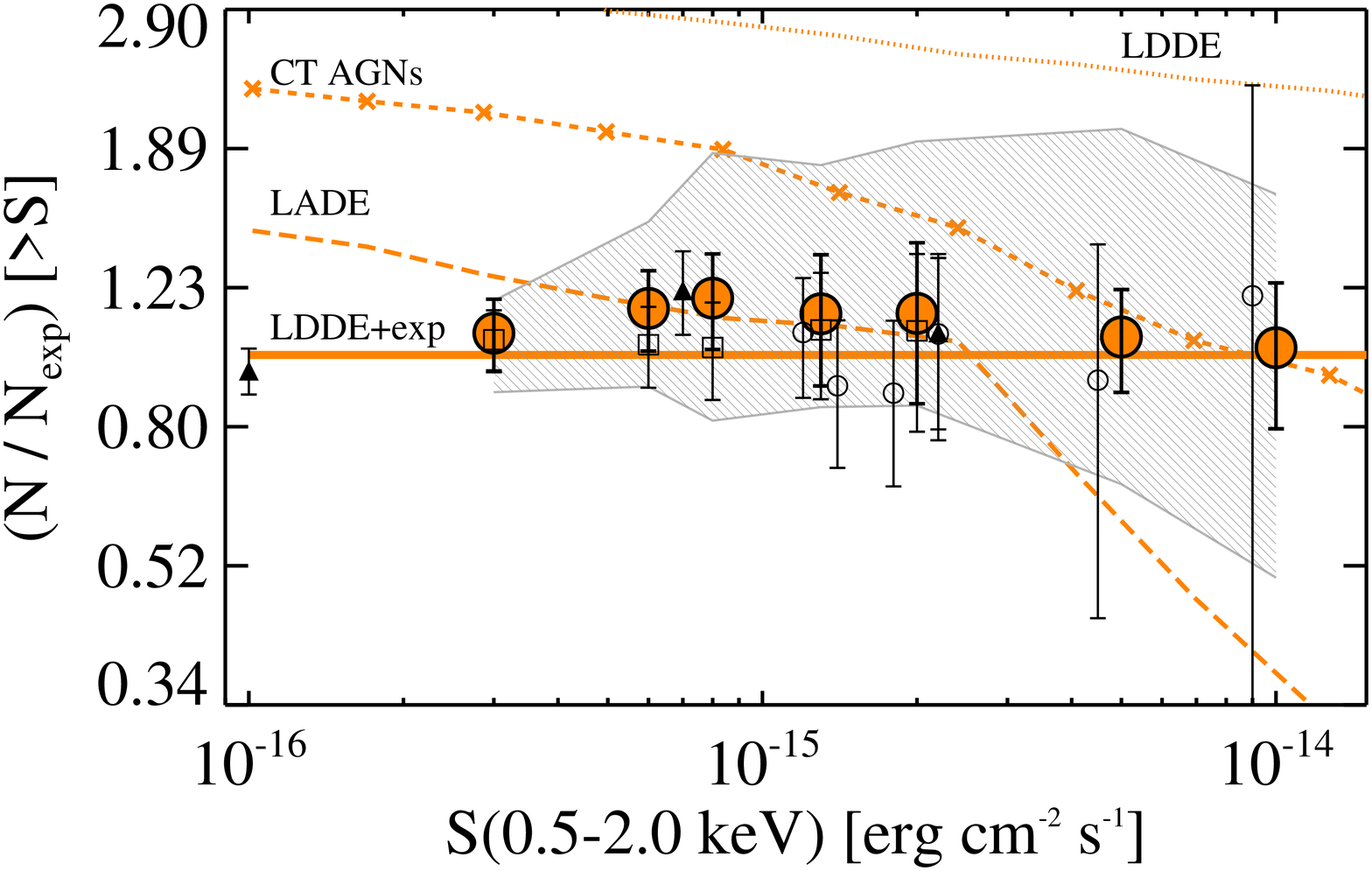}
\includegraphics[trim=55 1 1 60, clip, scale=0.26]{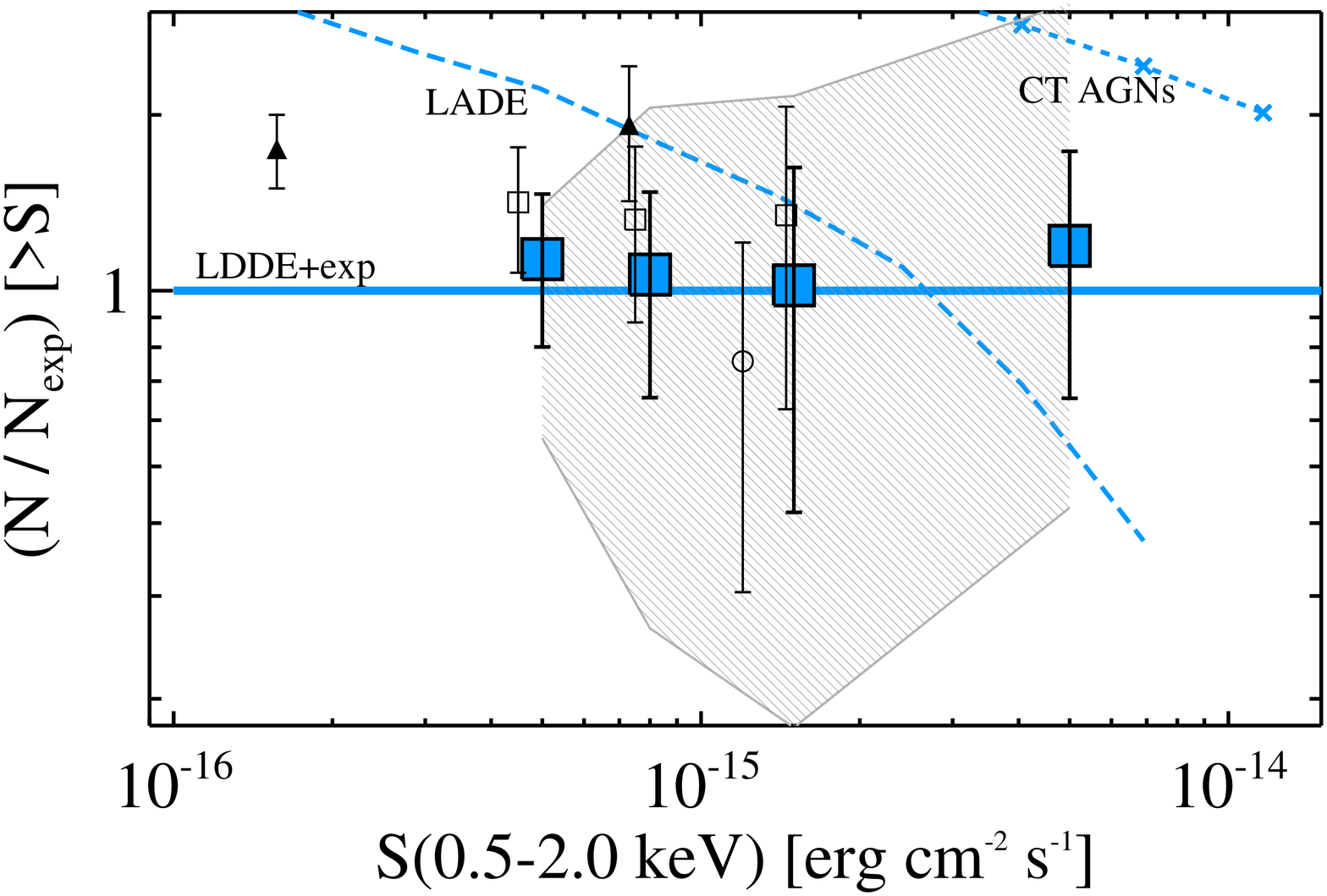}
\caption{Top: The binned logN-logS relation (with associated errors)
  of the $z > 3$ (orange circles) and $z > 4$ (blue squares) QSOs
  population. The grey shaded area represents the maximum and minimum
  number counts under the assumptions described in
  Section~\ref{section:logNlogS}. The blue and orange curves
  correspond to the prediction based on the LDDE+exp (thick solid),
  LDDE (dotted), LADE (dashed) and CT AGNs (dash cross)models for each
  redshift range, respectively. The small open circles represents the
  number counts estimated by Brusa et al. (2009), the small open
  squares  are from Civano et al. (2011) work and small filled
  triangles from Vito et al. (2013). Middle: The ratio of the observed
  number counts for $z>3$ relative to the LDDE+exp model (thick solid
  line at the top panel). The thick solid line represents the LDDE+exp
  model $N/N_{\rm exp}=1$, the dotted line represents the LDDE
  relative to the LDDE+exp model, the dashed lines the LADE relative
  to the LDDE+exp model and the dash cross lines the CT AGNs model
  relative to the LDDE+exp model. Bottom: The ratio of the observed
  number counts for $z>4$ relative to the LDDE+exp model. Symbols are
  similar to the medium panel. The ratio of the LDDE relative to the
  LDDE+exp model is $N/N_{\rm exp}>8.0 $ and is not presented.} 
\label{fig:logN_logS}
\end{figure}

The binned logN-logS relations for two redshift ranges ($z > 3$; orange points and $z > 4$; blue points, with associated errors) are plotted in Figure~\ref{fig:logN_logS} (top panel). In integral form, the cumulative source distribution is represented by:
\begin{eqnarray}
N(>S)=\sum_{i=1}^{N_{S}} \frac{1}{\Omega_{i}}
\end{eqnarray}
\\
where $N(>S)$ is the number of sources with a flux greater than S and $\Omega_{i}$, is the limiting sky coverage associated with the $i$th source. The associated error is the variance:

\begin{eqnarray}
\sigma^{2}=\sum_{i=1}^{N_{S}} (\frac{1}{\Omega_{i}})^{2}
\end{eqnarray}
\\
The grey shaded area represents an estimate of the maximum and minimum number counts relation at $z >3$ obtained by considering three different effects: 

1) the $1\sigma$ uncertainty in the sky coverage area for each source using the sky coverage as a function of flux (see Figure~\ref{fig:Area_flux}) and the $1\sigma$ uncertainty in the flux; 

2) the 14 sources from C-COSMOS with no-optical detection (seen in the soft band);

3) the sources with photometric redshift $z_{photo}<3$ but $z_{photo} + \sigma_{zphoto} > 3$.

To compute the upper boundary of the shaded area, we included all the sources in the main sample plus the sources with no optical detection at their flux$+1\sigma$ error. For the lower boundary, we used the flux$-1\sigma$ error only for the sources with $z_{spec}>3$. Under these assumptions, the number counts estimation would follow the lower boundary curve only in the very unlikely case that all the photometric redshifts are overestimated, while the observations would be described by the upper boundary if the  selection of high redshift X-ray sources based  on the lack of optical detection was a 100\% reliable proxy.

We have compared our number counts with previous X-ray surveys that span the range from deep, small area (CDFS, at 464.5 arcmin$^{2}$ with a soft band flux limit of $\sim 9.1 \times 10^{-18} ~{\rm  erg~cm^{-2}~s^{-1}}$, \citealp{Xue2011}),  to moderate area and moderate depth (Chandra-COSMOS at 0.9 deg$^{2}$ with a 0.5-2 keV flux limit of $\sim1.9\time 10^{-16} ~{\rm erg~cm^{-2}~s^{-1}}$, \citealp{Elvis2009}) and finally to moderate area and shallower depth (XMM-COSMOS, at 2 deg$^{2}$ and a soft band flux limit of $\sim 1.7 \times 10^{-15}  ~{\rm erg~cm^{-2}~s^{-1}}$,
\citealp{Cappelluti2009}). The binned logN-logS relations are plotted in Figure~\ref{fig:logN_logS} (top panel), together with the XMM-COSMOS (\citealp{Brusa2009}, open circles), C-COSMOS (\citealp{Civano2011}, open squares) and 4~Ms CDF-S number counts (\citealp{Vito2013}, filled triangles). 

Good agreement exists among the comparison surveys presented here. At $z>3$ and fainter X-ray fluxes ($S_{\rm 0.5-2keV}<2\times 10^{-15}~{\rm erg~cm^{-2}~s^{-1}}$) our points confirm the agreement with the model predictions, previously found by \cite{Brusa2009,  Civano2011,Vito2013}. At the same redshift range but brighter X-ray fluxes, where only XMM-COSMOS sample have available points \citep{Brusa2009} based on 4 sources, we reduce the uncertainties by
factor of 4 using a sample of 66 sources with ($S_{\rm  0.5-2keV}>2\times 10^{-15}~{\rm erg~cm^{-2}~s^{-1}}$). Notably it is
the first time that data points at the bright end at $z>4$ are
included. At redshift $z > 4$, where the XMM-COSMOS sample had only 4
sources,  the 4~Ms CDF-S 9 sources and the C-COSMOS 14 sources, the
C-COSMOS \& ChaMP sample has 27 sources (2-5 times larger), making it
possible to compare the slope of the counts with models.  

A comparison with the AGN number counts from three different phenomenological model predictions is also presented in Figure~\ref{fig:logN_logS} with the different types of curves (orange
color for $z >3$ and blue color for $z> 4$): 

1) The thick solid lines correspond to the predictions of the XRB synthesis model of \cite{Gilli2007}, based on the X-ray luminosity function observed at low redshift \citep[e.g.][]{Hasinger2005}, parametrized with a luminosity dependent density evolution (LDDE) and a high redshift exponential decline with the same functional form adopted by \citep{Schmidt1995} ($\Phi (z)= \Phi (z_{0}) \times 10^{-0.43(z-z_{0})}$ and $z_{0}=2.7$) to fit the optical luminosity function between $z\sim2.5 - 6$ \citep{Fan2001}, corresponding to one e-–folding per unit redshift (hereafter referred to as LDDE+exp). 

2) The dotted curves correspond to the predictions of the LDDE model without the high-$z$ decline \citep{Gilli2007}, obtained extrapolating to high-z the best-fit parameters of \citep{Hasinger2005}. 

3) The dashed line is the luminosity and density evolution model (LADE; \citealp{Aird2010}) which fits the hard X-ray luminosity function derived by \citep{Aird2010} using the 2Ms Chandra Deep Fields and the AEGIS-X (200 ksec) survey to probe the faint end ($\log L_{X} < 43~{\rm  erg~ s^{-1}}$) and the high-z ($z\sim3$) range. 

4) The dash-crossed lines correspond to the \cite{Treister2009} X-ray background population synthesis predictions (CT AGNs).\footnote{Model predictions from the work of \cite{Treister2009} for a range of input values are publicly available at http://agn.astroudec.cl/j\_agn/main.html}

While at $z >3$ the two model predictions are very close, at $z > 4$,
where the models have different slopes, the errors on the data of the
previous studies do not allow a firm preference of one of the two
models, highlighting the advantage of our sample with respect to
previous surveys. In this work we find that the LDDE model
\citep{Gilli2007} without decline clearly overestimates the observed
counts even in the most optimistic scenario (upper boundary) in both
the $z>3$ and $z>4$ redshift ranges. Our results for the $z > 3$
sample (orange color) are in good agreement with both LDDE+exp (thick
solid line) and LADE (dashed line) model predictions but only up to
flux $\sim 2\times 10^{-15}~{\rm erg~cm^{-2}~s^{-1}}$, where the
difference of the two models is $<20$\%. However, the main advantage
of our sample becomes clear at brighter fluxes; our results strongly exclude the LADE model. This is in contrast to previous studies which could not distinguish between the two models due to their large uncertainties. 

At $z>4$ (blue color) our results are in good agreement with LDDE+exp predictions. We can not clearly exclude the LADE model if we take into account the upper and lower boundaries. However, we can point out for first time that there is no sign of the expected decline to higher fluxes. LDDE is fully ruled out at $z>4$. While fainter samples would be also useful for a better description of the model, considering only the $z>4$ subsample by \cite{Vito2013} (4~Ms CDF-S, filled triangles), the data lie between the LDDE+exp and LADE models prediction. \cite{Vito2013} have included the presence of 3 sources at $5 < z < 7.7$, whose redshifts are determined on the basis of relatively uncertain photometric information. If these sources were placed at $3 < z < 4$, a good agreement would be obtained with the LDDE+exp model (see \citealp{Vito2013}, Figure 9). In this case, the \citealp{Vito2013} $z>4$ sample would be consisted only by 5 sources.

\section{2-10 keV COMOVING SPACE DENSITY} \label{section:space_density}

To investigate the cosmological evolution of AGN at $z>3$ the comoving space densities were calculated from our sample utilizing the $1/V_{max}$ method \citep{Schmidt1968}. This method takes into account the fact that more luminous objects are detectable over a larger volume and is readily adapted to the case in which the survey area depends on flux. 

The maximum available volume, over which each source can be detected, was computed by using the formula: 

\begin{eqnarray}
V_{max}=\int_{z_{min}}^{z_{max}} \Omega(f(L_X,z,N_{H}))\frac{dV}{dz}dz
\end{eqnarray}
\\
where $\Omega(f(L_X,z,N_{H}))$ is the sky coverage at the flux $f(L_{X},z)$ corresponding to a source with absorption column density ${\rm N}_{H}$ and observed luminosity $L_{X}$, and $z_{max}$ is the maximum redshift at which the source can be observed at the flux limit of the survey. If $z_{max}>z_{up,bin}$, where $z_{up,bin}$ is the maximum redshift of the redshift bin, then the $z_{max}$ is the upper boundary of the redshift bin used for computing $V_{max}$. In the case of the ChaMP sample, $z_{max}$ is estimated using both the X-ray and optical survey limits and is selected to be the minimum of the two estimates so the source can be observed at the flux limit of both surveys. We computed the space density using the luminosities derived with $\Gamma=1.8$. The contribution of sources with photometric redshift to the space density is weighted for the fraction of their $P(z)$ at $z > 3$. 

\begin{figure}
\hspace{-0.55cm}
\includegraphics[scale=0.45]{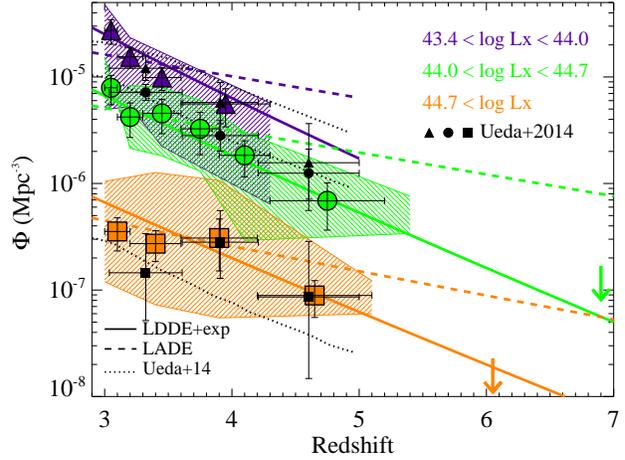}
\caption{The comoving space density in 3 different 2-10 keV X-ray luminosity ranges. The solid lines corresponds to the X-ray selected AGN space density computed for the same luminosity limit from the Gilli et al. 2007 LDDE+exp model. The dashed curve corresponds to the space density derived from the LADE model of Aird et al. 2010. The colors respond to the shaded areas in Figure~\ref{fig:Lx_z} and the shaded area represents the maximum and minimum space density under the assumptions described in the text. When only one source is included in the bin it has been plotted as an upper limit (at $3\sigma$). The small black symbols and dotted lines correspond to the comoving space density data points and model derived from Ueda et al. 2014 for similar X-ray luminosity ranges.}
\label{fig:space}
\end{figure}

After calculating the $V_{max}$ for each source, we sum the values in each redshift bin:

\begin{eqnarray}
\phi=\sum_{i=1}^{z_{min}<z<z_{max}} (\frac{1}{V_{{\rm max,}i}})
\end{eqnarray}
\\
where $\phi$ is the comoving space density in the redshift range of each bin ($z_{\rm min}- z_{\rm max}$), and $i$ is the index of the sample AGN falling into the redshift bin.  The $1\sigma$ uncertainty is given by: 

\begin{eqnarray}
\sigma \phi=\sqrt{\sum_{i=1}^{z_{min}<z<z_{max}} (\frac{1}{V_{{\rm max,}i}})^{2}		}.
\end{eqnarray}
\\

Including soft, hard and full band detected sources allows us to compute a space density which takes into account both unobscured sources, which emit more at softer energies, and obscured sources, which emit more at harder energies, without having to introduce any further correction or assumption. 

The resulting comoving space densities are shown in Figure~\ref{fig:space}. To reduce the effects of incompleteness and to have a complete sample over a given redshift range, we divided the sample in three luminosity intervals (see Figure~\ref{fig:Lx_z}, shaded areas):

1) At low luminosities (purple shading) we computed the space density in 3 redshift bins ($z = 3 - 4.3$) at $43.3<\log (L_{X}/{\rm erg~s^{-1}})\leq44.0$.

2) At intermediate luminosities (green shading) we computed the space density in 3 redshift bins ($z = 3 - 5.4$) at $44.0<\log (L_{X}/{\rm erg~s^{-1}})\leq44.7$ 

3) At high luminosities (orange shading) we computed the space density in 5 redshift bins ($z = 3 - 5.0$) at $\log (L_{X}/{\rm erg~s^{-1}})>44.7$. 

The X-ray flux errors have been taken into account for the estimation of the space density upper and lower limits. For example, if the flux of a source is lower than the applied flux limit but its flux$+1\sigma$ is higher than this limit, then the source will be included in the upper boundary sample. Similarly, if the flux of a source exceed the applied flux limit but its flux$-1\sigma$ is lower than this limit, then taking into account the lower limit, this source will be by the lower boundary sample. The shaded areas include the above uncertainties affecting the computation of the space density, i.e., the flux errors and thus errors on the maximum volume associated to each source.

As explained in Section~\ref{section:Cosmos_sample}, the 15 sources with no optical band detection from C-COSMOS survey have not been included in the space density boundaries. However, we computed the space density assuming that all the 15 sources were at the redshift corresponding to the first bin, then to the second bin and so on \citep{Civano2011}. The space density values computed in this case, in the first three bins, are within the shaded areas.  

The space density in the three luminosity ranges is compared with the predictions, at the same luminosity threshold, from the same three models discussed in Section 4. The LDDE+exp model \citep{Gilli2007} used for the logN-logS (solid lines), including in the model all the sources up to a column density of $10^{25}{\rm cm^{-2}}$. We also compare with the LADE model (\citealp{Aird2010}; dashed line). The LDDE model is fully ruled so it is not included in the following comparisons. In agreement with the results obtained from the number counts, the LDDE+exp model provides an excellent representation of the observed data, although the LADE model cannot be rejected taking into account the upper and lower boundaries. We confirm that the shape of the space-density evolution of X-ray selected luminous AGN is consistent with that derived from optical quasar surveys within current uncertainties.  

The results from the \cite{Ueda2014} for AGNs with the same X-ray luminosity ranges are also plotted for comparison (small black symbols). As can be seen, our results are consistent with those of \cite{Ueda2014} within the statistical errors, indicating a significant decline in the AGN space density from $z = 3$ to higher redshifts. To take into account the observed decline in their LDDE model (Figure~\ref{fig:space}; black dotted lines), \cite{Ueda2014} introduced another (luminosity-dependent) cutoff redshift above which the model declines. Their model indicates an `up-sizing' evolution instead of the global `downsizing' evolution, where more luminous AGN have their number density peak at higher redshifts compared with less luminous ones.  In the cases of lower and intermediate X-ray luminosities the \cite{Ueda2014} LDDE model slightly overestimates our results but it is within the upper boundaries. In the case of the higher X-ray luminosities, where our data significantly reduce the uncertainties of the \cite{Ueda2014} sample, their model underestimates our result while a our data may indicate a flatting up to redshift $z\sim4$.

\subsection{Type-1 vs Type-2 AGN}

Comparing the high redshift evolution of optical- and X-ray- selected AGN samples, the same decline profile has been revealed. Considering that X-ray selected samples provide reduced obscuration bias in comparison with optically selected AGN, this similarity suggests no significant cosmological evolution of obscured AGN fraction at least at higher redshifts. Supporting these result, previous studies have been concluded that the obscured AGN fraction increases with redshift $z = 1$ to $z = 2$  \citep{Ballantyne2006}  but decreases at higher redshifts \citep{Hasinger2008}. To investigate further the cosmological evolution of type-1 and type-2 AGN and their fraction with in the redshift range of $z=3-7$, we calculate the co-moving space density for the two sub-classes of AGN in our sample following the same method as described in Section~\ref{section:space_density}. 

\begin{figure*}
\includegraphics[trim=72 10 30 1, clip, scale=0.40]{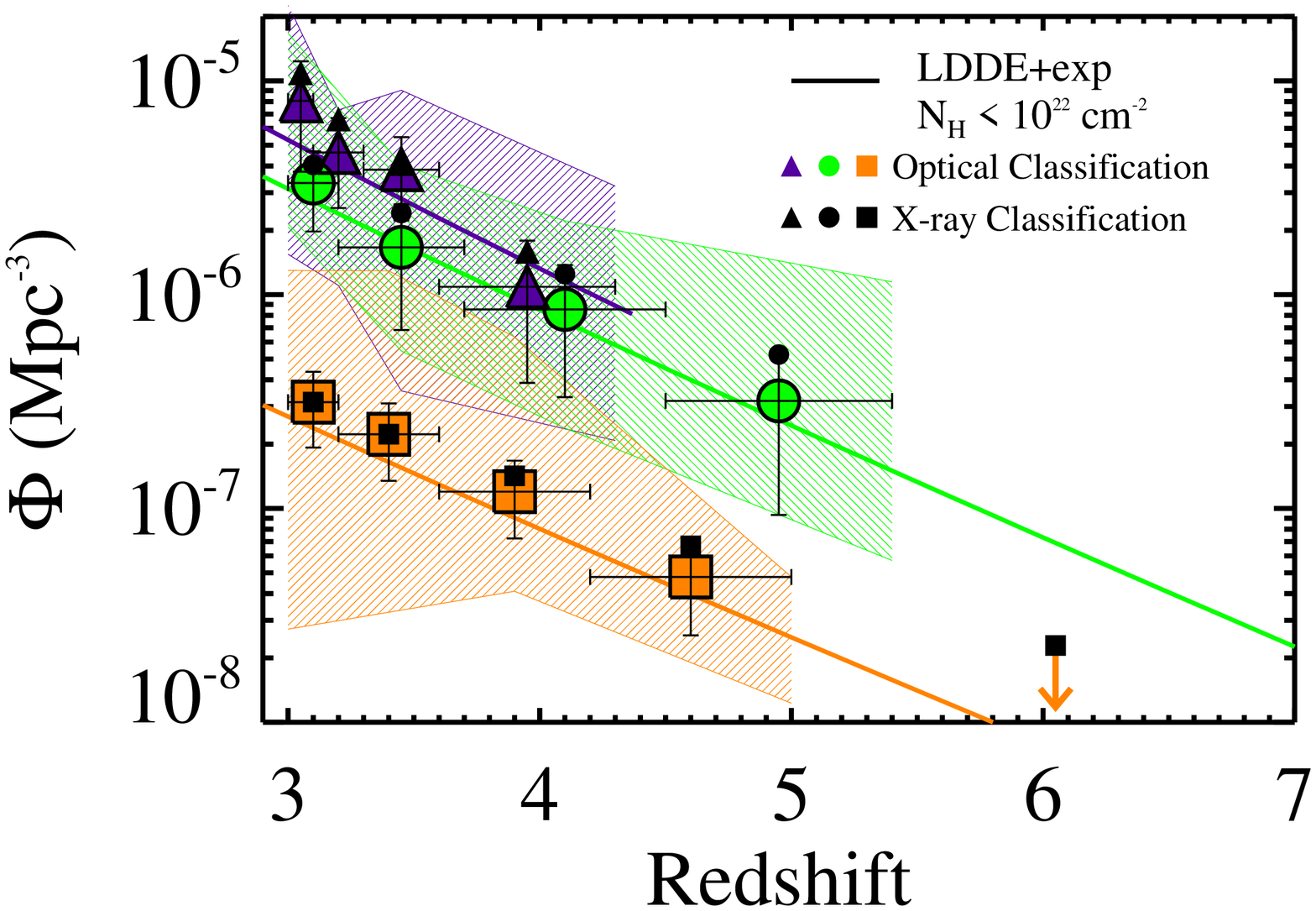}
\includegraphics[trim=72 10 20 1, clip, scale=0.40]{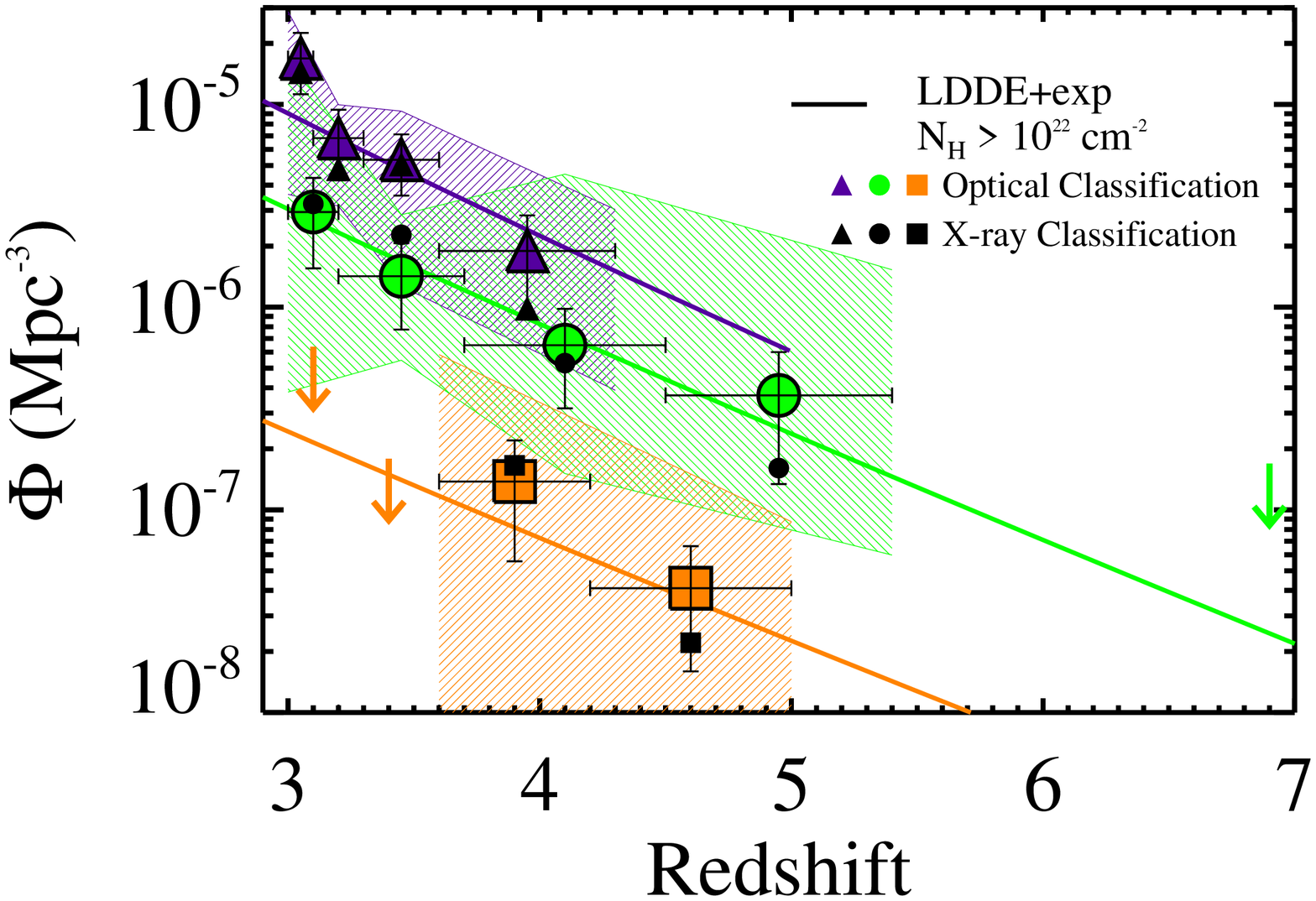}
\caption{The comoving space density in 3 different 2-10 keV X-ray luminosity ranges for type-1 (left) and type-2 (right) AGN. The solid lines corresponds to the X-ray selected AGN space density computed for the same luminosity limit from the Gilli et al. 2007 model and for $N_{H}\leq 10^{22}~{\rm cm^{2}}$ in the case of type-1 AGN and $N_{H}> 10^{22}~{\rm cm^{2}}$ for type-2 AGN. The colors respond to the shaded areas in Figure~\ref{fig:Lx_z} and the shaded area represents the maximum and minimum space density under the assumptions described in Section~\ref{section:space_density}. The small black symbols represent the comoving space density in the 3 different 2-10 keV X-ray luminosity ranges for sources classified as unobscured (left) and obscured (right) AGN based on the X-ray classification and the $N_{H}= 10^{22}~{\rm cm^{2}}$ limit.}
\label{fig:space_types}
\end{figure*}

The co-moving space density is shown in Figure~\ref{fig:space_types}. The upper and lower boundaries are estimated similarly to Section~\ref{section:space_density} taking into account the X-ray flux errors and the 15 C-COSMOS source with no optical band detection. We have used both optical (large symbols) and X-ray (small symbols) AGN classification and they seem to be in a good agreement. To estimate the upper and lower boundaries, we also take into account the $\sim 29$\% of mismatches into the optical and the X-ray classification of AGN types, in order to estimate the upper and lower boundaries. Specifically, in the case of optically classified type-1 AGN, for the upper boundary we also include in each bin the sources classified as unobscured ($N_{H}<10^{22}{\rm cm^{-2}}$) even if they are defined as type-2 AGN based on their optical classification. For the lower boundary, we exclude from each bin the type-1 AGN which have $N_{H}>10^{22}{\rm cm^{-2}}$. The same method is also applied for the type-2 AGN. So, the upper and lower boundaries include also the uncertainties due to the mis-classification of sources. 

The space density in the three luminosity ranges is compared with the predictions, at the same luminosity threshold, from the same LDDE+exp \cite{Gilli2007} model, including in the model all the sources with $N_{H}\leq 10^{22}{\rm cm^{-2}}$ for the case of type-1 AGN and all the sources with $N_{H}=10^{22} -10^{24} {\rm cm^{-2}}$ for the case of type-2 AGN. In agreement with our previous results, the LDDE+exp model provides an excellent representation of both type-1 and type-2 AGN. Since both type-1 and type-2 AGN follow the same decline profile it suggests that there is no significant cosmological evolution for their fraction above $z>3$. The same results are obtained even if we follow the X-ray classification of the AGN sample into obscured and unobscured sources based (Figure~\ref{fig:space_types}; small black symbols). 

The LDDE+exp model for these three luminosity ranges and the $N_{H}=10^{22}{\rm cm^{-2}}$ division predicts a fraction of objects classified as type-2 over the  type-1 AGN sources of 0.63, 0.5 and 0.48 at low luminosities $43.3<\log (L_{X}/{\rm erg~s^{-1}})\leq44.0$, intermediate luminosities $44.0<\log (L_{X}/{\rm erg~s^{-1}})\leq44.7$ and high luminosities $\log (L_{X}/{\rm erg~s^{-1}})>44.7$ respectively. For the same luminosity bins, we have calculated the mean co-moving space density ratio for the two types $\Phi_{type2}/ \Phi_{type1} = 0.59\pm0.02, 0.48\pm0.04~{\rm and}~ 0.31\pm0.18$ from low to high luminosities. These are in good agreement with the LDDE+exp model predictions.  

Recently, \cite{Hiroi2012} estimated the X-ray type-2 AGN fraction, classified based on the $N_{H}>10^{22}{~\rm cm^{-2}}$ criterion, to be $0.54^{+0.17}_{-0.19}$ at $z=3.0-5.0$ and in the luminosity range of $\log L_{X}=44.0-45.0$ while for their optical selection of type-2 AGN they found a fraction of $0.59\pm0.09$. Their estimates are somewhat larger than our result of $0.48\pm0.04$, although they agree within the errors. This difference could be easily explained from the fact that their sample is significantly smaller (30 sources) than ours and contains only 4 sources with $z>4$ from which the 2 classified as type-2 AGN have only photometric redshifts.   

\section{Summary}

We have presented the results of the largest X-ray selected sample of $z>3$ AGN to date, compiled from the C-COSMOS and ChaMP surveys. The large body of C-COSMOS and ChaMP data and their combination allowed us to devise a robust method to build a sizable sample of X-ray selected AGN and control the selection effects including both type-1 (unobscured) and type-2 (obscured) AGN, at $z>3$. Our sample consists of 209 detections in the soft and/or hard  and/or full band. We find:

1) The average HR of the type-1 and type-2 AGN samples is consistent with the $N_{H}<10^{22}{~\rm cm^{-2}}$ criterion for the classification of the X-ray unobscured AGN and the $N_{H}>10^{22}{~\rm cm^{-2}}$ criterion for the classification of the X-ray obscured AGN, respectively.

2) For the individual sources there is a mis-match of $\sim 26$\% between optical and X-ray classification. The contribution from starburst emission in the soft band, or misclassification of faint type-1 with strong optical/IR contamination from host galaxy light could possibly explain the differences between the two classifications.

3) The number counts derived in this work (Figure~\ref{fig:logN_logS}) are consistent with previous determinations from the literature, yet significantly reduce the uncertainties especially at bright fluxes and at high redshifts ($z>4$). The number counts of our combined C-COSMOS/ChaMP sources are consistent with the trend that the space density significantly declines at higher redshifts, similarly to XMM \citep{Brusa2009} and  C-COSMOS \citep{Civano2011} results at similar fluxes at least within their errors and CDF-S \citep{Vito2013} at fainter fluxes, and are better described by the LDDE+exp model \citep{Gilli2007}. 

In contrast to the previous studies, and due mainly to the large sample and wide flux coverage, our results exclude the \cite{Aird2010} LADE model, at the brighter fluxes. At fluxes $2\times 10^{-16}\leq F_{0.5-2~{\rm keV}}\leq 2\times 10^{-15}~{\rm erg~cm^{-2}~s^{-1}}$ the predictions of this model are very similar to the \cite{Gilli2007} LDDE+exp model, but at fainter and brighter fluxes the two models deviate significantly. The \cite{Vito2013} results trace the \cite{Gilli2007} LDDE+exp model well at fainter fluxes, but only for $z>3$, while our sample give the same results at the brighter fluxes where (due to the low expected counts) a large sample is required.   

4) In agreement with the number counts, the space density is well-described with the LDDE+exp model at all X-ray luminosity bins and redshifts, while the LADE model fails to fit the data. These results confirm the declining space density as observed in the optical wavelengths. 

5) Taking into account both optical and X-ray classifications, we derived the space density for type-1 and type-2 (obscured and unobscured) AGN separately. In both cases the results are in agreement with the LDDE+exp model suggesting that the high-redshift evolution of obscured AGNs is similar to that of unobscured AGN. For each luminosity bin, we derived the type-2 AGN fraction among the total AGN sample to be $0.59\pm0.02$, $0.48\pm0.04$ and $0.31\pm0.18$ at $z>3$ in the luminosity ranges of $L_{X}=10^{43.3-44.0},~10^{44.0-44.7},~10^{44.7-46.0}~{\rm  erg~s^{-1}}$. 

The last result should have a significant impact on our understanding of the galaxy and black hole co-evolution. Either or both line-of-sight orientation or evolutionary phase (e.g., covering factor of high column obscuration) can affect the apparent obscuration (and therefore classification) of AGN.  Given that orientation does not evolve preferentially towards us, this work allows us to to say that the obscured and unobscured evolutionary phases do evolve similarly. 

\section*{Acknowledgments}
The authors thank J. Aird and E. Glikman for sharing their luminosity functions and Y. Ueda for providing his space density estimations. The authors would like to thank the referee F. Bauer for the helpful and constructive report. This work was supported by NASA Chandra archival grant, designated code 617817 and contract number AR2-13010X.

\bibliography{biblio.bib}

\begin{thebibliography}{108}
\expandafter\ifx\csname natexlab\endcsname\relax\def\natexlab#1{#1}\fi

\bibitem[{{Aird} {et~al}\mbox{.}(2010){Aird}, {Nandra}, {Laird}, {Georgakakis},
  {Ashby}, {Barmby}, {Coil}, {Huang}, {Koekemoer}, {Steidel}, \&
  {Willmer}}]{Aird2010}
{Aird} J. {et~al.}, 2010, \mnras, 401, 2531

\bibitem[{{Alexander} {et~al}\mbox{.}(2001){Alexander}, {Brandt},
  {Hornschemeier}, {Garmire}, {Schneider}, {Bauer}, \&
  {Griffiths}}]{Alexander2001}
{Alexander} D.~M., {Brandt} W.~N., {Hornschemeier} A.~E., {Garmire} G.~P.,
  {Schneider} D.~P., {Bauer} F.~E., {Griffiths} R.~E., 2001, \aj, 122, 2156

\bibitem[{{Alexandroff} {et~al}\mbox{.}(2013){Alexandroff}, {Strauss},
  {Greene}, {Zakamska}, {Ross}, {Brandt}, {Liu}, {Smith}, {Ge}, {Hamann},
  {Myers}, {Petitjean}, {Schneider}, {Yesuf}, \& {York}}]{Alexandroff2013}
{Alexandroff} R. {et~al.}, 2013, \mnras, 435, 3306

\bibitem[{{Antonucci}(1993)}]{Antonucci1993}
{Antonucci} R., 1993, \araa, 31, 473

\bibitem[{{Ballantyne}(2008)}]{Ballantyne2008}
{Ballantyne} D.~R., 2008, \apj, 685, 787

\bibitem[{{Ballantyne} {et~al}\mbox{.}(2006){Ballantyne}, {Everett}, \&
  {Murray}}]{Ballantyne2006}
{Ballantyne} D.~R., {Everett} J.~E., {Murray} N., 2006, \apj, 639, 740

\bibitem[{{Barger} \& {Cowie}(2005)}]{Barger2005}
{Barger} A.~J., {Cowie} L.~L., 2005, \apj, 635, 115

\bibitem[{{Brandt} \& {Hasinger}(2005)}]{Brandt2005}
{Brandt} W.~N., {Hasinger} G., 2005, \araa, 43, 827

\bibitem[{{Brusa} {et~al}\mbox{.}(2010){Brusa}, {Civano}, {Comastri}, {Miyaji},
  {Salvato}, {Zamorani}, {Cappelluti}, {Fiore}, {Hasinger}, {Mainieri},
  {Merloni}, {Bongiorno}, {Capak}, {Elvis}, {Gilli}, {Hao}, {Jahnke},
  {Koekemoer}, {Ilbert}, {Le Floc'h}, {Lusso}, {Mignoli}, {Schinnerer},
  {Silverman}, {Treister}, {Trump}, {Vignali}, {Zamojski}, {Aldcroft},
  {Aussel}, {Bardelli}, {Bolzonella}, {Cappi}, {Caputi}, {Contini},
  {Finoguenov}, {Fruscione}, {Garilli}, {Impey}, {Iovino}, {Iwasawa},
  {Kampczyk}, {Kartaltepe}, {Kneib}, {Knobel}, {Kovac}, {Lamareille},
  {Leborgne}, {Le Brun}, {Le Fevre}, {Lilly}, {Maier}, {McCracken}, {Pello},
  {Peng}, {Perez-Montero}, {de Ravel}, {Sanders}, {Scodeggio}, {Scoville},
  {Tanaka}, {Taniguchi}, {Tasca}, {de la Torre}, {Tresse}, {Vergani}, \&
  {Zucca}}]{Brusa2010}
{Brusa} M. {et~al.}, 2010, \apj, 716, 348

\bibitem[{{Brusa} {et~al}\mbox{.}(2009){Brusa}, {Comastri}, {Gilli},
  {Hasinger}, {Iwasawa}, {Mainieri}, {Mignoli}, {Salvato}, {Zamorani},
  {Bongiorno}, {Cappelluti}, {Civano}, {Fiore}, {Merloni}, {Silverman},
  {Trump}, {Vignali}, {Capak}, {Elvis}, {Ilbert}, {Impey}, \&
  {Lilly}}]{Brusa2009}
{Brusa} M. {et~al.}, 2009, \apj, 693, 8

\bibitem[{{Brusa} {et~al}\mbox{.}(2003){Brusa}, {Comastri}, {Mignoli}, {Fiore},
  {Ciliegi}, {Vignali}, {Severgnini}, {Cocchia}, {La Franca}, {Matt}, {Perola},
  {Maiolino}, {Baldi}, \& {Molendi}}]{Brusa2003}
{Brusa} M. {et~al.}, 2003, \aap, 409, 65

\bibitem[{{Burlon} {et~al}\mbox{.}(2011){Burlon}, {Ajello}, {Greiner},
  {Comastri}, {Merloni}, \& {Gehrels}}]{Burlon2011}
{Burlon} D., {Ajello} M., {Greiner} J., {Comastri} A., {Merloni} A., {Gehrels}
  N., 2011, \apj, 728, 58

\bibitem[{{Cappelluti} {et~al}\mbox{.}(2009){Cappelluti}, {Brusa}, {Hasinger},
  {Comastri}, {Zamorani}, {Finoguenov}, {Gilli}, {Puccetti}, {Miyaji},
  {Salvato}, {Vignali}, {Aldcroft}, {B{\"o}hringer}, {Brunner}, {Civano},
  {Elvis}, {Fiore}, {Fruscione}, {Griffiths}, {Guzzo}, {Iovino}, {Koekemoer},
  {Mainieri}, {Scoville}, {Shopbell}, {Silverman}, \& {Urry}}]{Cappelluti2009}
{Cappelluti} N. {et~al.}, 2009, \aap, 497, 635

\bibitem[{{Casali} {et~al}\mbox{.}(2007){Casali}, {Adamson}, {Alves de
  Oliveira}, {Almaini}, {Burch}, {Chuter}, {Elliot}, {Folger}, {Foucaud},
  {Hambly}, {Hastie}, {Henry}, {Hirst}, {Irwin}, {Ives}, {Lawrence}, {Laidlaw},
  {Lee}, {Lewis}, {Lunney}, {McLay}, {Montgomery}, {Pickup}, {Read}, {Rees},
  {Robson}, {Sekiguchi}, {Vick}, {Warren}, \& {Woodward}}]{Casali2007}
{Casali} M. {et~al.}, 2007, \aap, 467, 777

\bibitem[{{Civano} {et~al}\mbox{.}(2011){Civano}, {Brusa}, {Comastri}, {Elvis},
  {Salvato}, {Zamorani}, {Capak}, {Fiore}, {Gilli}, {Hao}, {Ikeda}, {Kakazu},
  {Kartaltepe}, {Masters}, {Miyaji}, {Mignoli}, {Puccetti}, {Shankar},
  {Silverman}, {Vignali}, {Zezas}, \& {Koekemoer}}]{Civano2011}
{Civano} F. {et~al.}, 2011, \apj, 741, 91

\bibitem[{{Civano} {et~al}\mbox{.}(2005){Civano}, {Comastri}, \&
  {Brusa}}]{Civano2005}
{Civano} F., {Comastri} A., {Brusa} M., 2005, \mnras, 358, 693

\bibitem[{{Civano} {et~al}\mbox{.}(2012){Civano}, {Elvis}, {Brusa}, {Comastri},
  {Salvato}, {Zamorani}, {Aldcroft}, {Bongiorno}, {Capak}, {Cappelluti},
  {Cisternas}, {Fiore}, {Fruscione}, {Hao}, {Kartaltepe}, {Koekemoer}, {Gilli},
  {Impey}, {Lanzuisi}, {Lusso}, {Mainieri}, {Miyaji}, {Lilly}, {Masters},
  {Puccetti}, {Schawinski}, {Scoville}, {Silverman}, {Trump}, {Urry},
  {Vignali}, \& {Wright}}]{Civano2012}
{Civano} F. {et~al.}, 2012, \apjs, 201, 30

\bibitem[{{Comastri} {et~al}\mbox{.}(2003){Comastri}, {Brunetti}, {Dallacasa},
  {Bondi}, {Pedani}, \& {Setti}}]{Comastri2003}
{Comastri} A., {Brunetti} G., {Dallacasa} D., {Bondi} M., {Pedani} M., {Setti}
  G., 2003, \mnras, 340, L52

\bibitem[{{Cowie} {et~al}\mbox{.}(1996){Cowie}, {Songaila}, {Hu}, \&
  {Cohen}}]{Cowie1996}
{Cowie} L.~L., {Songaila} A., {Hu} E.~M., {Cohen} J.~G., 1996, \aj, 112, 839

\bibitem[{{Croton} {et~al}\mbox{.}(2006){Croton}, {Springel}, {White}, {De
  Lucia}, {Frenk}, {Gao}, {Jenkins}, {Kauffmann}, {Navarro}, \&
  {Yoshida}}]{Croton2006}
{Croton} D.~J. {et~al.}, 2006, \mnras, 365, 11

\bibitem[{{Damen} {et~al}\mbox{.}(2009){Damen}, {Labb{\'e}}, {Franx}, {van
  Dokkum}, {Taylor}, \& {Gawiser}}]{Damen2009}
{Damen} M., {Labb{\'e}} I., {Franx} M., {van Dokkum} P.~G., {Taylor} E.~N.,
  {Gawiser} E.~J., 2009, \apj, 690, 937

\bibitem[{{Draper} \& {Ballantyne}(2010)}]{Draper2010}
{Draper} A.~R., {Ballantyne} D.~R., 2010, \apjl, 715, L99

\bibitem[{{Dye} {et~al}\mbox{.}(2006){Dye}, {Warren}, {Hambly}, {Cross},
  {Hodgkin}, {Irwin}, {Lawrence}, {Adamson}, {Almaini}, {Edge}, {Hirst},
  {Jameson}, {Lucas}, {van Breukelen}, {Bryant}, {Casali}, {Collins}, {Dalton},
  {Davies}, {Davis}, {Emerson}, {Evans}, {Foucaud}, {Gonzales-Solares},
  {Hewett}, {Kendall}, {Kerr}, {Leggett}, {Lodieu}, {Loveday}, {Lewis}, {Mann},
  {McMahon}, {Mortlock}, {Nakajima}, {Pinfield}, {Rawlings}, {Read}, {Riello},
  {Sekiguchi}, {Smith}, {Sutorius}, {Varricatt}, {Walton}, \&
  {Weatherley}}]{Dye2006}
{Dye} S. {et~al.}, 2006, \mnras, 372, 1227

\bibitem[{{Elvis} {et~al}\mbox{.}(2009){Elvis}, {Civano}, {Vignali},
  {Puccetti}, {Fiore}, {Cappelluti}, {Aldcroft}, {Fruscione}, {Zamorani},
  {Comastri}, {Brusa}, {Gilli}, {Miyaji}, {Damiani}, {Koekemoer}, {Finoguenov},
  {Brunner}, {Urry}, {Silverman}, {Mainieri}, {Hasinger}, {Griffiths},
  {Carollo}, {Hao}, {Guzzo}, {Blain}, {Calzetti}, {Carilli}, {Capak}, {Ettori},
  {Fabbiano}, {Impey}, {Lilly}, {Mobasher}, {Rich}, {Salvato}, {Sanders},
  {Schinnerer}, {Scoville}, {Shopbell}, {Taylor}, {Taniguchi}, \&
  {Volonteri}}]{Elvis2009}
{Elvis} M. {et~al.}, 2009, \apjs, 184, 158

\bibitem[{{Fan} {et~al}\mbox{.}(2001){Fan}, {Strauss}, {Schneider}, {Gunn},
  {Lupton}, {Becker}, {Davis}, {Newman}, {Richards}, {White}, {Anderson},
  {Annis}, {Bahcall}, {Brunner}, {Csabai}, {Hennessy}, {Hindsley}, {Fukugita},
  {Kunszt}, {Ivezi{\'c}}, {Knapp}, {McKay}, {Munn}, {Pier}, {Szalay}, \&
  {York}}]{Fan2001}
{Fan} X. {et~al.}, 2001, \aj, 121, 54

\bibitem[{{Ferrarese} \& {Merritt}(2000)}]{Ferrarese2000}
{Ferrarese} L., {Merritt} D., 2000, \apjl, 539, L9

\bibitem[{{Fiore} {et~al}\mbox{.}(2003){Fiore}, {Brusa}, {Cocchia}, {Baldi},
  {Carangelo}, {Ciliegi}, {Comastri}, {La Franca}, {Maiolino}, {Matt},
  {Molendi}, {Mignoli}, {Perola}, {Severgnini}, \& {Vignali}}]{Fiore2003}
{Fiore} F. {et~al.}, 2003, \aap, 409, 79

\bibitem[{{Freeman} {et~al}\mbox{.}(2001){Freeman}, {Doe}, \&
  {Siemiginowska}}]{Freeman2001}
{Freeman} P., {Doe} S., {Siemiginowska} A., 2001, in Society of Photo-Optical
  Instrumentation Engineers (SPIE) Conference Series, Vol. 4477, Astronomical
  Data Analysis, {Starck} J.-L., {Murtagh} F.~D., eds., pp. 76--87

\bibitem[{{Garmire} {et~al}\mbox{.}(2003){Garmire}, {Bautz}, {Ford}, {Nousek},
  \& {Ricker}}]{Garmire2003}
{Garmire} G.~P., {Bautz} M.~W., {Ford} P.~G., {Nousek} J.~A., {Ricker}, Jr.
  G.~R., 2003, in Society of Photo-Optical Instrumentation Engineers (SPIE)
  Conference Series, Vol. 4851, X-Ray and Gamma-Ray Telescopes and Instruments
  for Astronomy., {Truemper} J.~E., {Tananbaum} H.~D., eds., pp. 28--44

\bibitem[{{Gebhardt} {et~al}\mbox{.}(2000){Gebhardt}, {Bender}, {Bower},
  {Dressler}, {Faber}, {Filippenko}, {Green}, {Grillmair}, {Ho}, {Kormendy},
  {Lauer}, {Magorrian}, {Pinkney}, {Richstone}, \& {Tremaine}}]{Gebhardt2000}
{Gebhardt} K. {et~al.}, 2000, \apjl, 539, L13

\bibitem[{{Gilli} {et~al}\mbox{.}(2007){Gilli}, {Comastri}, \&
  {Hasinger}}]{Gilli2007}
{Gilli} R., {Comastri} A., {Hasinger} G., 2007, \aap, 463, 79

\bibitem[{{Glikman} {et~al}\mbox{.}(2011){Glikman}, {Djorgovski}, {Stern},
  {Dey}, {Jannuzi}, \& {Lee}}]{Glikman2011}
{Glikman} E., {Djorgovski} S.~G., {Stern} D., {Dey} A., {Jannuzi} B.~T., {Lee}
  K.-S., 2011, \apjl, 728, L26

\bibitem[{{Granato} {et~al}\mbox{.}(2004){Granato}, {De Zotti}, {Silva},
  {Bressan}, \& {Danese}}]{Granato2004}
{Granato} G.~L., {De Zotti} G., {Silva} L., {Bressan} A., {Danese} L., 2004,
  \apj, 600, 580

\bibitem[{{Granato} {et~al}\mbox{.}(2001){Granato}, {Silva}, {Monaco},
  {Panuzzo}, {Salucci}, {De Zotti}, \& {Danese}}]{Granato2001}
{Granato} G.~L., {Silva} L., {Monaco} P., {Panuzzo} P., {Salucci} P., {De
  Zotti} G., {Danese} L., 2001, \mnras, 324, 757

\bibitem[{{Green} {et~al}\mbox{.}(2009){Green}, {Aldcroft}, {Richards},
  {Barkhouse}, {Constantin}, {Haggard}, {Karovska}, {Kim}, {Kim}, {Vikhlinin},
  {Anderson}, {Mossman}, {Kashyap}, {Myers}, {Silverman}, {Wilkes}, \&
  {Tananbaum}}]{Green2009}
{Green} P.~J. {et~al.}, 2009, \apj, 690, 644

\bibitem[{{Hambly} {et~al}\mbox{.}(2008){Hambly}, {Collins}, {Cross}, {Mann},
  {Read}, {Sutorius}, {Bond}, {Bryant}, {Emerson}, {Lawrence}, {Rimoldini},
  {Stewart}, {Williams}, {Adamson}, {Hirst}, {Dye}, \& {Warren}}]{Hambly2008}
{Hambly} N.~C. {et~al.}, 2008, \mnras, 384, 637

\bibitem[{{Hasinger}(2008)}]{Hasinger2008}
{Hasinger} G., 2008, \aap, 490, 905

\bibitem[{{Hasinger} {et~al}\mbox{.}(2001){Hasinger}, {Altieri}, {Arnaud},
  {Barcons}, {Bergeron}, {Brunner}, {Dadina}, {Dennerl}, {Ferrando},
  {Finoguenov}, {Griffiths}, {Hashimoto}, {Jansen}, {Lumb}, {Mason}, {Mateos},
  {McMahon}, {Miyaji}, {Paerels}, {Page}, {Ptak}, {Sasseen}, {Schartel},
  {Szokoly}, {Tr{\"u}mper}, {Turner}, {Warwick}, \& {Watson}}]{Hasinger2001}
{Hasinger} G. {et~al.}, 2001, \aap, 365, L45

\bibitem[{{Hasinger} {et~al}\mbox{.}(2005){Hasinger}, {Miyaji}, \&
  {Schmidt}}]{Hasinger2005}
{Hasinger} G., {Miyaji} T., {Schmidt} M., 2005, \aap, 441, 417

\bibitem[{{Hewett} {et~al}\mbox{.}(2006){Hewett}, {Warren}, {Leggett}, \&
  {Hodgkin}}]{Hewett2006}
{Hewett} P.~C., {Warren} S.~J., {Leggett} S.~K., {Hodgkin} S.~T., 2006, \mnras,
  367, 454

\bibitem[{{Hiroi} {et~al}\mbox{.}(2012){Hiroi}, {Ueda}, {Akiyama}, \&
  {Watson}}]{Hiroi2012}
{Hiroi} K., {Ueda} Y., {Akiyama} M., {Watson} M.~G., 2012, \apj, 758, 49

\bibitem[{{Hodgkin} {et~al}\mbox{.}(2009){Hodgkin}, {Irwin}, {Hewett}, \&
  {Warren}}]{Hodgkin2009}
{Hodgkin} S.~T., {Irwin} M.~J., {Hewett} P.~C., {Warren} S.~J., 2009, \mnras,
  394, 675

\bibitem[{{Hopkins} {et~al}\mbox{.}(2006){Hopkins}, {Somerville}, {Hernquist},
  {Cox}, {Robertson}, \& {Li}}]{Hopkins2006}
{Hopkins} P.~F., {Somerville} R.~S., {Hernquist} L., {Cox} T.~J., {Robertson}
  B., {Li} Y., 2006, \apj, 652, 864

\bibitem[{{Ikeda} {et~al}\mbox{.}(2011){Ikeda}, {Nagao}, {Matsuoka},
  {Taniguchi}, {Shioya}, {Trump}, {Capak}, {Comastri}, {Enoki}, {Ideue},
  {Kakazu}, {Koekemoer}, {Morokuma}, {Murayama}, {Saito}, {Salvato},
  {Schinnerer}, {Scoville}, \& {Silverman}}]{Ikeda2011}
{Ikeda} H. {et~al.}, 2011, \apjl, 728, L25

\bibitem[{{Ilbert} {et~al}\mbox{.}(2009){Ilbert}, {Capak}, {Salvato}, {Aussel},
  {McCracken}, {Sanders}, {Scoville}, {Kartaltepe}, {Arnouts}, {Le Floc'h},
  {Mobasher}, {Taniguchi}, {Lamareille}, {Leauthaud}, {Sasaki}, {Thompson},
  {Zamojski}, {Zamorani}, {Bardelli}, {Bolzonella}, {Bongiorno}, {Brusa},
  {Caputi}, {Carollo}, {Contini}, {Cook}, {Coppa}, {Cucciati}, {de la Torre},
  {de Ravel}, {Franzetti}, {Garilli}, {Hasinger}, {Iovino}, {Kampczyk},
  {Kneib}, {Knobel}, {Kovac}, {Le Borgne}, {Le Brun}, {F{\`e}vre}, {Lilly},
  {Looper}, {Maier}, {Mainieri}, {Mellier}, {Mignoli}, {Murayama}, {Pell{\`o}},
  {Peng}, {P{\'e}rez-Montero}, {Renzini}, {Ricciardelli}, {Schiminovich},
  {Scodeggio}, {Shioya}, {Silverman}, {Surace}, {Tanaka}, {Tasca}, {Tresse},
  {Vergani}, \& {Zucca}}]{Ilbert2009}
{Ilbert} O. {et~al.}, 2009, \apj, 690, 1236

\bibitem[{{Jiang} {et~al}\mbox{.}(2009){Jiang}, {Fan}, {Bian}, {Annis}, {Chiu},
  {Jester}, {Lin}, {Lupton}, {Richards}, {Strauss}, {Malanushenko},
  {Malanushenko}, \& {Schneider}}]{Jiang2009}
{Jiang} L. {et~al.}, 2009, \aj, 138, 305

\bibitem[{{Jiang} {et~al}\mbox{.}(2007){Jiang}, {Fan}, {Vestergaard}, {Kurk},
  {Walter}, {Kelly}, \& {Strauss}}]{Jiang2007}
{Jiang} L., {Fan} X., {Vestergaard} M., {Kurk} J.~D., {Walter} F., {Kelly}
  B.~C., {Strauss} M.~A., 2007, \aj, 134, 1150

\bibitem[{{Kalfountzou} {et~al}\mbox{.}(2012){Kalfountzou}, {Jarvis},
  {Bonfield}, \& {Hardcastle}}]{Kalfountzou2012}
{Kalfountzou} E., {Jarvis} M.~J., {Bonfield} D.~G., {Hardcastle} M.~J., 2012,
  \mnras, 427, 2401

\bibitem[{{Kalfountzou} {et~al}\mbox{.}(2014){Kalfountzou}, {Stevens},
  {Jarvis}, {Hardcastle}, {Smith}, {Bourne}, {Dunne}, {Ibar}, {Eales},
  {Ivison}, {Maddox}, {Smith}, {Valiante}, \& {de Zotti}}]{Kalfountzou2014}
{Kalfountzou} E. {et~al.}, 2014, \mnras, 442, 1181

\bibitem[{{Kalfountzou} {et~al}\mbox{.}(2011){Kalfountzou}, {Trichas},
  {Rowan-Robinson}, {Clements}, {Babbedge}, \& {Seiradakis}}]{Kalfountzou2011}
{Kalfountzou} E., {Trichas} M., {Rowan-Robinson} M., {Clements} D., {Babbedge}
  T., {Seiradakis} J.~H., 2011, \mnras, 413, 249

\bibitem[{{Kim} {et~al}\mbox{.}(2007){Kim}, {Kim}, {Wilkes}, {Green}, {Kim},
  {Anderson}, {Barkhouse}, {Evans}, {Ivezi{\'c}}, {Karovska}, {Kashyap}, {Lee},
  {Maksym}, {Mossman}, {Silverman}, \& {Tananbaum}}]{Kim2007a}
{Kim} M. {et~al.}, 2007, \apjs, 169, 401

\bibitem[{{Kodama} {et~al}\mbox{.}(2004){Kodama}, {Yamada}, {Akiyama}, {Aoki},
  {Doi}, {Furusawa}, {Fuse}, {Imanishi}, {Ishida}, {Iye}, {Kajisawa}, {Karoji},
  {Kobayashi}, {Komiyama}, {Kosugi}, {Maeda}, {Miyazaki}, {Mizumoto},
  {Morokuma}, {Nakata}, {Noumaru}, {Ogasawara}, {Ouchi}, {Sasaki}, {Sekiguchi},
  {Shimasaku}, {Simpson}, {Takata}, {Tanaka}, {Ueda}, {Yasuda}, \&
  {Yoshida}}]{Kodama2004}
{Kodama} T. {et~al.}, 2004, \mnras, 350, 1005

\bibitem[{{Koekemoer} {et~al}\mbox{.}(2004){Koekemoer}, {Alexander}, {Bauer},
  {Bergeron}, {Brandt}, {Chatzichristou}, {Cristiani}, {Fall}, {Grogin},
  {Livio}, {Mainieri}, {Moustakas}, {Padovani}, {Rosati}, {Schreier}, \&
  {Urry}}]{Koekemoer2004}
{Koekemoer} A.~M. {et~al.}, 2004, \apjl, 600, L123

\bibitem[{{La Franca} {et~al}\mbox{.}(2005){La Franca}, {Fiore}, {Comastri},
  {Perola}, {Sacchi}, {Brusa}, {Cocchia}, {Feruglio}, {Matt}, {Vignali},
  {Carangelo}, {Ciliegi}, {Lamastra}, {Maiolino}, {Mignoli}, {Molendi}, \&
  {Puccetti}}]{LaFranca2005}
{La Franca} F. {et~al.}, 2005, \apj, 635, 864

\bibitem[{{Laird} {et~al}\mbox{.}(2009){Laird}, {Nandra}, {Georgakakis},
  {Aird}, {Barmby}, {Conselice}, {Coil}, {Davis}, {Faber}, {Fazio},
  {Guhathakurta}, {Koo}, {Sarajedini}, \& {Willmer}}]{Laird2009}
{Laird} E.~S. {et~al.}, 2009, \apjs, 180, 102

\bibitem[{{Lanzuisi} {et~al}\mbox{.}(2013){Lanzuisi}, {Civano}, {Elvis},
  {Salvato}, {Hasinger}, {Vignali}, {Zamorani}, {Aldcroft}, {Brusa},
  {Comastri}, {Fiore}, {Fruscione}, {Gilli}, {Ho}, {Mainieri}, {Merloni}, \&
  {Siemiginowska}}]{Lanzuisi2013}
{Lanzuisi} G. {et~al.}, 2013, \mnras, 431, 978

\bibitem[{{Lawrence} \& {Elvis}(1982)}]{Lawrence1982}
{Lawrence} A., {Elvis} M., 1982, \apj, 256, 410

\bibitem[{{Lawrence} \& {Elvis}(2010)}]{Lawrence2010}
{Lawrence} A., {Elvis} M., 2010, \apj, 714, 561

\bibitem[{{Lawrence} {et~al}\mbox{.}(2007){Lawrence}, {Warren}, {Almaini},
  {Edge}, {Hambly}, {Jameson}, {Lucas}, {Casali}, {Adamson}, {Dye}, {Emerson},
  {Foucaud}, {Hewett}, {Hirst}, {Hodgkin}, {Irwin}, {Lodieu}, {McMahon},
  {Simpson}, {Smail}, {Mortlock}, \& {Folger}}]{Lawrence2007}
{Lawrence} A. {et~al.}, 2007, \mnras, 379, 1599

\bibitem[{{Lehmann} {et~al}\mbox{.}(2001){Lehmann}, {Hasinger}, {Schmidt},
  {Giacconi}, {Tr{\"u}mper}, {Zamorani}, {Gunn}, {Pozzetti}, {Schneider},
  {Stanke}, {Szokoly}, {Thompson}, \& {Wilson}}]{Lehmann2001}
{Lehmann} I. {et~al.}, 2001, \aap, 371, 833

\bibitem[{{Lilly} {et~al}\mbox{.}(2009){Lilly}, {Le Brun}, {Maier}, {Mainieri},
  {Mignoli}, {Scodeggio}, {Zamorani}, {Carollo}, {Contini}, {Kneib}, {Le
  F{\`e}vre}, {Renzini}, {Bardelli}, {Bolzonella}, {Bongiorno}, {Caputi},
  {Coppa}, {Cucciati}, {de la Torre}, {de Ravel}, {Franzetti}, {Garilli},
  {Iovino}, {Kampczyk}, {Kovac}, {Knobel}, {Lamareille}, {Le Borgne}, {Pello},
  {Peng}, {P{\'e}rez-Montero}, {Ricciardelli}, {Silverman}, {Tanaka}, {Tasca},
  {Tresse}, {Vergani}, {Zucca}, {Ilbert}, {Salvato}, {Oesch}, {Abbas},
  {Bottini}, {Capak}, {Cappi}, {Cassata}, {Cimatti}, {Elvis}, {Fumana},
  {Guzzo}, {Hasinger}, {Koekemoer}, {Leauthaud}, {Maccagni}, {Marinoni},
  {McCracken}, {Memeo}, {Meneux}, {Porciani}, {Pozzetti}, {Sanders},
  {Scaramella}, {Scarlata}, {Scoville}, {Shopbell}, \& {Taniguchi}}]{Lilly2009}
{Lilly} S.~J. {et~al.}, 2009, \apjs, 184, 218

\bibitem[{{Lilly} {et~al}\mbox{.}(2007){Lilly}, {Le F{\`e}vre}, {Renzini},
  {Zamorani}, {Scodeggio}, {Contini}, {Carollo}, {Hasinger}, {Kneib}, {Iovino},
  {Le Brun}, {Maier}, {Mainieri}, {Mignoli}, {Silverman}, {Tasca},
  {Bolzonella}, {Bongiorno}, {Bottini}, {Capak}, {Caputi}, {Cimatti},
  {Cucciati}, {Daddi}, {Feldmann}, {Franzetti}, {Garilli}, {Guzzo}, {Ilbert},
  {Kampczyk}, {Kovac}, {Lamareille}, {Leauthaud}, {Borgne}, {McCracken},
  {Marinoni}, {Pello}, {Ricciardelli}, {Scarlata}, {Vergani}, {Sanders},
  {Schinnerer}, {Scoville}, {Taniguchi}, {Arnouts}, {Aussel}, {Bardelli},
  {Brusa}, {Cappi}, {Ciliegi}, {Finoguenov}, {Foucaud}, {Franceschini},
  {Halliday}, {Impey}, {Knobel}, {Koekemoer}, {Kurk}, {Maccagni}, {Maddox},
  {Marano}, {Marconi}, {Meneux}, {Mobasher}, {Moreau}, {Peacock}, {Porciani},
  {Pozzetti}, {Scaramella}, {Schiminovich}, {Shopbell}, {Smail}, {Thompson},
  {Tresse}, {Vettolani}, {Zanichelli}, \& {Zucca}}]{Lilly2007}
{Lilly} S.~J. {et~al.}, 2007, \apjs, 172, 70

\bibitem[{{Maccacaro} {et~al}\mbox{.}(1988){Maccacaro}, {Gioia}, {Wolter},
  {Zamorani}, \& {Stocke}}]{Maccacaro1988}
{Maccacaro} T., {Gioia} I.~M., {Wolter} A., {Zamorani} G., {Stocke} J.~T.,
  1988, \apj, 326, 680

\bibitem[{{Maddox} {et~al}\mbox{.}(2008){Maddox}, {Hewett}, {Warren}, \&
  {Croom}}]{Maddox2008}
{Maddox} N., {Hewett} P.~C., {Warren} S.~J., {Croom} S.~M., 2008, \mnras, 386,
  1605

\bibitem[{{Magorrian} {et~al}\mbox{.}(1998){Magorrian}, {Tremaine},
  {Richstone}, {Bender}, {Bower}, {Dressler}, {Faber}, {Gebhardt}, {Green},
  {Grillmair}, {Kormendy}, \& {Lauer}}]{Magorrian1998}
{Magorrian} J. {et~al.}, 1998, \aj, 115, 2285

\bibitem[{{McConnell} \& {Ma}(2013)}]{McConnell2013}
{McConnell} N.~J., {Ma} C.-P., 2013, \apj, 764, 184

\bibitem[{{Menci} {et~al}\mbox{.}(2008){Menci}, {Fiore}, {Puccetti}, \&
  {Cavaliere}}]{Menci2008}
{Menci} N., {Fiore} F., {Puccetti} S., {Cavaliere} A., 2008, \apj, 686, 219

\bibitem[{{Merloni} {et~al}\mbox{.}(2014){Merloni}, {Bongiorno}, {Brusa},
  {Iwasawa}, {Mainieri}, {Magnelli}, {Salvato}, {Berta}, {Cappelluti},
  {Comastri}, {Fiore}, {Gilli}, {Koekemoer}, {Le Floc'h}, {Lusso}, {Lutz},
  {Miyaji}, {Pozzi}, {Riguccini}, {Rosario}, {Silverman}, {Symeonidis},
  {Treister}, {Vignali}, \& {Zamorani}}]{Merloni2014}
{Merloni} A. {et~al.}, 2014, \mnras, 437, 3550

\bibitem[{{Nandra} \& {Pounds}(1994)}]{Nandra1994}
{Nandra} K., {Pounds} K.~A., 1994, \mnras, 268, 405

\bibitem[{{Noterdaeme} {et~al}\mbox{.}(2012){Noterdaeme}, {Petitjean},
  {Carithers}, {P{\^a}ris}, {Font-Ribera}, {Bailey}, {Aubourg}, {Bizyaev},
  {Ebelke}, {Finley}, {Ge}, {Malanushenko}, {Malanushenko},
  {Miralda-Escud{\'e}}, {Myers}, {Oravetz}, {Pan}, {Pieri}, {Ross},
  {Schneider}, {Simmons}, \& {York}}]{Noterdaeme2012}
{Noterdaeme} P. {et~al.}, 2012, \aap, 547, L1

\bibitem[{{P{\^a}ris} {et~al}\mbox{.}(2014){P{\^a}ris}, {Petitjean}, {Aubourg},
  {Ross}, {Myers}, {Streblyanska}, {Bailey}, {Hall}, {Strauss}, {Anderson},
  {Bizyaev}, {Borde}, {Brinkmann}, {Bovy}, {Brandt}, {Brewington},
  {Brownstein}, {Cook}, {Ebelke}, {Fan}, {Filiz Ak}, {Finley}, {Font-Ribera},
  {Ge}, {Hamann}, {Ho}, {Jiang}, {Kinemuchi}, {Malanushenko}, {Malanushenko},
  {Marchante}, {McGreer}, {McMahon}, {Miralda-Escud{\'e}}, {Muna},
  {Noterdaeme}, {Oravetz}, {Palanque-Delabrouille}, {Pan}, {Perez-Fournon},
  {Pieri}, {Riffel}, {Schlegel}, {Schneider}, {Simmons}, {Viel}, {Weaver},
  {Wood-Vasey}, {Y{\`e}che}, \& {York}}]{Paris2014}
{P{\^a}ris} I. {et~al.}, 2014, \aap, 563, A54

\bibitem[{{Park} {et~al}\mbox{.}(2006){Park}, {Kashyap}, {Siemiginowska}, {van
  Dyk}, {Zezas}, {Heinke}, \& {Wargelin}}]{Park2006}
{Park} T., {Kashyap} V.~L., {Siemiginowska} A., {van Dyk} D.~A., {Zezas} A.,
  {Heinke} C., {Wargelin} B.~J., 2006, \apj, 652, 610

\bibitem[{{Puccetti} {et~al}\mbox{.}(2009){Puccetti}, {Vignali}, {Cappelluti},
  {Fiore}, {Zamorani}, {Aldcroft}, {Elvis}, {Gilli}, {Miyaji}, {Brunner},
  {Brusa}, {Civano}, {Comastri}, {Damiani}, {Fruscione}, {Finoguenov},
  {Koekemoer}, \& {Mainieri}}]{Puccetti2009}
{Puccetti} S. {et~al.}, 2009, \apjs, 185, 586

\bibitem[{{Richards} {et~al}\mbox{.}(2002){Richards}, {Fan}, {Newberg},
  {Strauss}, {Vanden Berk}, {Schneider}, {Yanny}, {Boucher}, {Burles},
  {Frieman}, {Gunn}, {Hall}, {Ivezi{\'c}}, {Kent}, {Loveday}, {Lupton},
  {Rockosi}, {Schlegel}, {Stoughton}, {SubbaRao}, \& {York}}]{Richards2002}
{Richards} G.~T. {et~al.}, 2002, \aj, 123, 2945

\bibitem[{{Richards} {et~al}\mbox{.}(2009){Richards}, {Myers}, {Gray},
  {Riegel}, {Nichol}, {Brunner}, {Szalay}, {Schneider}, \&
  {Anderson}}]{Richards2009}
{Richards} G.~T. {et~al.}, 2009, \apjs, 180, 67

\bibitem[{{Richards} {et~al}\mbox{.}(2006){Richards}, {Strauss}, {Fan}, {Hall},
  {Jester}, {Schneider}, {Vanden Berk}, {Stoughton}, {Anderson}, {Brunner},
  {Gray}, {Gunn}, {Ivezi{\'c}}, {Kirkland}, {Knapp}, {Loveday}, {Meiksin},
  {Pope}, {Szalay}, {Thakar}, {Yanny}, {York}, {Barentine}, {Brewington},
  {Brinkmann}, {Fukugita}, {Harvanek}, {Kent}, {Kleinman}, {Krzesi{\'n}ski},
  {Long}, {Lupton}, {Nash}, {Neilsen}, {Nitta}, {Schlegel}, \&
  {Snedden}}]{Richards2006}
{Richards} G.~T. {et~al.}, 2006, \aj, 131, 2766

\bibitem[{{Ross} {et~al}\mbox{.}(2013){Ross}, {McGreer}, {White}, {Richards},
  {Myers}, {Palanque-Delabrouille}, {Strauss}, {Anderson}, {Shen}, {Brandt},
  {Y{\`e}che}, {Swanson}, {Aubourg}, {Bailey}, {Bizyaev}, {Bovy}, {Brewington},
  {Brinkmann}, {DeGraf}, {Di Matteo}, {Ebelke}, {Fan}, {Ge}, {Malanushenko},
  {Malanushenko}, {Mandelbaum}, {Maraston}, {Muna}, {Oravetz}, {Pan},
  {P{\^a}ris}, {Petitjean}, {Schawinski}, {Schlegel}, {Schneider}, {Silverman},
  {Simmons}, {Snedden}, {Streblyanska}, {Suzuki}, {Weinberg}, \&
  {York}}]{Ross2013}
{Ross} N.~P. {et~al.}, 2013, \apj, 773, 14

\bibitem[{{Ruiz} {et~al}\mbox{.}(2010){Ruiz}, {Miniutti}, {Panessa}, \&
  {Carrera}}]{Ruiz2010}
{Ruiz} A., {Miniutti} G., {Panessa} F., {Carrera} F.~J., 2010, \aap, 515, A99

\bibitem[{{Salvato} {et~al}\mbox{.}(2009){Salvato}, {Hasinger}, {Ilbert},
  {Zamorani}, {Brusa}, {Scoville}, {Rau}, {Capak}, {Arnouts}, {Aussel},
  {Bolzonella}, {Buongiorno}, {Cappelluti}, {Caputi}, {Civano}, {Cook},
  {Elvis}, {Gilli}, {Jahnke}, {Kartaltepe}, {Impey}, {Lamareille}, {Le Floc'h},
  {Lilly}, {Mainieri}, {McCarthy}, {McCracken}, {Mignoli}, {Mobasher},
  {Murayama}, {Sasaki}, {Sanders}, {Schiminovich}, {Shioya}, {Shopbell},
  {Silverman}, {Smol{\v c}i{\'c}}, {Surace}, {Taniguchi}, {Thompson}, {Trump},
  {Urry}, \& {Zamojski}}]{Salvato2009}
{Salvato} M. {et~al.}, 2009, \apj, 690, 1250

\bibitem[{{Salvato} {et~al}\mbox{.}(2011){Salvato}, {Ilbert}, {Hasinger},
  {Rau}, {Civano}, {Zamorani}, {Brusa}, {Elvis}, {Vignali}, {Aussel},
  {Comastri}, {Fiore}, {Le Floc'h}, {Mainieri}, {Bardelli}, {Bolzonella},
  {Bongiorno}, {Capak}, {Caputi}, {Cappelluti}, {Carollo}, {Contini},
  {Garilli}, {Iovino}, {Fotopoulou}, {Fruscione}, {Gilli}, {Halliday}, {Kneib},
  {Kakazu}, {Kartaltepe}, {Koekemoer}, {Kovac}, {Ideue}, {Ikeda}, {Impey}, {Le
  Fevre}, {Lamareille}, {Lanzuisi}, {Le Borgne}, {Le Brun}, {Lilly}, {Maier},
  {Manohar}, {Masters}, {McCracken}, {Messias}, {Mignoli}, {Mobasher}, {Nagao},
  {Pello}, {Puccetti}, {Perez-Montero}, {Renzini}, {Sargent}, {Sanders},
  {Scodeggio}, {Scoville}, {Shopbell}, {Silvermann}, {Taniguchi}, {Tasca},
  {Tresse}, {Trump}, \& {Zucca}}]{Salvato2011}
{Salvato} M. {et~al.}, 2011, \apj, 742, 61

\bibitem[{{Schmidt}(1968)}]{Schmidt1968}
{Schmidt} M., 1968, \apj, 151, 393

\bibitem[{{Schmidt} {et~al}\mbox{.}(1998){Schmidt}, {Hasinger}, {Gunn},
  {Schneider}, {Burg}, {Giacconi}, {Lehmann}, {MacKenty}, {Trumper}, \&
  {Zamorani}}]{Schmidt1998}
{Schmidt} M. {et~al.}, 1998, \aap, 329, 495

\bibitem[{{Schmidt} {et~al}\mbox{.}(1995){Schmidt}, {Schneider}, \&
  {Gunn}}]{Schmidt1995}
{Schmidt} M., {Schneider} D.~P., {Gunn} J.~E., 1995, \aj, 110, 68

\bibitem[{{Silverman} {et~al}\mbox{.}(2008){Silverman}, {Green}, {Barkhouse},
  {Kim}, {Kim}, {Wilkes}, {Cameron}, {Hasinger}, {Jannuzi}, {Smith}, {Smith},
  \& {Tananbaum}}]{Silverman2008}
{Silverman} J.~D. {et~al.}, 2008, \apj, 679, 118

\bibitem[{{Silverman} {et~al}\mbox{.}(2010){Silverman}, {Mainieri}, {Salvato},
  {Hasinger}, {Bergeron}, {Capak}, {Szokoly}, {Finoguenov}, {Gilli}, {Rosati},
  {Tozzi}, {Vignali}, {-a-Alexander}, {Brandt}, {Lehmer}, {Luo}, {Rafferty},
  {Xue}, {Balestra}, {Bauer}, {Brusa}, {Comastri}, {Kartaltepe}, {Koekemoer},
  {Miyaji}, {Schneider}, {Treister}, {Wisotski}, \& {Schramm}}]{Silverman2010}
{Silverman} J.~D. {et~al.}, 2010, \apjs, 191, 124

\bibitem[{{Spergel} {et~al}\mbox{.}(2003){Spergel}, {Verde}, {Peiris},
  {Komatsu}, {Nolta}, {Bennett}, {Halpern}, {Hinshaw}, {Jarosik}, {Kogut},
  {Limon}, {Meyer}, {Page}, {Tucker}, {Weiland}, {Wollack}, \&
  {Wright}}]{Spergel2003}
{Spergel} D.~N. {et~al.}, 2003, \apjs, 148, 175

\bibitem[{{Steidel} {et~al}\mbox{.}(1996){Steidel}, {Giavalisco}, {Pettini},
  {Dickinson}, \& {Adelberger}}]{Steidel1996}
{Steidel} C.~C., {Giavalisco} M., {Pettini} M., {Dickinson} M., {Adelberger}
  K.~L., 1996, \apjl, 462, L17

\bibitem[{{Stern} \& {Laor}(2012)}]{Stern2012}
{Stern} J., {Laor} A., 2012, \mnras, 426, 2703

\bibitem[{{Stocke} {et~al}\mbox{.}(1991){Stocke}, {Morris}, {Gioia},
  {Maccacaro}, {Schild}, {Wolter}, {Fleming}, \& {Henry}}]{Stocke1991}
{Stocke} J.~T., {Morris} S.~L., {Gioia} I.~M., {Maccacaro} T., {Schild} R.,
  {Wolter} A., {Fleming} T.~A., {Henry} J.~P., 1991, \apjs, 76, 813

\bibitem[{{Treister} \& {Urry}(2006)}]{Treister2006}
{Treister} E., {Urry} C.~M., 2006, \apjl, 652, L79

\bibitem[{{Treister} {et~al}\mbox{.}(2009){Treister}, {Urry}, \&
  {Virani}}]{Treister2009}
{Treister} E., {Urry} C.~M., {Virani} S., 2009, \apj, 696, 110

\bibitem[{{Trichas} {et~al}\mbox{.}(2009){Trichas}, {Georgakakis},
  {Rowan-Robinson}, {Nandra}, {Clements}, \& {Vaccari}}]{Trichas2009}
{Trichas} M., {Georgakakis} A., {Rowan-Robinson} M., {Nandra} K., {Clements}
  D., {Vaccari} M., 2009, \mnras, 399, 663

\bibitem[{{Trichas} {et~al}\mbox{.}(2013){Trichas}, {Green}, {Constantin},
  {Aldcroft}, {Kalfountzou}, {Sobolewska}, {Hyde}, {Zhou}, {Kim}, {Haggard}, \&
  {Kelly}}]{Trichas2013}
{Trichas} M. {et~al.}, 2013, \apj, 778, 188

\bibitem[{{Trichas} {et~al}\mbox{.}(2012){Trichas}, {Green}, {Silverman},
  {Aldcroft}, {Barkhouse}, {Cameron}, {Constantin}, {Ellison}, {Foltz},
  {Haggard}, {Jannuzi}, {Kim}, {Marshall}, {Mossman}, {P{\'e}rez},
  {Romero-Colmenero}, {Ruiz}, {Smith}, {Smith}, {Torres}, {Wik}, {Wilkes}, \&
  {Wolfgang}}]{Trichas2012}
{Trichas} M. {et~al.}, 2012, \apjs, 200, 17

\bibitem[{{Trichas} {et~al}\mbox{.}(2010){Trichas}, {Rowan-Robinson},
  {Georgakakis}, {Valtchanov}, {Nandra}, {Farrah}, {Morrison}, {Clements}, \&
  {Waddington}}]{Trichas2010}
{Trichas} M. {et~al.}, 2010, \mnras, 405, 2243

\bibitem[{{Ueda} {et~al}\mbox{.}(2014){Ueda}, {Akiyama}, {Hasinger}, {Miyaji},
  \& {Watson}}]{Ueda2014}
{Ueda} Y., {Akiyama} M., {Hasinger} G., {Miyaji} T., {Watson} M.~G., 2014,
  \apj, 786, 104

\bibitem[{{Ueda} {et~al}\mbox{.}(2003){Ueda}, {Akiyama}, {Ohta}, \&
  {Miyaji}}]{Ueda2003}
{Ueda} Y., {Akiyama} M., {Ohta} K., {Miyaji} T., 2003, \apj, 598, 886

\bibitem[{{Urry} \& {Padovani}(1995)}]{Urry1995}
{Urry} C.~M., {Padovani} P., 1995, \pasp, 107, 803

\bibitem[{{Vignali} {et~al}\mbox{.}(2003){Vignali}, {Brandt}, \&
  {Schneider}}]{Vignali2003}
{Vignali} C., {Brandt} W.~N., {Schneider} D.~P., 2003, \aj, 125, 433

\bibitem[{{Vito} {et~al}\mbox{.}(2013){Vito}, {Vignali}, {Gilli}, {Comastri},
  {Iwasawa}, {Brandt}, {Alexander}, {Brusa}, {Lehmer}, {Bauer}, {Schneider},
  {Xue}, \& {Luo}}]{Vito2013}
{Vito} F. {et~al.}, 2013, \mnras, 428, 354

\bibitem[{{Warren} {et~al}\mbox{.}(2000){Warren}, {Hewett}, \&
  {Foltz}}]{Warren2000}
{Warren} S.~J., {Hewett} P.~C., {Foltz} C.~B., 2000, \mnras, 312, 827

\bibitem[{{Willott} {et~al}\mbox{.}(2010){Willott}, {Delorme}, {Reyl{\'e}},
  {Albert}, {Bergeron}, {Crampton}, {Delfosse}, {Forveille}, {Hutchings},
  {McLure}, {Omont}, \& {Schade}}]{Willott2010}
{Willott} C.~J. {et~al.}, 2010, \aj, 139, 906

\bibitem[{{Wright} {et~al}\mbox{.}(2010){Wright}, {Eisenhardt}, {Mainzer},
  {Ressler}, {Cutri}, {Jarrett}, {Kirkpatrick}, {Padgett}, {McMillan},
  {Skrutskie}, {Stanford}, {Cohen}, {Walker}, {Mather}, {Leisawitz}, {Gautier},
  {McLean}, {Benford}, {Lonsdale}, {Blain}, {Mendez}, {Irace}, {Duval}, {Liu},
  {Royer}, {Heinrichsen}, {Howard}, {Shannon}, {Kendall}, {Walsh}, {Larsen},
  {Cardon}, {Schick}, {Schwalm}, {Abid}, {Fabinsky}, {Naes}, \&
  {Tsai}}]{Wright2010}
{Wright} E.~L. {et~al.}, 2010, \aj, 140, 1868

\bibitem[{{Wu} {et~al}\mbox{.}(2012){Wu}, {Hao}, {Jia}, {Zhang}, \&
  {Peng}}]{Wu2012}
{Wu} X.-B., {Hao} G., {Jia} Z., {Zhang} Y., {Peng} N., 2012, \aj, 144, 49

\bibitem[{{Wu} \& {Jia}(2010)}]{Wu2010}
{Wu} X.-B., {Jia} Z., 2010, \mnras, 406, 1583

\bibitem[{{Wu} {et~al}\mbox{.}(2004){Wu}, {Zhang}, \& {Zhou}}]{Wu2004}
{Wu} X.-B., {Zhang} W., {Zhou} X., 2004, ChjAA, 4, 17

\bibitem[{{Xue} {et~al}\mbox{.}(2011){Xue}, {Luo}, {Brandt}, {Bauer}, {Lehmer},
  {Broos}, {Schneider}, {Alexander}, {Brusa}, {Comastri}, {Fabian}, {Gilli},
  {Hasinger}, {Hornschemeier}, {Koekemoer}, {Liu}, {Mainieri}, {Paolillo},
  {Rafferty}, {Rosati}, {Shemmer}, {Silverman}, {Smail}, {Tozzi}, \&
  {Vignali}}]{Xue2011}
{Xue} Y.~Q. {et~al.}, 2011, \apjs, 195, 10

\bibitem[{{Young} {et~al}\mbox{.}(2010){Young}, {Elvis}, \&
  {Risaliti}}]{Young2010}
{Young} M., {Elvis} M., {Risaliti} G., 2010, \apj, 708, 1388

\end{thebibliography}

\appendix

\section{} 

QSO candidates at redshift $\simeq3.0-3.5$ satisfied the following cuts: 
\begin{eqnarray}
\begin{array}{cclc}
\sigma_r & < & 0.13 & {\rm AND} \\
u        & > & 20.6 & {\rm AND} \\
u-g      & > & 1.5  & {\rm AND} \\
g-r      & < & 1.2  & {\rm AND} \\
r-i      & < & 0.3  & {\rm AND} \\
i-z      & > & -1.0 & {\rm AND} \\
g-r      & < & 0.44(u-g)-0.76.\\
\end{array}
\end{eqnarray}
For the redshift range $\simeq3.5-4.5$ this selection becomes
\begin{eqnarray}
\begin{array}{ccclccclc}
{\rm A}) \;\, &\sigma_r & < & 0.2 & \\
{\rm B}) \;\, &u-g      & > & 1.5  & OR & u & > & 20.6 \\
{\rm C}) \;\, &g-r      & > & 0.7  & \\
{\rm D}) \;\, &g-r      & > & 2.8  & OR & r-i & < & 0.44(g-r)-0.558\\
{\rm E}) \;\, &i-z      & < & 0.25 & {\rm AND} & i-z & > & -1.0,\\
\end{array}
\end{eqnarray}
in the combination A AND B AND C AND D AND E.  For the redshifts above
$\simeq4.5$ we use 
\begin{eqnarray}
\begin{array}{cclc}
u        & > & 21.5 & {\rm AND} \\
g        & > & 21.0  & {\rm AND} \\
r-i      & > & 0.6  & {\rm AND} \\
i-z      & > & -1.0 & {\rm AND} \\
i-z      & < & 0.52(r-i)-0.762.\\
\end{array}
\end{eqnarray}

\begin{table*}
\caption{Properties of the high - redshift AGN sample.}
\begin{minipage}{18.3cm}
\begin{tabular} {c c c c c c c c c c c c c c c c}

\hline
\hline

Survey\footnote{1=ChaMP; 2=C-COSMOS} & R.A.  & Dec. & $z_{spec}$ & $z_{phot}$ & $z_{phot}$\footnote{Photometric redshift lower limit}	&	$z_{phot}$\footnote{Photometric redshift upper limit}  &  Optical & X-ray &  $S_{\rm soft}$\footnote{The 0.5-2 keV flux in units of ${10^{-13}\rm erg~cm^{-2}~s^{-1}}$ converted to $\Gamma=1.8$. For sources undetected in soft band a negative symbol is given and the flux estimated by the hard or full band detection.} & log$Lx$\footnote{The 2-10 keV luminosity in units of ${\rm erg ~s^{-1}}$.} & HR\footnote{Hardness ratio defined as HR = (H + S)/(H - S), S: 0.5-2.0 keV count rate and H: 2.0-10 keV count rate.} & HR &  HR & $N_{\rm H}$\footnote{The absorption column density in units of ${10^{22}\rm cm^{-2}}$}\\

	& (deg)  & (deg) & &  & lower& upper  &  Type\footnote{1: optical type-1 AGNs, and 2: optical type-2 AGNs} & Type\footnote{1: unobscured AGNs, and 2: obscured AGNs; based on $N_{\rm H}=10^{22}~{\rm cm^{-2}}$ limit.}  &  & & &  low\footnote{Hardness ratio lower limit} &up\footnote{Hardness ratio upper limit} & \\

\hline
\hline

1&	1.682	&	-0.200	&	3.109	&	...		&	...	&	...		&	1	&	1	&	0.071	&	44.65	& -0.50	&	-0.77	&	-0.12 & 0.1\\
1&	5.826	&	-1.048	&	...		&	4.5		&	4.3	&	4.7		& 	1	&	1	&	0.094	&	45.13	& ...		&	...		& -0.38	 &	$<1.0$\\
1&	18.1925	&	-1.2188	&	3.592	&	3.545	&	3.35	&	4.37		&	1	&	1	&	 0.090	&	44.90	&	-0.84	&	-0.98	&-0.57	& $<0.1$\\
1&	29.8140	&	0.4529	&	...		&	3.675	&	3.51	&	4.05		&	1	&	1	&	0.039	&	44.56	&	-0.92	&	...	&	-0.84	& $<0.1$\\
1&	29.886	&	0.482	&	...		&	4.6		&	4.3	&	4.9		&	2	&	1	&	0.037	&	44.75	&	-0.70	&	...	& -0.53	& $<0.1$\\
1&	32.680	&	-0.3051	&	4.733	&	4.565	&	4.05	&	4.98		&	1	&	1	&	0.277	&	45.65	&	-0.67	&	-0.85	&-0.56	&$<0.1$\\
1&	38.906	&	-0.722	&	...		&	4.15		&	3.72	&	4.58		&	2	&	1	&	0.203	&	45.39	&	-0.58	&	-0.76	& -0.41	& $<0.1$\\
1&	116.8620	&	27.6253	&	3.152	&	2.905	&	2.58	&	3.25		&	1	&	1	&	0.072	&	44.67	&	-0.42	&	-0.70	&	-0.09 & 0.1\\
1& 	119.1940	&	41.1201	&	3.734	&	3.135	&	2.88	&	4.29		&	1	&	1	&	0.130	&	45.09	&	-0.41	&	-0.66	&-0.14	& 0.1\\
1&	119.218	&	45.044	&	3.185	&	...		&	...	&	...		&	1	&	1	&	0.105	&	44.85	&	-0.85	&	-0.98	&-0.62	&$<0.1$\\
1&	120.2877	&	36.1514	&	...		&	3.675	&	3.59	&	4.06		&	1	&	1	&	0.156	&	45.16	&	-0.64	&	-0.69	&	-0.56& $<0.1$\\
1&	125.4118	&	12.2922	&	3.112	&	...		&	...	&	...		&	1	&	1	&	0.053	&	44.53	&	-0.64	&	-0.76	&	-0.47& $<0.1$\\
1&	125.4123	&	12.2917	&	3.114	&	2.995	&	2.78	&	3.23		&	1	&	1	&	0.443	&	45.45	&	-0.48	&	-0.51 	& -0.42	& $<0.1$\\
1&	127.837	&	19.151	& 	....		&	4.6		& 	4.2	&	5.0		&	2	&	1	&	0.065	&	44.99	&	-0.57	&	-0.73 	& -0.42 & $<0.1$\\
1&	130.2606	&	13.2202	&	...		&	2.945	&	2.76	&	3.29		&	1	&	1	&	0.078	&	44.66	&	-0.54	&	-0.66	&	-0.44 & $<0.1$\\
1&	132.169	&	44.959	& 	3.093	&	...		&	...	&	...	&	1	&	1	&	0.030	&	44.28	&	-0.31	&	-0.41	& -0.22 & 4.0\\
1&	133.446	&	57.987	& 	....		& 	4.1		&	3.4	&	4.8	&	2	&	1	&	0.043	&	44.70	&	-0.36	&	-0.55	&-0.19 & 4.0\\
1&	134.4131	&	9.0255	& 	3.130	&	3.065	&	2.76	&	3.39	&	1	&	1	&	0.076	&	44.69	&	-0.57	&	-0.89	&-0.05 & 0.1\\
1&	137.6868	&	17.7420	& 	4.098	&	4.105	&	3.35	&	4.47	&	1	&	1	&	0.044	&	44.71	&	-0.50	&	-0.67	&-0.35 & 0.1\\
1&	137.786	&	54.298	& 	3.234	&	...		&	...	&	...	&	2	&	1	&	0.036	&	44.40	&	-0.44	&	-0.55	&-0.35 & 0.1\\
1&	138.0425	&	5.7952	& 	3.246	&	3.345	&	3.09	&	3.63	&	1	&	1	&	0.065	&	44.64	&	-0.50	&	-0.70	&-0.29 & 0.1\\
1&	139.086	&	29.522	& 	3.098	&	...		&	...	&	...	&	1	&	1	&	0.054	&	44.53	&	-0.40	&	-0.57	&-0.19 & 0.1\\
1&	141.0521	&	31.2774	& 	...		& 	2.795	&	2.21	&	3.03	&	1	&	1	&	0.367	&	45.34	&	-0.65	&	-0.79	&-0.48 & $<0.1$\\
1&	143.3646	&	55.4032	& 	...		& 	3.045	&	2.78	&	3.33	&	1	&	1	&	0.086	&	44.72	&	-0.31	&	-0.43	&-0.20 & 4.0\\
1&	149.425	&	33.251	& 	3.001	& 	...		&	...	&	...	&	1	&	1	&	0.128	&	44.87	&	-0.77	&	-0.93	&-0.49 & $<0.1$\\
1&	149.5942	&	7.7968	& 	3.220	&	3.145	&	2.8	&	3.47	&	1	&	1	&	0.086	&	44.77	&	...		&	...		&	-0.88 	& $<0.1$\\
1&	150.7009	&	32.7668	& 	...		&	3.405	& 	3.11	&	4.37	&	1	&	1	&	 0.085	&	44.82	 &	 -0.45 & -0.61 & -0.34 & 0.1\\
1&	155.4521	&	13.1647	&	3.055	&    2.905	&	2.55	&	3.26	&	1	&	1	&	 0.075	&	44.66	 &	 -0.47 & -0.67 & -0.21 & 0.1\\
1&	156.5952	&	47.3187	&	 4.941	&	4.895	&	4.77	&	5.1	&	1	&	1	&	 0.102	&	45.25	 &	 -0.44 & -0.90 & -0.24 & 0.1\\
1&	162.590	&	58.625	& 	...		&	3.65		& 	3.47	&	3.83	&	2	&	1	&	 0.077	&	44.84	 &	 -0.56 & -0.82 & -0.22 & 0.1\\
1&	162.938	&	57.468	&	 3.409	&	...		&	...	&	...	&	1	&	1	&	 0.051	&	44.60	 &	 -0.99 & -1.0 & -0.73 & $<0.1$\\
1&	163.275	&	57.5735	&	...		&	2.755	&	2.52	&	3.1	&	1	&	1	&	0.045	&	44.44	&	-0.75  & -0.87 & -0.59 & $<0.1$ \\	
1&	168.5505	&	40.5661	&	 3.597	&	3.495	&	3.09	&	4.35	&	1	&	1	&	 0.053	&	44.67	 &	 -0.44 & -0.6 & -0.32 & 0.1\\
1&	168.7335	&	40.6372	&	 4.913	&	1.305	&	0.9	&	1.52	&	2	&	1	&	 0.253	&	45.64	 &	 -0.69 & -0.73 & -0.64 & $<0.1$\\
1&	169.4807	&	48.0722	& 	...		&	2.905	& 	2.54	&	3.23	&	1	&	1	&	 0.101	&	44.84	 &	 -0.68 & -0.77 & -0.57 & $<0.1$\\
1&	169.607	&	7.7262	& 	...		& 	3.75		& 	3.45	&	4.05	&	1	&	2	&	 0.035	&	44.53	 &	 -0.24 & -0.41 & -0.09 & 13.0\\
1&	170.0871	&	43.4292	&	 3.552	&	3.665	&	2.92	&	4.12	&	1	&	1	&	 0.073	&	44.79	 &	 -0.46 & -0.61 & -0.32 & 0.1\\
1&	170.5414	&	24.2938	&	 3.336	&	3.045	&	2.73	&	3.39	&	1	&	1	&	 0.054	&	44.61	 &	 -0.51 & -0.66 & -0.38 & 0.1\\
1&	171.748	&	56.772	& 	...		& 	3.6		& 	3.5	&	3.7	&	1	& 	1	&	 0.033	&	44.46	 &	 -0.77 & -0.96 & -0.53 & $<0.1$\\
1&	175.151	&	66.221	&	 3.337	&	...		&	...	&	...	&	1	&	1	&	 0.065	&	44.68	 &	 -0.67 & -0.74 & -0.61 & $<0.1$\\
1&	178.8811	&	-1.770	&	 3.202 	&	3.495	&	3.16	&	4.21	&	1	&	1	&	 0.055	&	44.57	 &	 -0.35 & -0.47 & -0.21 & 2.0\\
1&	182.037	&	0.129 	& 	...		&	3.5		& 	2.4	&	4.6	&	2	&	2	&	 0.056	&	44.66	 &	 -0.17 & -0.44 & 0.08 & 17.0\\
1&	183.2733	&	2.8584	&	...		&	3.145	&	2.76	&	3.83	&	1	&	1	&	0.165	&	45.11	&	-0.64 & -0.71 & -0.53 & $<0.1$ \\
1&	183.366	&	2.960 	& 	...		&	3.5 		&	2.4	&	4.6	&	1	&	1	&	 0.090	&	44.87	 &	 -0.46 & -0.65 & -0.24 & 0.1\\
1&	188.1259	&	47.6153	&	 3.041	&	3.205	&	2.88	&	3.83	&	1	&	1	&	 0.175	&	45.02	 &	 -0.99 & -0.99 & -0.76 & $<0.1$\\
1&	190.4594	&	9.6307 	& 	...		&	3.045	& 	2.66	&	3.37	&	2	&	2	&	 0.057	&	44.54	 &	 -0.18 & -0.44 & 0.04 & 13.0\\
1&	192.551	&	33.855	&	 3.6 		&	...		&	...	&	...	&	1	&	1	&	 0.074	&	44.81	 &	 -0.54 & -0.83 & -0.35 & 0.1\\
1&	192.9299 &	0.1240	&	...		&	2.905	&	2.69	&	3.23	&	1	&	1	&	0.220	&	45.38	&	-0.5 & -0.68 & -0.27 & 0.1\\
1&	195.0080	&	1.3067	&	 4.612	&	4.485	&	3.88	&	4.69	&	1	&	1	&	 0.454	&	45.84	 &	 -0.37 & -0.73 & -0.02 & 4.0\\
\hline
\end{tabular}
\end{minipage}
\end{table*}

\begin{table*}
\begin{minipage}{20.3cm}
\begin{tabular} {c c c c c c  c c c c c c c c c c}
\hline
\hline
Survey & R.A.  & Dec. & $z_{spec}$ & $z_{phot}$ & $z_{phot}$	&	$z_{phot}$  &  Optical & X-ray &  $S_{\rm soft}$ & log$Lx$& HR& HR &  HR & $N_{\rm H}$\\

	& (deg)  & (deg) & &  & lower& upper  &  Type& Type &  & & &  lower &upper & \\
\hline
\hline

1&	195.445	&	-0.112	&	 3.7 		& 	...		&	...	&	...	&	1	&	2	&	 0.160	&	45.17	 &	 -0.16 & -0.41 & 0.09 & 20.0\\
1&	196.534	&	3.940 	&	 6.016	&	...		&	...	&	...	&	1	&	1	&	 0.031	&	44.93	 &	 -0.69 & -0.75 & -0.61 & $<0.1$\\
1&	196.6521	&	46.4961	& 	...		&	3.065	&	2.76	&	3.23	&	2	&	1	&	 0.055	&	44.53	 &	 -0.26 & -0.53 & 0.06 & 8.0\\
1&	197.4348	&	57.6389	& 	...		&	2.905	&	2.67	&	3.25	&	1	&	1	&	 0.068	&	44.69	 &	 ...  &  ... & -0.64 &$<0.1$\\
1&	198.2170	&	42.4716	&	3.181	&	2.945	&	2.75	&	3.3	&	1	&	1	&	0.072	&	44.81	&	-0.66 & -0.82	& -0.52 & $<0.1$	\\
1&	199.322	&	1.1538	& 	...		&	3.4 		& 	3.275&	3.525&	1	&	1	&	 0.090	&	44.84	 &	 -0.59 & -0.94 & -0.20 & 0.1\\
1&	201.424	&	11.477	& 	...		&	4.6 		& 	4.3	&	4.9	&	2	&	2	&	 0.071	&	45.03	 &	 0.01 & -0.33 & 0.4 & 50.0\\
1&  201.8283  &	29.1482	&	...		&	3.675	&	3.6	&	4.08	&	1	&	1	&	0.030	&	44.45	&	-0.97 & -0.99	&	-0.53 & $<0.1$ \\
1&  202.9168  &	11.2840	&	...		&	2.755	&	2.21	&	3.03	&	1	&	1	&	0.057	&	44.53	&	-0.56 & -0.68	&  -0.40  & $<0.1$ \\
1&  203.5933  &	-1.4221	&	3.827	&	3.775	&	3.01	&	4.33	&	1	&	1	&	0.032	&	44.51	&	-0.51 & -0.90	&  -0.34  & 0.1 \\
1&	 203.6122 & 	-1.4816 	& 	...		& 	2.905 	& 	2.57	&	3.26	& 	1	&	1	& 	0.074 	& 	44.69	 &	 -0.66 & -0.94 & -0.54 & $<0.1$\\
1&	 206.078	 & 	-0.511 	& 	3.070 	&	...		&	...	&	...	& 	1	&	1	& 	0.060 	& 	44.57	 &	 -0.64 & -0.84 & -0.51 & $<0.1$\\
1&	 207.7399 & 	60.1377	& 	...		& 	3.285 	& 	3.05	& 	3.95	&	1	&	1	& 	0.110 	& 	44.89	 &	 -0.68 & -0.75 & -0.63 & $<0.1$1\\
1&	 208.2111 & 	33.4822	& 	...		& 	2.755 	& 	2.32	&	3.07	& 	1	&	1	& 	0.056 	& 	44.54	 &	 -0.462 & -0.57 & -0.30 & 0.1\\
1&	212.7671	&	52.2988	&	...		&	2.815	&	2.66	&	3.13	&	1	&	1	&	0.097	&	44.99	&	-0.42 & -0.48 & -0.36 & 0.1 \\
1&	 214.2243 & 	44.6294	& 	...		& 	2.795 	& 	2.52	& 	3.12	&	1	&	1	& 	0.134 	& 	44.94	 &	 -0.54 & -0.66 & -0.43 & $<0.1$\\
1&	 214.4241 & 	53.0886	& 	...		& 	3.495 	& 	3.15	&	4.37	& 	1	&	1	& 	0.037 	& 	44.49	 &	 -0.75 & -0.85 & -0.61 & $<0.1$\\
1&	 214.8523 & 	53.5323	& 	...		& 	3.535 	& 	3.41	&	3.9 	& 	1	&	1	& 	0.185 	& 	45.20	 &	 -0.74 & -0.95 & -0.47 &$<0.1$\\
1&	 215.1504 & 	47.2415	& 	3.237 	&	2.905	&	2.71	&	3.26	& 	1	&	1	& 	0.300 	& 	45.32	 &	 -0.75 & -0.90 & -0.58 &$<0.1$\\
1&	 215.7710 & 	24.0856	& 	4.112 	& 	4.165	&	3.36	&	4.54	& 	1	&	2	& 	0.032 	& 	44.58	 &	 -0.26 & -0.54 & 0.01 & 14.0\\
1&	 216.0719 & 	22.8406	& 	...		& 	3.675 	& 	3.58	&	4.06	& 	1	&	1	& 	0.734 	& 	45.83	 &	 -0.60 & -0.64 & -0.56 &$<0.1$\\
1&	 216.447	 & 	35.455	& 	3.53 	&	...		&	...	&	...	& 	1	&	1	& 	0.053 	& 	44.65	 &	 -0.56 & -0.66 & -0.41 & $<0.1$\\
1&	 218.114	 & 	-1.160	& 	...		& 	3.3 		& 	3.0	&	3.6	& 	2	&	1	& 	0.079 	& 	44.76	 &	 -0.41 & -0.56 & -0.28 & 0.1\\
1&	 219.5024 & 	3.6388	& 	3.306 	&	3.285	&	2.92	&	4.08	& 	1	&	1	& 	0.058 	& 	44.62	 &	 -0.58 & -0.68 & -0.48 & $<0.1$\\
1&	 219.633	 & 	3.641	& 	...		& 	3.275 	& 	3.05	&	3.5	& 	1	&	1	& 	0.070 	& 	44.70	 &	 -0.81 & -0.89 & -0.72 & $<0.1$\\
1&	 219.6653 & 	3.6288	& 	3.194 	&	3.205	&	2.88	&	3.94 & 	1	&	1	& 	0.040 	& 	44.44	 &	 -0.68 & -0.77 & -0.56 & $<0.1$\\
1&	 220.8577 & 	58.8995 	& 	...		& 	2.905 	& 	2.68	&	3.22	&	1	&	1	& 	0.030 	& 	44.32	 &	 -0.84 & -0.96 & -0.35 & 0.1\\
1&	 221.3084 & 	-0.4237 	& 	3.142 	&	3.065	&	2.68	&	3.36	& 	1	&	1	& 	0.058 	& 	44.58	 &	 -0.24 & -0.54 & 0.0 & 10.0\\
1&	 225.0836 & 	22.7196	& 	3.268 	&	3.475	&	3.18	&	4.21	& 	1	&	1	& 	0.051 	& 	44.56	 &	 -0.79 & -0.97 & -0.41 & $<0.1$\\
1&	 230.1945 & 	52.9896	& 	3.385 	&	3.495	&	3.15	&	4.35	& 	1	&	1	& 	0.060 	& 	44.66	 &	 -0.63 & -0.83 & -0.34 & $<0.1$\\
1&	 230.8973 & 	28.6609	& 	...		& 	2.715 	& 	2.44	&	3.02	& 	1	&	1	& 	0.062 	& 	44.56	 &	 -0.61 & -0.72 & -0.48 & $<0.1$\\
1&	 240.6529 & 	42.5518	& 	3.889	& 	4.125 	& 	3.27	&	4.46	& 	1	&	1	& 	0.045 	& 	44.72	 &	 -0.82 & -1.0 & -0.71 & $<0.1$\\
1&	 244.0288 & 	47.2661	& 	...		& 	3.205 	& 	2.92	&	3.97	& 	1	&	1	& 	0.058 	& 	44.60	 &	 -0.58 & -0.95 & 0.01 & 0.1\\
1&	 258.2694 & 	61.3793	& 	3.150 	&	3.005	&	2.76	&	3.33	& 	1	&	1	& 	0.126 	& 	44.91	 &	  ...	  & ... & -0.89 & $<0.1$\\
1&	 258.956	 & 	63.246	& 	...		& 	3.6 		& 	3.5	&	3.7	& 	1	& 	1	&	0.139 	& 	45.09	 &	 -0.61 & -0.86 & -0.27 & $<0.1$\\
1&	 260.0728 & 	26.5628	& 	3.057 	&	3.145	&	2.77	&	3.5	& 	1	&	1	&	0.053 	& 	44.51	 &	 -0.85 & -0.95 & -0.71 & $<0.1$\\
1&	 340.9458 & 	-9.6855	& 	...		& 	4.125 	& 	3.36	&	4.56	& 	1	&	1	& 	0.078 	& 	44.97	 &	 -0.71 & -0.88 & -0.48 & $<0.1$\\
1&	 359.3544 & 	0.6643	& 	...		& 	3.285 	& 	2.98	& 	4.19	&	1	& 	1	&	0.039 	& 	44.45	 &	  ...	  & ... & -0.75 & $<0.1$\\
2&	 149.5438 &	 1.9537    &	 ...		 &	2.951	 & 	2.91	 &	3.01	&	2	&	2	&	 0.024	&	44.28	&	 0.49 & 0.32 	& 0.79 & 80.0\\
2&	 149.6242 &	 1.8854	 &	 ... 		& 	2.264 	&    2.1	&	3.86	 &	1	&	1	&	 0.007	&	43.98	&	 ...		 &	...	 & -0.34 & $<3.0$\\
2&	149.6908	&	2.2641	& 	...		&	...		&	$>$3.0 &	...	&	...	&	2	&	0.031	&	45.46	&	0.22		&	 -0.17	 & 0.53 & 50.0\\
2&	 149.6696 &	 2.1677	 &	 3.089	 &	3.094 	 & 	2.75	&	3.21	& 2		&	2	&	 0.018	&	44.21	&	 -0.04	 & -0.18	 & 0.10 & 22.0\\
2&	 149.7362 &	 2.1799	 &	 4.255	 &	4.245	& 	4.23	 &	4.26	& 	1	&	1	&	0.014	&	44.40	& 	-0.41 	& -0.59	 & -0.23 & 0.1\\
2&	 149.7425 &	 2.5354	 &	 ...		 &	2.942	 & 	2.9	 & 	3.01	&	2	&	2	 &    -0.015 	&    44.30	 & 	...		 & 0.31 & ... & 	$\geq$55.0\\
2&	 149.7617 &	 2.4358	 &	 ...		 &	 3.647	 &  	3.64	&	3.66	 & 1		 &	2	&	0.023	 & 	44.46	 & 	-0.16 	& -0.39 	& -0.11 & 22.0\\
2&	 149.7710 &	 2.3658	 &	 ...		 &	 3.447	 &  3.35	&	3.53	& 2 		&	2	&	0.002	&	43.38	 & 	... 		& 	...	 & 0.73 & $<1000.0$\\
2&	 149.7821 &	 2.4713	 &	 ...		 &	 3.246	 &   3.23	&	3.26	&	1	&	2	& 	-0.014	 &  44.334	 & ...		 & -0.05 	& ... & $\geq$25.0\\
2&	 149.7837 &	 2.4521	 &	 5.07	 &	1.939	 & 	1.9	&	1.98	& 	2	&	2	&	0.007	 & 44.31		 & ... 		& ...		 & 0.11 & $<75.0$\\
2&	149.7975	&	2.2897	&	...		&	...		&	$>$3.0 &	...	&	...	&	2	&	-0.002	&	44.33	&	...	&	0.69	& ... & $<1000.0$\\
2&	 149.8035 &	 2.1417	 &	 ...		 &	 2.838	 &  2.24	&	3.09	 & 	1	&	2	 & 	0.009	&	43.87	&	 -0.37 	& 	-0.58	 & -0.12	& 1.5\\
2& 	149.8046	 &	 2.1189	 &	 ...		 & 	3.791	& 	1.59	 & 	4.58	&	2	&	2	&	 0.004	 & 43.68		 & ...	 &	...	 & 0.60	 &	 $<300.0$\\
2& 	149.8079	 &	 1.8105	 &	 ...		 & 	2.382	&	2.33	&	3.260& 	1	&	2	& 	0.006	 & 43.79		 & -0.22 & -0.65	 & 0.20	 & 11.0\\
2& 	149.8085	 &	 2.3148	 &	 ...		 & 3.471 		& 	3.41	&	3.79	 &	2	 & 	1	&	0.025	 & 44.44	 	& -0.34	 &	 -0.46	 & -0.23	 & 4.0\\
2& 	149.8123	 &	 2.2830	 &	 ...		 & 3.297 		& 	3.24	 &	3.35	& 	1	 & 	1	&	0.009	 & 43.95		 & 	...	 & 	... & 	-0.69	 & 0.1\\
2&	149.8144	&	2.7348	&	...		&	...		&	$>$3.0 &	...	&	...	&	2	&	0.021	&	45.29	&	...	&	...	&	0.51	&	$<85$\\	
2&	149.8233	&	2.5398	&	...		&	...		&	$>$3.0 &	...	&	...	&	2	&	-0.010	&	44.97	&	...	&	0.52	& ...	&	$>85$\\	
2& 	149.8458	 &	 2.4817	 &	 ...		 & 3.375 		& 	3.36	&	3.39	 & 1		&	1	 & 	0.007	 & 43.87		 & ...	 & 	... 	& -0.498 	& $<0.1$\\
2& 	149.8515	 &	 2.2764	 &	 3.371 	& 3.317		&	3.3	&	3.33	 & 1		&	1	 & 0.012		 & 44.09		 & -0.53 	& -0.70	 & -0.35		 & 0.1\\
2& 	149.8517	 &	 2.4269	 &	 ...		 & 3.35		 & 	3.16	&	3.58	 & 	2	&	2	 & 0.004		 & 43.69		 & 0.33 & 	0.15	 & 0.52	 & 68.0\\
2& 	149.8585	 &	 2.4093	 &	 ...		 & 4.108 		&    3.68	&	4.29	& 	2	 &	2	&   0.049		 & 44.91		 & -0.27		& -0.34	 & -0.21	 & 11.0\\
2& 	149.8605	 &	 2.3876	 &	 ...		 & 2.949 		& 	2.33	&	3.97	 & 	1	&	2	&	 0.002 	& 43.51		 & 0.22	 & -0.13		 & 0.57 & 65.0\\
2& 	149.8697	 &	 2.2941	 &	 3.345 	& 3.4		 & 	3.39	&	3.41	 & 	1	 & 	2	&	0.024	 & 44.42		 & 0.06	 & 	-0.02	 & 0.15	 & 37.0\\
2& 	149.8744	 &	 2.3615	 &	 ...		 & 6.88		&	6.88	& 	7.0 	& 	2	&	2	&	 0.002	 & 44.10		 & ...	 & 	...	 & 	0.74 & $<1000.0$\\
2& 	149.8792	 &	 2.2258	 &	 3.65	 & 3.647		& 	3.64	&	3.66	 & 1		&	1	& 0.018		 & 44.35		 & -0.41 	& -0.54	 & -0.28 & 0.1\\
2& 	149.8825	 &	 2.5051	 &	 ...		 & 3.1		 &	3.02	&	3.19	&	2	&	2	 &  0.046	 	& 44.60		 & -0.22	 & -0.29	 & -0.16 & 10.0\\
2& 	149.8861	 &	 2.2759	 &	 3.335	 & 3.277		 & 	3.26	&	3.3	& 	2	&	2	 &  -0.004 	&  43.83 		&	...	 & 0.29	 & 	... & $>65.0$\\

\hline
\end{tabular}
\end{minipage}
\end{table*}

\begin{table*}
\begin{minipage}{20.3cm}
\begin{tabular} {c c c c c c c c c c c c c c c c}
\hline
\hline
Survey & R.A.  & Dec. & $z_{spec}$ & $z_{phot}$ & $z_{phot}$	&	$z_{phot}$  &  Optical & X-ray &  $S_{\rm soft}$ & log$Lx$& HR& HR &  HR & $N_{\rm H}$\\

	& (deg)  & (deg) & &  & lower& upper  &  Type& Type &  & & &  lower &upper & \\

\hline
\hline
2& 	149.8894	 &	 1.9662	 &	 ...		 & 3.053		 &   2.2	&	3.12	& 	2	 &	1	&	 0.004	 & 43.47		 & ...	 & 	...	 & -0.32 &  $<5.0$\\
2& 	149.8942	 &	 2.4330	 &	 3.382	 & 3.393		 & 	3.38	&	3.41	& 1		 &	2	&	 0.003	 & 43.43		 & -0.19 	& -0.70 	& 0.31	 & 14.0\\
2& 	149.8977	 &	 2.3244	 &	 ...		 & 2.998		 & 	2.9	&	3.16	&	2	&	2	 &  0.006		 & 43.72		 & ...	 & ...	 & 0.45	 & $<75.0$\\
2&	149.9084	&	2.5723	&	...		&	...		&	$>$3.0 &	...	&	...	&	2	&	0.004	&	43.64	&	...	&	...	&	0.88	&	$<1000.0$\\
2& 	149.9093	 &	 2.6199	 &	 ...		 & 2.569		 &   2.13	&	4.23	 & 1		&	1	& 0.006		 & 44.02		 & ...	 & ...	 & -0.29 & $<6.0$\\
2& 	149.9108	 &	 1.8996	 &	 ...		 & 3.063		 &   2.585 &	3.194 & 2		&	2	&	 0.002	 & 43.24		 & ...	 & ...	 & 0.62 & $<200.0$\\
2& 	149.9111	 &	 1.8427	 &	 ...		 & 3.295		 & 	2.81	&	3.45	 & 2		&	...	& -0.003		 & 43.50		 & ... 	& ... & ... & ...\\
2&	149.9137	&	2.2465	&	...		&	...		&	$>$3.0 &	...	&	...	&	2	&	0.010	&	44.02	&	0.46	&	0.23	 & 0.52	 & 71.0\\
2&	 149.9195 &	 2.3454	 & 3.021 		& 3.043		 & 	3.03	&	3.05	 & 	1	&	1 & 0.009 & 43.85		 & ...	& ... & 	-0.69 & $<0.1$\\
2&	 149.9249 & 	1.8441	 & ...	 	& 2.936		 & 	1.79	&	3.05	 & 	2	 &	2	& 0.004	 & 43.55	 & ...	 & 	...	 & -0.21		 & $<13.0$\\
2&	149.9316	&	2.3519	&	...		&	...		&	$>$3.0 &	...	&	...	&	2	&	0.004	&	43.62	&	...	&	0.87	&	...	&	1000.0\\
2&	 149.9445 & 	1.7404	 & 	...		& 2.044		 & 	2.0	&	3.66	 &	1	&	1	 & 0.006 	& 43.88	 & ...	 & ...	 & -0.28 & $<6.0$\\
2&	 149.9522 & 	2.6514	 & 3.08 		& 	3.083	 & 	3.07	 &	3.09	& 	2	&	2	 & -0.011	 & 	44.16	& ...	 & 0.411 & ... &	$>70.0$ \\
2&	 149.9666	& 2.4325	 & 	...	 & 2.893	 &  2.23		 &	3.3	& 2		&		2	 & -0.013	 & 	44.33	 & ... & 0.753 & ... & $>1000.0$\\
2&	 149.9692	& 2.3048	 & 3.155 	& 	3.099	 & 	3.076	&	3.13		& 	1	&	2		&	 0.014	 & 44.12	 & 0.06	 & -0.07	 & 0.19	 & 32.0\\
2&	 149.9728	& 1.9415	 & 	...	 & 2.843 & 2.63	&	3.06	& 	2	&	1		&	0.013	 & 44.02		 & -0.26	 & -0.41	 & -0.12	 & 7.0\\
2&	 149.9816	& 2.3150	 & 	...	 & 3.003 & 	2.97	&	3.03	 &	2	 &	2	&	 0.023		 & 44.28		 & -0.06	 & -0.15	 & 0.04	& 22.0\\
2&	 149.9905	& 2.2973	& 3.026 	& 	2.97		&  2.93	&	3.05	 & 	2	&	1	&	0.009	 & 43.87			 & 	...	 & 	... & -0.72 & $<0.1$\\
2&	 149.9986	& 1.9745	& 	...	& 2.911 & 	2.88	 &	3.01	& 	2	&	1	 & 	0.013	 & 44.02		 & ...	 & 	...	 & -0.78 & $<0.1$\\
2&	 150.0044	& 2.0389	& 3.515 & 	3.5	 &	3.49	 &	3.51	& 	1	 & 	1	 & 0.018		 & 44.31		 & -0.41	 & -0.54	 & -0.28		 & 0.1\\
2&	150.0070		&	1.9180	&	...	&	...	&	$>$3.0 &	...	&	...	&	2	&	0.006	&	43.76	&	0.60	&	 0.33 & 0.78 & 150.0\\
2&	 150.0078	& 2.1489	&  ...	& 2.851 & 	2.77	&	3.43	 & 	1	&	2	 & 0.005		 & 43.78		 & 0.17	 & -0.03	 & 0.37	 & 48.0\\
2&	 150.0094	& 1.8526	& ...		 & 4.032 & 	4.02	&	4.05	 & 	2	 &	1	&	 0.008	 & 44.11		 & ...	 & 	...	 & -0.63 &$<0.1$\\
2&	 150.0172	& 2.4979	& ...		 & 2.914 & 	2.5	 &	3.26	& 	2	&	1	 & 0.009		 & 43.95		 & ...	 & 	... & -0.71 &$<0.1$\\
2&	 150.0258	& 2.0038	& ...		 & 3.124 & 	2.67	&	3.17	& 	2	&	2	 & -0.006	 &	43.93	 & ...	 & 0.48 & 	... & $\ge$1000.0\\
2&	 150.0427	& 1.8722	& 3.371 	&  3.338	 & 	3.32	 &	3.35	& 	1	&	2	 & 0.008 		& 43.94		 & -0.32	 & -0.58	 & -0.06	 & 5.0\\
2&	 150.0465	& 2.3674	& 	...	 & 2.313	&	1.94	&	3.1	&	2	 &	1	& 0.004		& 43.50		 & ...	 & ...	 & -0.39 & 0.1\\
2&	150.0483		&	2.4816	& ...	&	...	&	$>$3.0 &	...	&	...	&	2	&	0.004	&	43.64	&	...	&	...	&	0.86	&	1000.0\\
2&	 150.0621	& 1.7226	& 	...	 & 3.033 & 	2.91	&	3.41	&	2	&	2	& 0.008	 & 43.84		 & 0.06	 & -0.21 & 0.33 & 32.0\\
2&	 150.0635	& 2.4219	& 	...	 & 3.087 & 	3.02	&	3.25	&	2	&	2	&	-0.006 & 43.93	 & ...	 & 0.48 & ...	 & $>80.0$\\
2&	 150.0637	& 1.7774	& ... 	& 3.426 & 	1.21	&	3.84	& 	2	&	2	 &	-0.009	 & 44.17	 & 	...	 & 0.32	 & ... & $>50.0$\\
2&	 150.0647	& 2.1910	& 	...	 & 2.887 & 	1.53	 &	3.06	&	1	&	2	 & 0.025		 & 44.35	 & 0.11	 & 0.03	 & 0.19 & 35.0\\
2&	 150.0673	& 2.0843	& ... 	&  3.01 & 	2.99	  &	3.03	& 	1	&	1	 & 0.006	 & 43.66	 & ...	 & ... & -0.57	 & $<0.1$\\
2&	 150.0862	& 2.139 	& ...	 	& 5.045 &  4.73	&	5.14	 & 	1	&	2	 & 0.003	 & 43.99	 & 0.25 & 0.01 & 0.49 & 100.0\\
2&	 150.0910	& 1.9292	& ...	 	& 2.864	&  	2.84	&	3.23	&	1	&	2	&	-0.007	&	44.04	 & ... & 0.54 & ... & $>90.0$\\
2&	 150.0969	& 2.0215	& 3.546 	& 	3.362	&	3.33	&	3.39 &	2	&	1	&	0.006 & 43.81	 &	...	 & 	...	 & -0.59 & $<0.1$\\
2&	 150.0999	& 2.1527	& 	...	 & 2.844	& 	2.6	&	3.06	& 2 &	1	&	 0.008	 & 43.81	 & ...	 & ... & -0.67 & $<0.1$\\
2&	 150.1010	& 2.4194	& 4.66 & 	4.545	&	4.53	&	4.56	 & 1	&	2	& 0.009	 & 44.33	 & -0.27	 & -0.49 & -0.05 & 14.0\\
2&	 150.1074	& 1.7592	& 4.16 & 	3.949	&	3.93	&	3.96	 & 1		&	2	 & 0.006	&	  44.01 & 0.01 & -0.28 & 0.3 & 43.0\\
2&	 150.1336	& 2.4574	& 3.189 & 	3.111	&	3.08	 &	3.15	& 	2	&	1	 & 0.006	 & 43.71	 & ...	 & 	...	 & -0.494 & $<0.1$\\
2& 	150.1522		& 2.3079	& 3.175 & 	3.163	&	3.08	 &	3.24	&	2	&	2	 & 0.003 & 43.55	 & 0.42 & 0.24	 & 0.60 & 75.0\\
2& 	150.1640		& 2.1793	&	...	 & 2.82 		& 	2.75	&	3.15	& 	1	 &	2	&	-0.004	&	43.57 & 	...	 & 0.407 & ... & 27.0\\
2& 	150.1763		& 2.5697	& 	...	 & 3.144 		& 	 3.09	&	3.24	 &	2	&	2	&	 0.005	 & 43.61	 & ...	 & ... & 0.65 & $<5000.0$\\
2& 	150.1807		& 2.0760	& 3.01 & 	3.015	&	3.0	&	3.03	& 1	&	1 & 0.011	 & 43.94		 & ...	 & 	... & -0.77 &$<0.1$\\
2& 	150.1926		& 2.2199	& 3.09 & 	3.077	& 	3.07	&	3.09	&	1	&	2	&	 0.007 & 43.76	 & -0.15 & -0.39 & 0.1 & 15.0\\
2& 	150.1997		& 1.7312	& 	...	 & 2.821 		& 	2.66	&	3.45	&	2	&	2	 &  0.004	 & 43.68	 & 0.18 & -0.10	 & 0.47 & 50.0\\
2& 	150.2050		& 1.7360	& 	...	 & 2.929 		&	2.4	 &	4.19	& 1	&	2	 & 0.006 & 44.02	 & 0.04	& -0.22 & 0.30 & 48.0\\
2& 	150.2053		& 2.5029	&	...	 & 3.072 		& 	3.06	&	3.09	&	1	&	2		&	0.014 & 44.07 & -0.08 & -0.28 & 0.12 & 20.0\\
2& 	150.2089		& 2.4819	& 3.333 &  3.378		&	3.37	&	3.39	& 	1	&	1	 & 0.024	 & 44.37		 & -0.49	 & -0.62		 & -0.35 & 0.1\\
2& 	150.2090		& 2.4385	& 3.715 & 3.563		&	3.55	&	3.57	& 1	&	1	& 0.010	 & 44.09		 & ...	 & ... & -0.717 & $<0.1$\\	
2& 	150.2108		 & 2.3915 & 3.095 &  3.085		&	3.07 &	3.1	& 	1	 & 	2	&	0.009 & 43.88	 & -0.17	 & -0.35	 & 0.02 & 13.0\\
2&	150.2142		 & 2.4750 & ...	 & 3.075 		&	3.06	&	3.09	 & 	1	&	2	 &	 0.020	 & 44.24		 & -0.07 & -0.17	 & 0.03 & 20.0\\
2& 150.2146	 & 2.5827	 & 5.3 & 5.201 & 	5.15	&	5.28	& 	2	&	2	 & 0.009	 & 44.42		 &	...	 & 	... &0.36 & $<500.0$\\
2& 150.2259	 & 1.7999	 & ...	 & 3.264 & 	3.12	&	3.46	 & 1		&	2	 & 0.004	 & 43.63	 	& 0.36 & 0.16	 & 0.55 & 68.0\\
2& 150.2478	 & 2.4422	 & 3.029 &  2.992  & 	2.97	 &	3.02	&	1	 &	2	 & 0.023	 & 44.31		 & 0.10 & 0.02	 & 0.18 & 35.0\\
2& 150.2595	 & 2.3761	 & 3.717 &  2.673	 & 2.66	&2.68	&	1 & 1	 & 0.008	 & 44.03		 & 	...	 & 	...	 & -0.68 & $<0.1$\\
2&	150.2608  & 2.1180	& ...	&	...	&	$>$3.0 &	...	&	...	&	...	&	-0.009	&	43.97	&	...	& ...	& ...	& ...	\\
2& 150.2618	 & 1.7316	 & ...	 & 3.126 & 	3.08	&	3.55	 & 2		 &	2	&	-0.005	 & 	43.67	 & 	... &  	... &	0.51 &  $>65.0$\\
2& 150.2674	 & 1.7009	 & ...      & 3.412 & 	3.33	&	3.46	 & 	1	&	1	 & 0.014 & 44.17 & ...	 & 	... & -0.45 & $<0.1$\\
2& 150.2684	&	1.7891	& ...	&	...	&	$>$3.0 &	...	&	...	&	2	&	0.010	&	44.02	&	0.72	& 0.48  & 0.87 & 1000.0\\
2& 150.2716	 & 1.6139	 & ...      & 3.466 & 	3.44	&	3.5		& 2	&	2	 & 0.011 & 44.14	 & 0.24 	& 0.0 & 0.49 & 60.0\\
2& 150.2848	&	1.8183	& ...	&	...	&	$>$3.0 &	...	&	...	&	2	&	0.011	&	44.06	&	-0.033	&  -0.18	&	0.25	&	12.0\\
2& 150.285	 & 2.309	 	& ...      & 2.897 & 	2.42	&	3.0		& 2	 &	1	&	 0.011	 & 43.94	 & -0.38 & -0.56	 & -0.20 & 1.2\\
2& 150.2881	 & 1.6511	 & ...      & 3.871 &  3.13	&	3.95	 & 1	 &	1	&	 0.018	 & 44.40	 & ...	 & 	... & -0.48 & $<0.1$\\
2& 150.2973	 & 2.1489	 & 3.328 & 3.459	&	3.45	&	3.47	 & 1	&	2	&	0.005 & 43.74	 & -0.11 & -0.42	 & 0.20 & 20.0\\
2& 150.2994	 & 1.6878	 & ...      & 2.976	& 	2.67	&	3.14	 &	1	&	1	 & 0.008	 & 43.83	 &	...	 & 	...  & -0.24 & $<9.0$\\

\hline
\end{tabular}
\end{minipage}
\end{table*}

\begin{table*}
\begin{minipage}{20.3cm}
\begin{tabular} {c c c c c c c c c c c c c c c c}
\hline
\hline
Survey & R.A.  & Dec. & $z_{spec}$ & $z_{phot}$ & $z_{phot}$	&	$z_{phot}$  &  Optical & X-ray &  $S_{\rm soft}$ & log$Lx$& HR& HR &  HR & $N_{\rm H}$\\

	& (deg)  & (deg) & &  & lower& upper  &  Type& Type &  & & &  lower &upper & \\

\hline
\hline
2& 150.3007	 & 2.3007	 & 3.498 & 3.434	&	3.41	&	3.46	& 2	 &	2	&	 0.003 & 43.55	& ...	 & 	... & 0.70 & $<1000.0$\\
2& 150.3048	 & 1.8242	 & ...      & 2.949 & 	2.49	&	3.35	&	1	 & 	1	&	0.006	 & 43.79	 & ...	 & 	... & -0.47 &$<0.1$\\
2& 150.3060	 & 1.7616	 & ...      & 3.265 & 	3.24	&	3.29	 & 	2	&	2	 & 0.003 & 43.46	 & ...	 & 	...  & 0.77 & $<1000.0$\\
2& 150.3100	&	2.5045	&  ...	&	...	&	$>$3.0 &	...	&	...	&	2	&	0.005	&	43.75	&	0.33	&	0.09	&  0.48	&	68.0\\
2& 150.3158	 & 2.3369	 & ...      & 4.216 & 	3.85	&	4.44	 & 2		 &	2	&	 0.002	 & 43.47	 & ...	 & 	...  & 0.76 & $<1000.0$\\
2& 150.3179	 & 2.0050	 & ...      & 3.428 & 	3.38	&	3.49	 & 	2	&	2	 & 	0.007	 & 43.86	 & ... & 	...	 & 0.03 & $<30.0$\\
2& 150.3187	&	2.2477	&	...	&	...	&	$>$3.0 &	...	&	...	&	2	&	0.008	&	43.88	&	0.06	&	-0.13	&	0.18	&	11.0\\
2& 150.3332	 & 2.4415	 & ...      & 2.481 &		2.12	&	3.07	 & 2		&	1	 & 0.015 & 44.09 & -0.54 & -0.76 & -0.33 & 0.1\\
2& 150.3344	 & 1.7882	 & ...      & 2.683 & 	2.34	&	3.02	 & 1		&	1	 & 0.005	 & 43.60		 & ... & ... & -0.24 & $<8.0$\\
2& 150.3443	 & 1.6361	 & ...      & 3.805 & 	3.74	&	3.85	 & 2		&	2	 & 0.019 & 44.41 & ... & ... & -0.08 & $<30.0$\\
2& 150.3451	 & 1.9579	 & ...      & 3.065 & 	2.98	&	3.17	& 2		&	2	 & 0.010	 & 43.93	 & 0.18 & -0.07 & 0.25 & 45.0\\
2& 150.3566	 & 2.2242	 & ...      & 2.701 & 	2.51	&	3.09	 & 2		& 	2	 &	-0.008	&	44.02	& ...	 & 0.56 & ...	 & $>75.0$\\
2& 150.3598	 & 2.0737	 & 4.917 & 1.991	&	1.96	&	2.01	 & 2	&2	&	 0.008	 & 44.30	 &	... & ... 	& 0.27 & $<70.0$\\
2& 150.3647	 & 2.1438	 & 3.328 & 3.361	&	3.35	&	3.37	& 1	&	1	 & 0.025	 & 44.40	 & -0.26 & -0.38 & -0.14 & 8.0\\
2& 150.3790	 & 1.8761	 & ...      & 3.33 & 	3.02	&	3.44	 & 	2	&	2	 & 0.009	& 43.96	 & -0.18 & -0.44 & 0.06 & 14.0\\
2& 150.3809	 & 2.0995	 & ...      & 4.498 & 4.43	&	4.56	 & 	2	&	2 	& 	-0.008	 & 	44.39	 & ... & 0.514 & ... & $>200.0$\\
2& 150.3836	 & 2.0748	 & ...      & 3.852 & 	3.83	&	3.88	 &2		&	2	 & 0.003	 & 43.62	 & 0.40 & 0.13 & 0.68 & 90.0\\
2& 150.3989	&	2.2709	& ...	&	...	&	$>$3.0 &	...	&	...	&	2	&	0.005	&	43.72	&	0.46	&	0.24	&	0.61	&	200.0\\
2& 150.4153	 & 1.9342	 & ...      & 3.681 & 1.28	&	3.75		 & 1		&	1	 & 0.011 & 44.15	 & 	...	&	...	 & -0.54 & $<0.1$\\
2& 150.4155	 & 2.3648	 & ...      & 2.423 	& 	2.24	&	3.29		 & 	1	&	1	 & 0.003 		& 43.47 	& 	...	&	...	 	& -0.29 & $<6.0$\\
2& 150.4251	 & 2.3120	 & ...      & 3.092 	& 	2.98	&	3.29	 	& 	2	&	2	 & 0.003		 & 43.42	 & 0.21 & -0.05 & 0.48 & 48.0\\
2& 150.4550	 & 1.9674	 & 3.485 & 	3.493	&	3.46	&	3.52	 	& 	1	&	1 	& 0.007		& 43.89	 & 	...	&	...	 & -0.46 & $<0.1$\\
2& 150.4679	 & 2.5315 	& ...      & 4.45 	& 	4.43	 	&	4.47	&	2	&	1	 & 0.014 		& 44.46	 & 	...	&	...	 & -0.563 & $<0.1$\\
2& 150.4851	 & 2.4135	 & ...      & 2.949	&	2.57 	& 	3.51	 	& 	2	&	2	 & -0.006	& 	44.04	 & 	...	& 0.14 & ...	 & $>25.0$\\
2& 150.4862	 & 2.4281	 & ...      & 3.464 	& 	2.95	&	3.61	 	&	2	 & 	2	 &	-0.005	&	43.77	 & ...	 & 0.07 & 	... & 32.0\\
2& 150.5467	 & 2.2243	 & ...      & 3.506 	& 	3.35	&	3.66	 	& 	2	&	2	&	0.029	 & 44.53	 & -0.08 & -0.21 & 0.03	 & 25.0\\
2& 150.5512	 & 2.1400	 & ...	 &3.028	&	2.97	 	&	3.07	&	2	 & 	2	 &  	-0.024	 & 44.50	 &	...	&	0.67	&	... & $\ge$1000.0\\
2& 150.6475	 & 2.4182	 & ...      & 3.034 	& 	2.6		&	3.32		& 	1	&	2	 & -0.052	& 	44.84	 & 	...	 & 0.69 & ... & $\ge$1000.0\\
\hline

\end{tabular}
\label{Table:Total_properties}
\end{minipage}
\end{table*}

\bsp

\label{lastpage}

\end{document}